\documentclass[structabstract]{my_style}  
\usepackage{graphicx}
\usepackage{subcaption}
\usepackage{color}
\usepackage{url}
\usepackage{amssymb}
\usepackage{amsmath}
\usepackage{bm}
\usepackage{txfonts}
\usepackage{multirow}
\usepackage{lscape}
\usepackage[flushleft]{threeparttable}
\usepackage{float}
\usepackage{natbib}
\usepackage{stmaryrd}
\usepackage{supertabular}
\usepackage{rotating}
\usepackage{longtable}
\usepackage{chngcntr}
\usepackage{changes}
\usepackage{pdflscape}
\usepackage{booktabs}
\usepackage{dcolumn}
\usepackage{gensymb}
\usepackage{tabularx}

\bibpunct{(}{)}{;}{a}{}{,}

\makeatletter
\def\hlinewd#1{%
\noalign{\ifnum0=`}\fi\hrule \@height #1 %
\futurelet\reserved@a\@xhline}
\makeatother

\newfont{\gwpfont}{cmssq8 scaled 1000}
\newcommand{\rexcess}{{\gwpfont REXCESS}}
\newcommand{\reflex}{{\gwpfont REFLEX}}

\newcolumntype{Y}{>{\centering\arraybackslash}X}

\begin{document}
   \title{Substructures in galaxy clusters: a comparative X-ray and photometric analysis of the REXCESS sample}

   \author{
          G. Fo\"ex\inst{1}
          }
   \institute{
          foex.gael[@]gmail.com
             }

  \abstract
  {}
  {The main goals of the present work are (i) to study the substructure content of a representative, X-ray selected sample of 31 galaxy clusters, as traced by the spatial distribution of their red-sequence galaxies, and (ii) to compare it to that observed in the intracluster medium distribution.}
    {Our substructure indicators are the asymmetry test, $\beta$, the residuals of the galaxy surface density map, $\Delta_\Sigma$, and the Fourier elongation, $FE$. We probe the clusters core with secondary tests: the offset between the central brightest cluster galaxy and the X-ray emission peak, $\Delta r_{\mathrm{BCG-X}}$, the magnitude offset between the first and second brightest galaxies, $\Delta m_{12}$, and their radial offset, $\Delta r_{12}$.}
    {The main indicators exhibit continuous distributions, allowing to define a ranking order but making a discrete classification difficult. A partition based on $\beta$ and $\Delta_\Sigma$ leads to a fraction $\sim35\%$ of disturbed systems; $\sim65\%$ of the clusters are disturbed according to at least one of these two quantities. The main indicators poorly correlate with the secondary quantities, likely due to substructures observed prior core crossing. Nine of the 12 X-ray disturbed clusters are also flagged as such by either $\beta$ or $\Delta_\Sigma$. Cool cores are hosted by systems with higher optical concentrations, smaller $\Delta r_{\mathrm{BCG-X}}$, and larger $\Delta r_{12}$; however, both populations do not differ significantly in terms of their overall morphology probed by the main indicators.}
  {Our results show the necessity of using a variety of independent tests and data sets to obtain a clear picture of a cluster's morphology. Furthermore, we find that a cluster's dynamical state, consequence of its recent merger history, is not necessarily representative of a future mass assembly via accretion of substructures. Conducting a similar multi-wavelength analysis on a larger sample could provide interesting constraints on the fraction of morphologically disturbed systems in the general population of galaxy clusters.}


   \maketitle

%

\section{Introduction}

Within the $\Lambda$CDM paradigm, galaxy clusters form hierarchically, through a smooth accretion of surrounding material and the merger of smaller scale structures (e.g. \citealt{colberg99,moore99,evrard02,springel05,springel06}). Their evolution via the second channel leads to systems characterised by a complex morphology of their mass, galaxy members, and intra cluster medium (hereafter ICM) spatial distributions. In the case of major mergers, massive substructures can produce significant departures from dynamical equilibrium once reaching the core of their host, a striking example being provided by the Bullet cluster (e.g. \citealt{markevitch04}).

Numerical simulations of structure formation and evolution make a series of predictions regarding the physical properties of substructures, such as their mass spectrum, e.g. \cite{delucia04,gao04,giocoli08,giocoli10,gao12,contini12}, their spatial distribution, e.g. \cite{ghigna00,delucia04,diemand04,nagai05}, or their contribution to the mass growth of their host, e.g. \cite{genel10}. These predictions must be compared to observational results, in order to test the validity of the different recipes involved in the modelling of structure formation. Observationally, \cite{haines17} in X-rays, \cite{lemze13} with dynamics, \cite{grillo15} in the strong-lensing regime, or \cite{guennou14} with weak lensing have provided valuable insight on the properties of substructures in galaxy clusters. However, this type of studies is rather limited due to the difficulty of identifying (e.g. projection effects or detection thresholds) and characterising (mass estimates) substructures. On the contrary, there has been numerous works focusing on estimating the occurrence of disturbed systems, either via X-ray (e.g. \citealt{mohr95,schuecker01,jeltema05,santos08,boehringer10,mann12}), optical (e.g. \citealt{plionis02,flin06,ramella07,einasto12,foex13,wen13}), lensing (e.g. \citealt{dahle02,smith05,martinet16}), or dynamical analyses (e.g. \citealt{girardi97,solanes99,oegerle01,aguerri10,einasto12}). The proportion of clusters hosting a substantial amount of substructures is found in the rather wide range $\sim20\%-70\%$, depending on the study. This variety results mostly from the fact that there is no unique way to introduce the parameter threshold for a discrete classification into relaxed and disturbed clusters. In addition, different methods highlight different morphologies and selection biases also play a role.

It is within the focus of understanding the origin of these discrepancies that we propose here to compare the substructure content of a cluster sample as derived from two different data sets, using various substructure indicators. To achieve this, we present in this work a comparative X-ray and photometric morphological analysis of \rexcess\ (Representative {\it XMM-Newton} Cluster Structure Survey, \citealt{bohringer07}). Our main objectives are (i) to apply several statistical tests on the spatial distribution of cluster galaxies to infer the substructure content of their hosts, and (ii) to compare these optical results with those obtained by \cite{boehringer10} from an X-ray analysis of the ICM morphology. Our work will contribute to obtain a comprehensive picture of an already very well studied cluster sample, with results obtained by \cite{pratt07} for the temperature profiles, \cite{croston08} for the gas density profiles, \cite{pratt09} for scaling relations based on the X-ray luminosity, \cite{arnaud10} for the pressure profiles and scaling relations based on the X-ray equivalent of the Sunyaev-Zeldovich Effect Comptonization parameter $Y_\mathrm{X}$, \cite{pratt10} for the entropy profiles, and \cite{haarsma10} for the properties of the brightest cluster galaxies (hereafter BCG). 

The paper is organised as follows. In Section 2 we briefly recall the main properties of \rexcess\ and we described the photometric data at the basis of our analysis. The definition of the different substructure indicators used to study the clusters, their observed distributions, and their cross-correlations are presented in Section 3. We compare the substructure content of the ICM and cluster galaxies distributions in Section 4, along with a discussion of the photometric properties of cool-core clusters. We summarise our conclusions in Section 5. All our results are scaled to a flat, $\Lambda$CDM cosmology with $\Omega_{m}=0.3$, $\Omega_{\Lambda}=0.7$ and a Hubble constant $H_{0}=70\,\mathrm{km\,s^{-1}\,Mpc^{-1}}$.

\section{Data}

\subsection{The \rexcess\ sample}
The 33 \rexcess\ galaxy clusters were drawn from the \reflex\ survey \citep{bohringer01,bohringer04} to homogeneously cover the luminosity range 0.4 to $20\times10^{44}\,h_{50}^{-1}\mathrm{erg\,s^{-1}}$ in the 0.1 to 2.4 keV band, corresponding to ICM temperatures $\sim2-10$ keV.  The redshift range, $z=0.0564$ to 0.1832, was chosen such that the clusters fit into the field of view of {\it XMM-Newton} while leaving a sufficient cluster-free area for a detailed background modelling. With its selection based on X-ray luminosities only, \rexcess\ should be representative of any local, unbiased X-ray survey. More details about the sample selection and the X-ray observations are available in \cite{bohringer07}. As in previous studies based on \rexcess, two targets were excluded from the original sample due to their complex morphology: the supercluster Abell 901/902 (RXCJ0956.4-1004) and the bimodal system RXCJ2152.2-1942. Another bimodal system, RXCJ2157.4-0747, was kept in the sample, since its two components can be well separated.

\subsection{Photometric observations}

The optical data used in this work were obtained with the Wide Field Imager (WFI; \citealt{baade99}) mounted on the Cassegrain focus of the ESO/MPG 2.2 m telescope at La Silla, Chile. The WFI is a mosaic camera composed of $4\times2$ CCD chips, each made of $2048\times4096$ pixels with an angular resolution of $0.238''$/pixel. The total FOV is $34'\times33'$, thus covering a circular area of diameter $\sim3.6$ Mpc at a redshift z=0.1. The clusters were observed in the B, V, and R passbands with exposure times of at least half an hour; the typical seeing in the R band is $0.7''-1''$.

The data reduction was conducted with the THELI pipeline \citep{schirmer13}, which performs internally the basic pre-processing steps (bias subtraction, flat-fielding, background modelling and sky subtraction) and uses third-party softwares for the astrometry ({\sc Scamp}, \citealt{bertin06}) and the co-addition of mosaic observations ({\sc SWarp}, \citealt{bertin10}). The photometric calibration was carried out for each cluster individually with observations of standard stars \citep{landolt92}. It provided the zero points and colour terms required to transform WFI's raw magnitudes into the Johnson photometric system.

The photometric catalogues were constructed with {\sc SExtractor} \citep{bertin96}, ran in dual mode, using the R band for the object detection. Stars, galaxies, and false detections were sorted according to their position in the magnitude/central flux diagram, their size with respect to that of the PSF, and their stellarity index (CLASS\_STAR parameter). In the following, magnitudes correspond to the MAG\_AUTO parameter, while colours were computed with MAG\_APER, measured in a fixed aperture of $3''$ on the PSF-matched images; the images were degraded to the worst seeing (usually the B band) with the IRAF task {\tt psfmatch}.

\begin{figure}
\center
\includegraphics[width=6.cm, angle=-90]{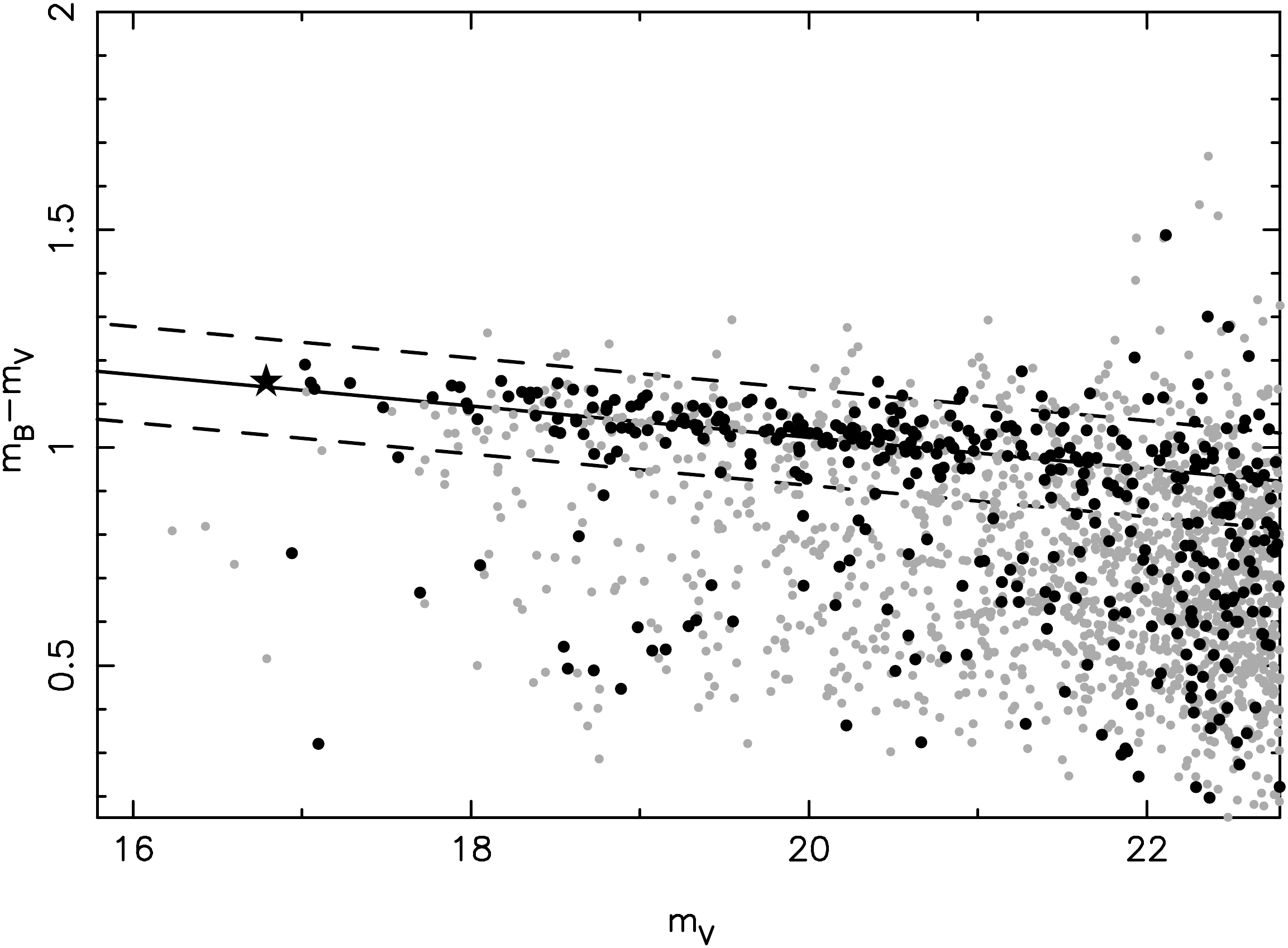}\\
\includegraphics[width=6.cm, angle=-90]{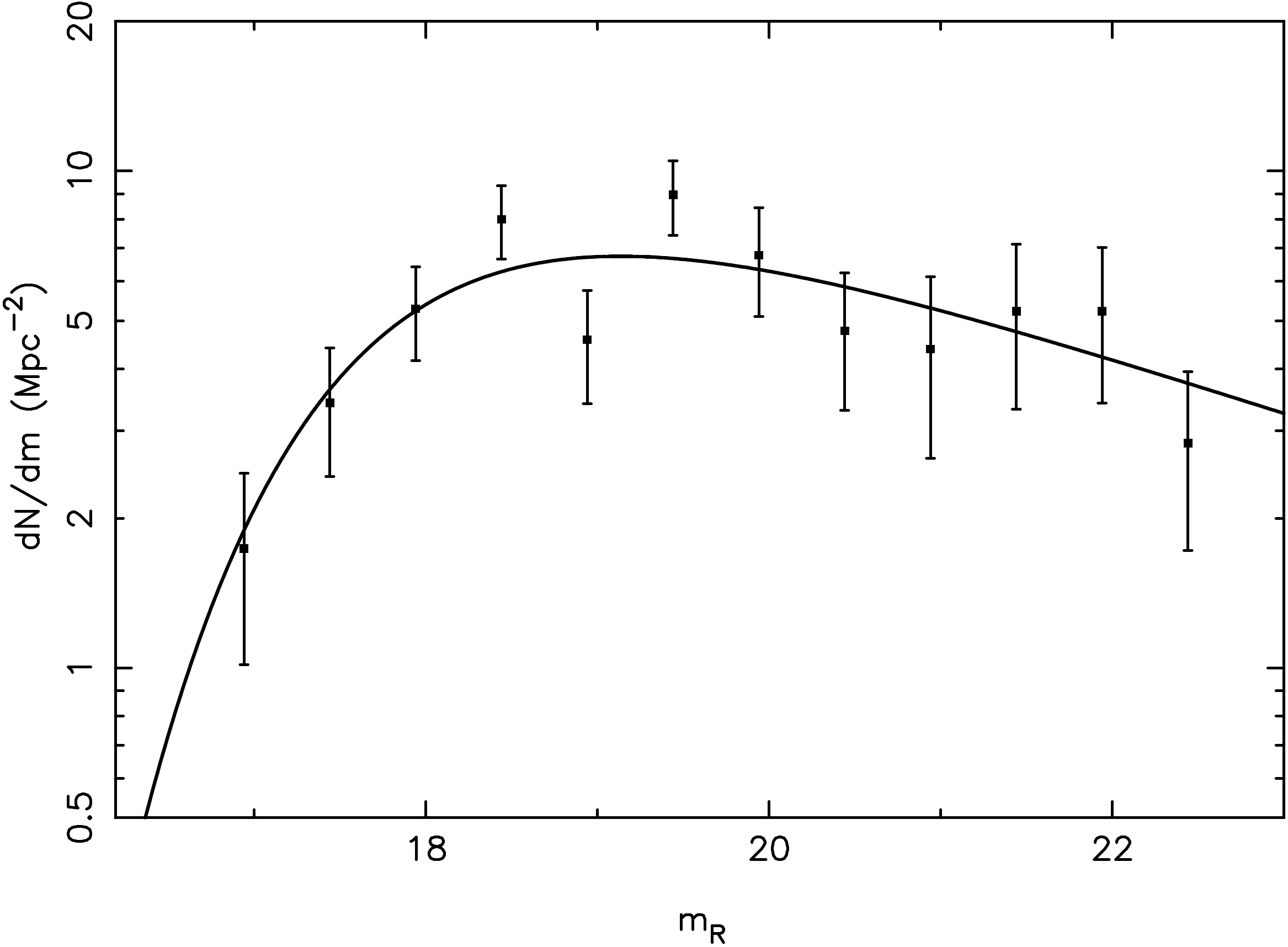}
\caption{Top panel: red sequence of RXCJ2234.5-3744. The grey and black points are the galaxies within $1.5R_{500}$ and $0.5R_{500}$ respectively. The star is the central BCG. The solid line traces the best-fit of the red sequence, whose width, equal to three times the dispersion in colour of the selected galaxies, is shown by the two dashed lines. Bottom panel: background-corrected number counts (with Poisson error bars) per magnitude bin for RXCJ2234.5-3744. The curve is the best-fit Schechter luminosity function, for which $m^*=17.8$.}
\label{fig:RS_LF} 
\end{figure}

To reduce the contamination of the photometric catalogues by foreground or background structures, we restricted our analysis to the red-sequence galaxies. The red sequences were fitted in the $(m_B-m_V)-m_V$ diagram using a $2\sigma$-clipping method; their width are typically in the range 0.07-0.15 mag. Furthermore, we cut the catalogues to $m<m^*+3$, since at faint magnitudes the red sequence no longer provides a secure criterion to select cluster members. The magnitudes $m^*$ were obtained by fitting a standard Schechter luminosity function to the galaxy counts per magnitude bins of size 0.5 mag, within $R_{500}$\footnote{$R_{500}$ is defined as the cluster radius within which its mean total mass density is 500 times the critical density of the Universe at the cluster redshift; the values used in this work were obtained by \cite{pratt09}.}, and corrected from a background contribution estimated beyond $1.5R_{500}$. We provide in Figure \ref{fig:RS_LF} an example of a red sequence and luminosity function. The number of red-sequence galaxies within $R_{500}$, including the background contribution, are given in Table \ref{table:sub}; they are in the range $N\sim50-370$ with a median value of $N\sim130$ galaxies. 

\section{Substructures in the cluster galaxies distribution}
\label{sec:indic}

\cite{Pinkney96} have studied the performance of several statistical tests to detect the presence of substructures in a cluster. According to their findings, the two most sensitive tests are the Fourier elongation (hereafter $FE$; \citealt{rhee91}) and the $\beta$ test \citep{west88}. We also performed an additional substructure test by computing the residuals between the galaxy surface density and its best-fit model, $\Delta_\Sigma$; a similar approach was used, for instance, by \cite{wen13}. These three statistics, which are described in more details below, are our primary tools to investigate the structure of the clusters; the results of these methods are summarised in Table \ref{table:sub}.

\begin{table*}
\centering 
\begin{threeparttable}
\caption{Values of the photometric substructure indicators for the 31 clusters.}
\label{table:sub}
\begin{tabular}{l c c c c c c c c c c}
\hline\hline\noalign{\smallskip}
Cluster & $N_{500}$ & $\beta$ & $\Delta_\Sigma$ & $FE$ & $\Delta r_{\mathrm{BCG-X}}$ & $\Delta m_{12}$ & $\Delta r_{12}$ & $c_{500}$ & CC & D\\
\noalign{\smallskip}\hline\noalign{\smallskip}
RXCJ0003.8+0203 & 83 & $0.001\pm0.028$ & $0.062\pm0.027$ & $-0.36\pm0.62$ & $0.004$ & $1.64$ & $0.24$ & $3.6\pm2.0$ & ... & ...\\
RXCJ0006.0-3443 & 219 & $0.041\pm0.014$ & $0.108\pm0.016$ & $2.72\pm0.69$ & $0.012$ & $1.39$ & $0.46$ & $1.9\pm0.5$ & ... & $\checkmark$\\
RXCJ0020.7-2542 & 216 & $0.025\pm0.013$ & $0.082\pm0.014$ & $3.61\pm0.70$ & $0.037$ & $0.00$ & $0.02$ & $3.6\pm0.9$ & ... & ...\\
RXCJ0049.4-2931 & 60 & $0.002\pm0.033$ & $0.048\pm0.021$ & $0.50\pm0.64$ & $0.005$ & $1.01$ & $0.01$ & $6.4\pm1.8$ & ... & ...\\
RXCJ0145.0-5300 & 173 & $0.005\pm0.015$ & $0.065\pm0.015$ & $4.57\pm0.67$ & $0.078$ & $0.23$ & $0.11$ & $3.4\pm0.8$ & ... & $\checkmark$\\
RXCJ0211.4-4017 & 57 & $0.032\pm0.033$ & $0.036\pm0.021$ & $0.29\pm0.66$ & $0.006$ & $1.18$ & $0.06$ & $7.7\pm2.5$ & ... & ...\\
RXCJ0225.1-2928 & 45 & $0.030\pm0.047$ & $0.060\pm0.029$ & $2.14\pm0.68$ & $0.007$ & $0.02$ & $0.06$ & $10.7\pm4.2$ & ... & $\checkmark$\\
RXCJ0345.7-4112 & 83 & $0.050\pm0.032$ & $0.061\pm0.023$ & $0.99\pm0.64$ & $0.004$ & $1.72$ & $0.41$ & $4.1\pm2.2$ & $\checkmark$ & ...\\
RXCJ0547.6-3152 & 292 & $0.013\pm0.011$ & $0.120\pm0.010$ & $1.01\pm0.64$ & $0.033$ & $0.60$ & $0.13$ & $1.8\pm0.4$ & ... & ...\\
RXCJ0605.8-3518 & 172 & $0.045\pm0.014$ & $0.095\pm0.016$ & $3.48\pm0.63$ & $0.005$ & $1.53$ & $0.49$ & $6.4\pm2.4$ & $\checkmark$ & ...\\
RXCJ0616.8-4748 & 91 & $0.026\pm0.027$ & $0.088\pm0.028$ & $1.03\pm0.65$ & $0.006$ & $2.19$ & $0.40$ & $1.2\pm0.2$ & ... & $\checkmark$\\
RXCJ0645.4-5413 & 255 & $0.043\pm0.013$ & $0.085\pm0.013$ & $2.64\pm0.62$ & $0.009$ & $0.92$ & $0.33$ & $3.5\pm0.6$ & ... & ...\\
RXCJ0821.8-0112 & 144 & $0.032\pm0.020$ & $0.072\pm0.017$ & $1.94\pm0.68$ & $0.010$ & $1.05$ & $0.31$ & $2.4\pm0.9$ & ... & ...\\
RXCJ0958.3-1103 & 129 & $0.020\pm0.018$ & $0.069\pm0.017$ & $3.11\pm0.65$ & $0.015$ & $0.51$ & $0.25$ & $2.2\pm0.7$ & $\checkmark$ & ...\\
RXCJ1044.5-0704 & 144 & $0.003\pm0.017$ & $0.087\pm0.021$ & $2.91\pm0.70$ & $0.006$ & $0.69$ & $0.37$ & $2.5\pm0.7$ & $\checkmark$ & ...\\
RXCJ1141.4-1216 & 85 & $0.013\pm0.026$ & $0.052\pm0.019$ & $0.67\pm0.63$ & $0.008$ & $0.80$ & $0.35$ & $5.6\pm2.1$ & $\checkmark$ & ...\\
RXCJ1236.7-3354 & 99 & $0.009\pm0.023$ & $0.038\pm0.018$ & $-0.29\pm0.64$ & $0.010$ & $0.37$ & $0.10$ & $2.5\pm0.7$ & ... & ...\\
RXCJ1302.8-0230 & 97 & $0.014\pm0.024$ & $0.063\pm0.018$ & $1.76\pm0.63$ & $0.021$ & $0.71$ & $0.50$ & $12.5\pm6.6$ & $\checkmark$ & $\checkmark$\\
RXCJ1311.4-0120 & 371 & $0.004\pm0.010$ & $0.058\pm0.010$ & $2.99\pm0.66$ & $0.001$ & $0.60$ & $0.13$ & $4.6\pm0.7$ & $\checkmark$ & ...\\
RXCJ1516.3+0005 & 146 & $0.026\pm0.019$ & $0.090\pm0.018$ & $2.76\pm0.68$ & $0.016$ & $0.69$ & $0.24$ & $3.0\pm0.9$ & ... & ...\\
RXCJ1516.5-0056 & 117 & $0.005\pm0.021$ & $0.121\pm0.023$ & $3.19\pm0.62$ & $0.007$ & $1.04$ & $0.44$ & $6.5\pm5.3$ & ... & $\checkmark$\\
RXCJ2014.8-2430 & 188 & $0.017\pm0.015$ & $0.092\pm0.018$ & $1.31\pm0.63$ & $0.006$ & $2.10$ & $0.28$ & $4.5\pm1.6$ & $\checkmark$ & ...\\
RXCJ2023.0-2056 & 57 & $0.030\pm0.042$ & $0.112\pm0.035$ & $1.23\pm0.65$ & $0.006$ & $0.73$ & $0.36$ & $2.0\pm1.0$ & ... & $\checkmark$\\
RXCJ2048.1-1750 & 223 & $0.033\pm0.014$ & $0.086\pm0.016$ & $1.50\pm0.62$ & $0.327$ & $0.26$ & $0.41$ & $1.6\pm0.3$ & ... & $\checkmark$\\
RXCJ2129.8-5048 & 110 & $0.076\pm0.025$ & $0.156\pm0.021$ & $0.97\pm0.67$ & $0.058$ & $0.49$ & $0.10$ & $3.0\pm1.3$ & ... & $\checkmark$\\
RXCJ2149.1-3041 & 123 & $0.003\pm0.021$ & $0.069\pm0.021$ & $1.32\pm0.63$ & $0.005$ & $0.95$ & $0.46$ & $2.5\pm0.7$ & $\checkmark$ & ...\\
RXCJ2157.4-0747 & 90 & $0.054\pm0.028$ & $0.069\pm0.028$ & $3.23\pm0.66$ & $0.016$ & $0.01$ & $0.09$ & $3.4\pm1.9$ & ... & $\checkmark$\\
RXCJ2217.7-3543 & 214 & $0.003\pm0.013$ & $0.056\pm0.014$ & $1.16\pm0.65$ & $0.011$ & $1.15$ & $0.01$ & $2.9\pm0.6$ & ... & ...\\
RXCJ2218.6-3853 & 192 & $0.006\pm0.015$ & $0.059\pm0.015$ & $5.04\pm0.62$ & $0.025$ & $0.96$ & $0.02$ & $4.6\pm1.2$ & ... & $\checkmark$\\
RXCJ2234.5-3744 & 305 & $0.065\pm0.011$ & $0.077\pm0.013$ & $-0.33\pm0.69$ & $0.110$ & $0.23$ & $0.03$ & $3.9\pm0.7$ & ... & ...\\
RXCJ2319.6-7313 & 68 & $0.024\pm0.030$ & $0.114\pm0.031$ & $1.14\pm0.67$ & $0.005$ & $0.33$ & $0.42$ & $9.8\pm7.7$ & $\checkmark$ & $\checkmark$\\
\noalign{\smallskip}\hline
\end{tabular}
    \begin{tablenotes}
      \small
      \item Columns: (1) Cluster name. (2) Number of red-sequence galaxies within $R_{500}$, including the background contribution. (3) Asymmetry. (4) Residuals of the galaxy surface density map. (5) Fourier elongation; negative values arise from the bias subtraction. (6) Offset between the central BCG and the X-ray emission peak, normalised by $R_{500}$. (7) Magnitude gap between the central BCG and the second brightest galaxy within $0.5R_{500}$. (8) Offset between the central BCG and the second brightest galaxy, normalised by $R_{500}$. (9) Optical concentration. (10) Systems classified as cool cores on the basis of central density vs. cooling time (see \citealt{pratt09}). (11) Systems classified as disturbed on the basis of the X-ray centroid shift parameter $\omega$ (see \citealt{pratt09}).
    \end{tablenotes}
  \end{threeparttable}
\end{table*}

\begin{figure*}
\center
\includegraphics[width=4.2cm, angle=-90]{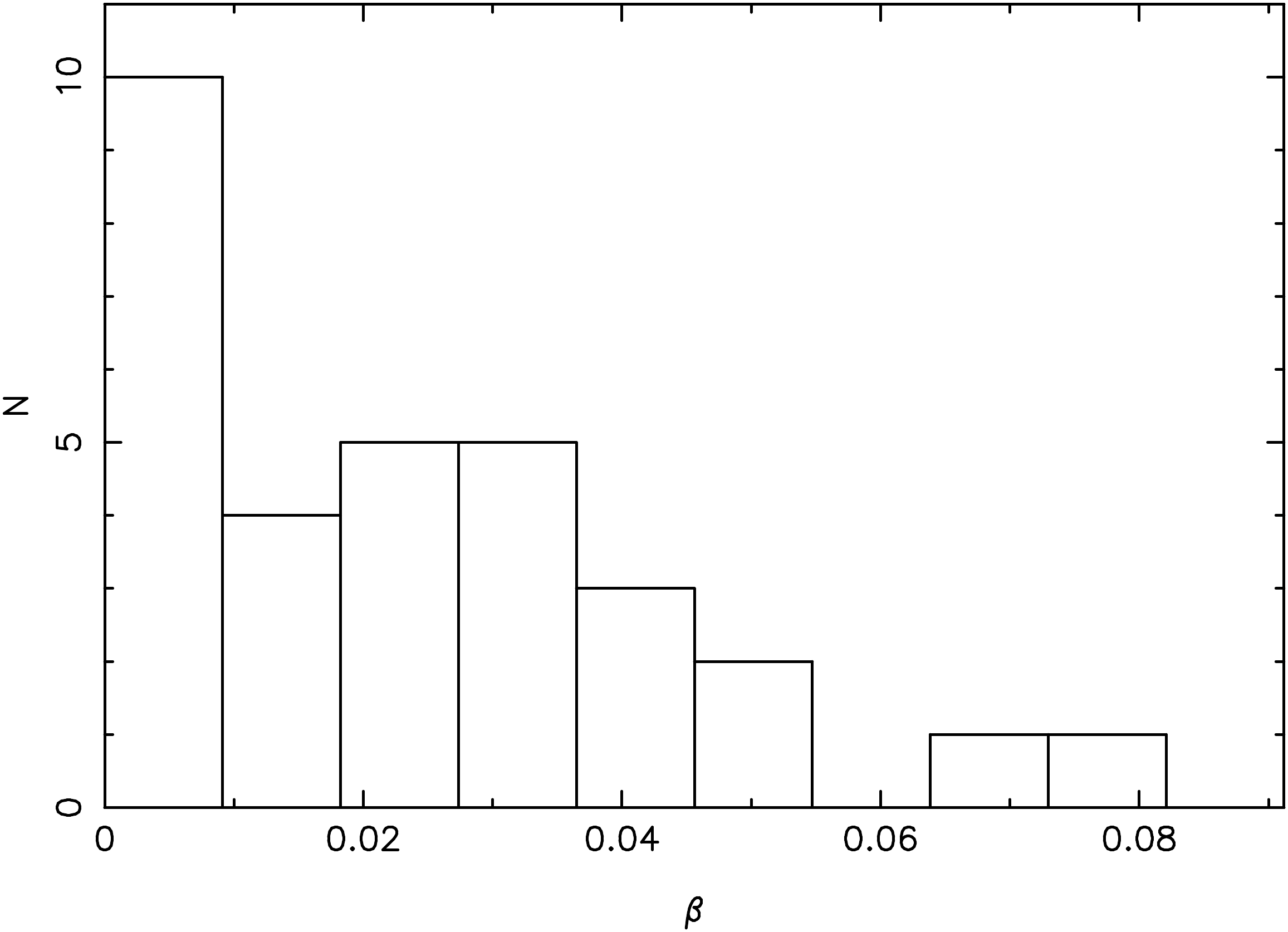}
\includegraphics[width=4.2cm, angle=-90]{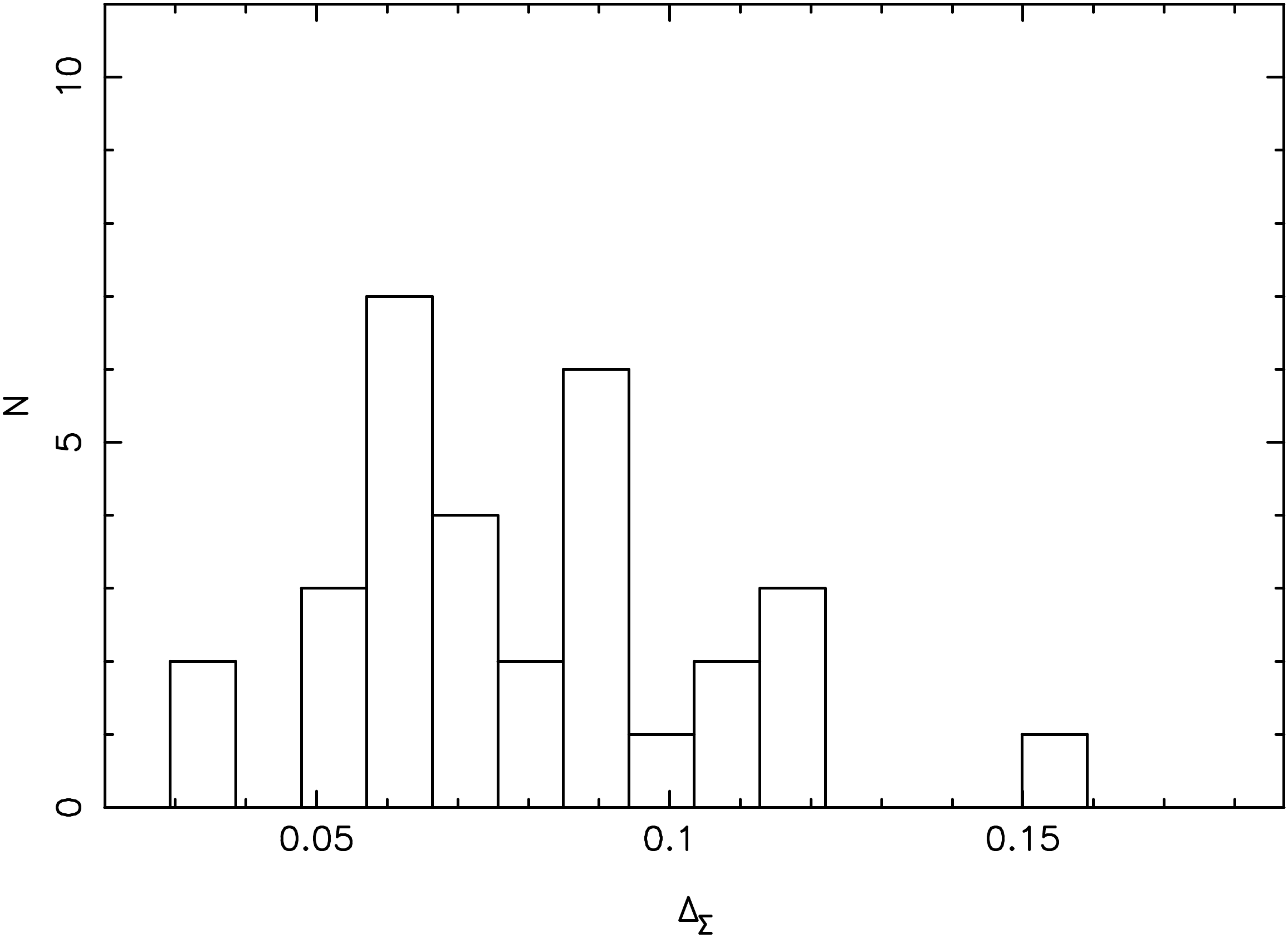}
\includegraphics[width=4.2cm, angle=-90]{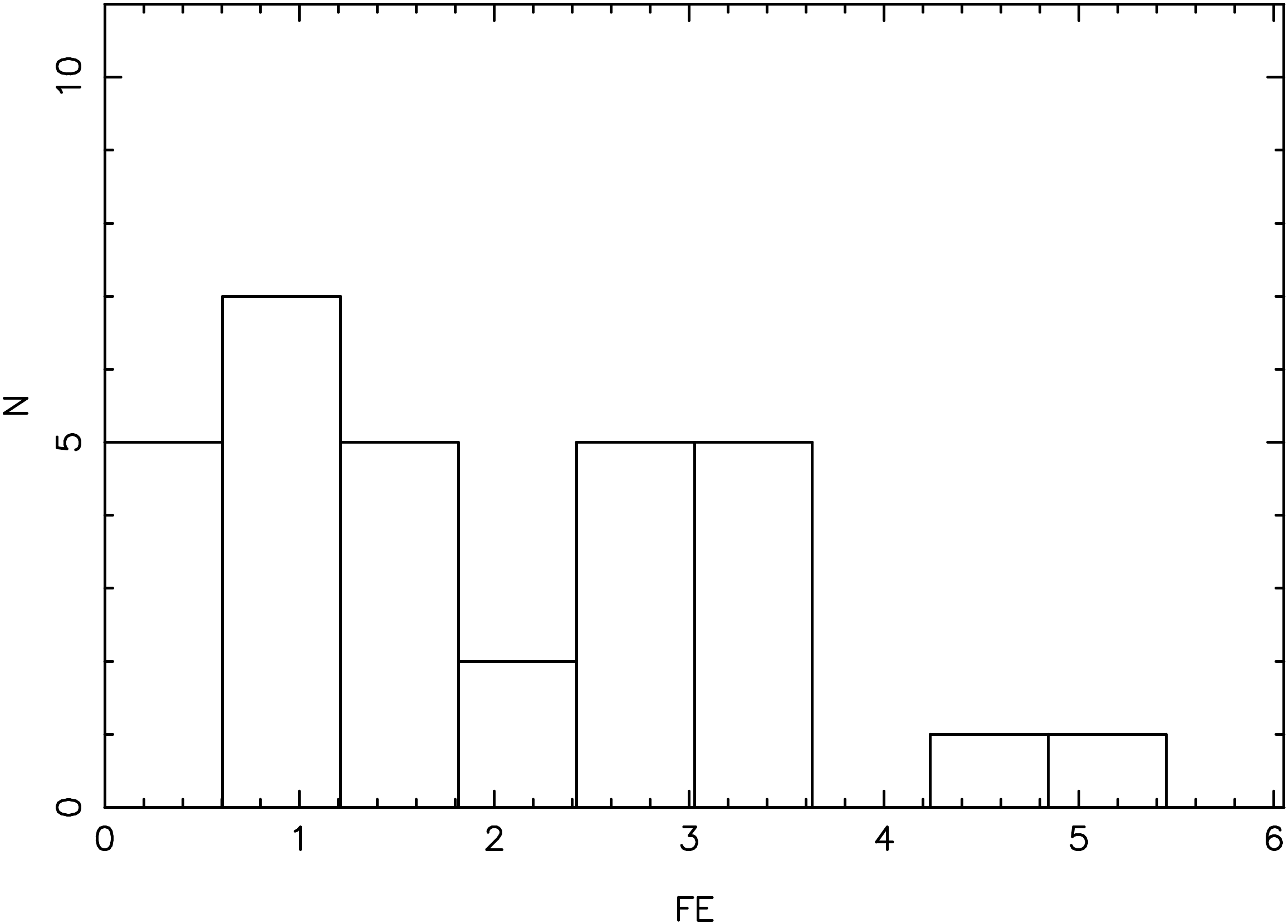}\\[3pt]
\includegraphics[width=4.2cm, angle=-90]{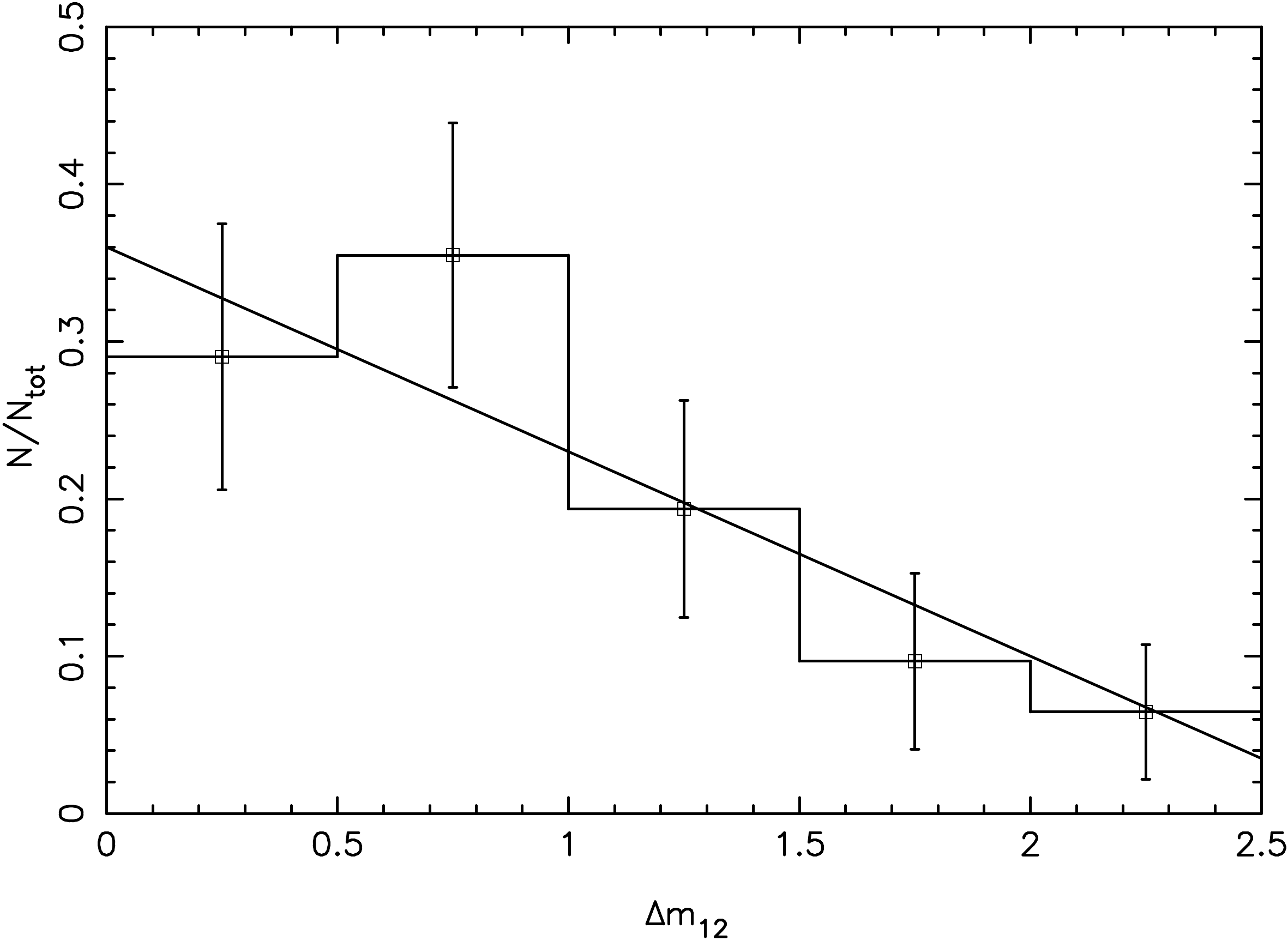}
\includegraphics[width=4.2cm, angle=-90]{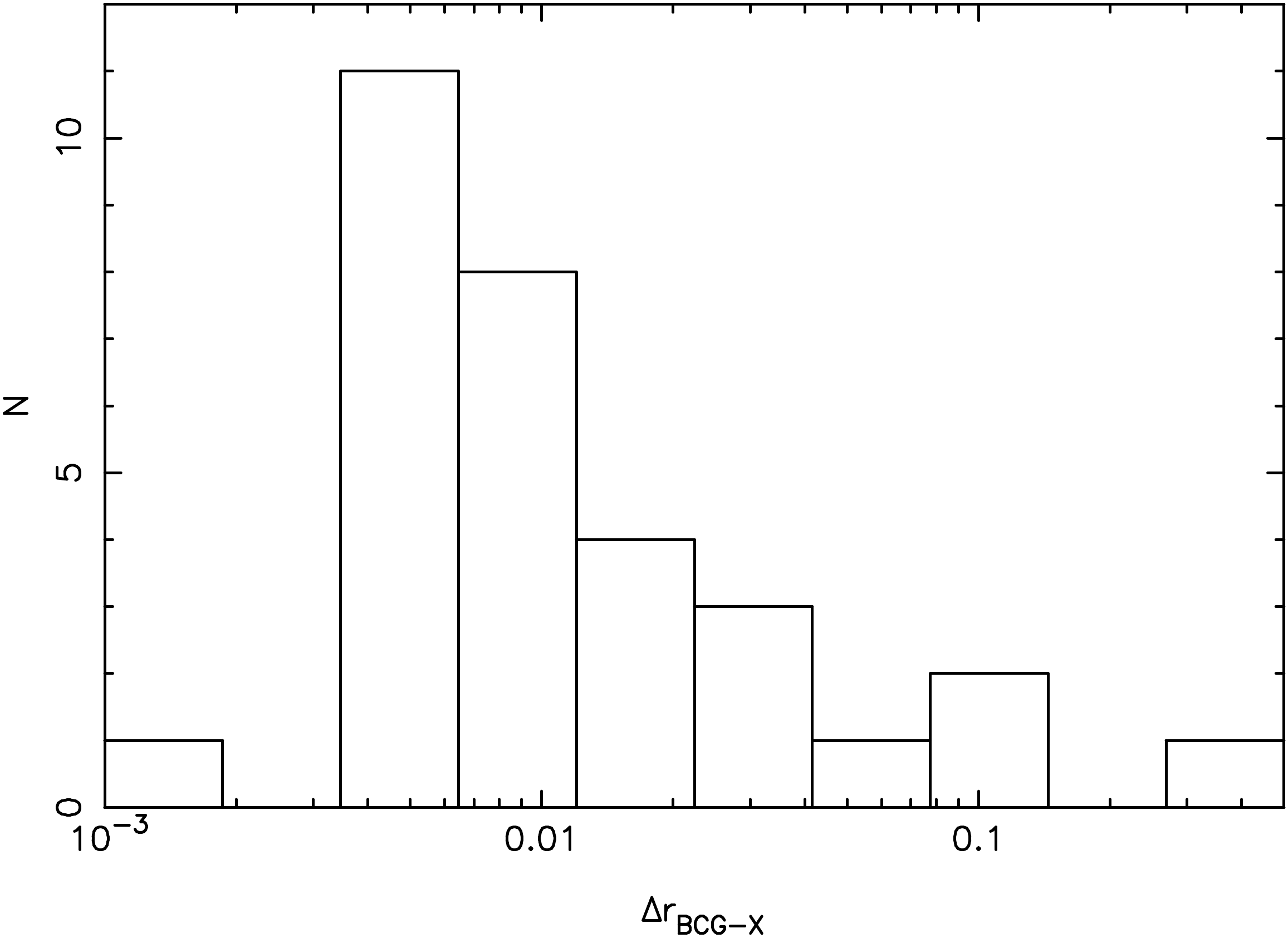}
\includegraphics[width=4.2cm, angle=-90]{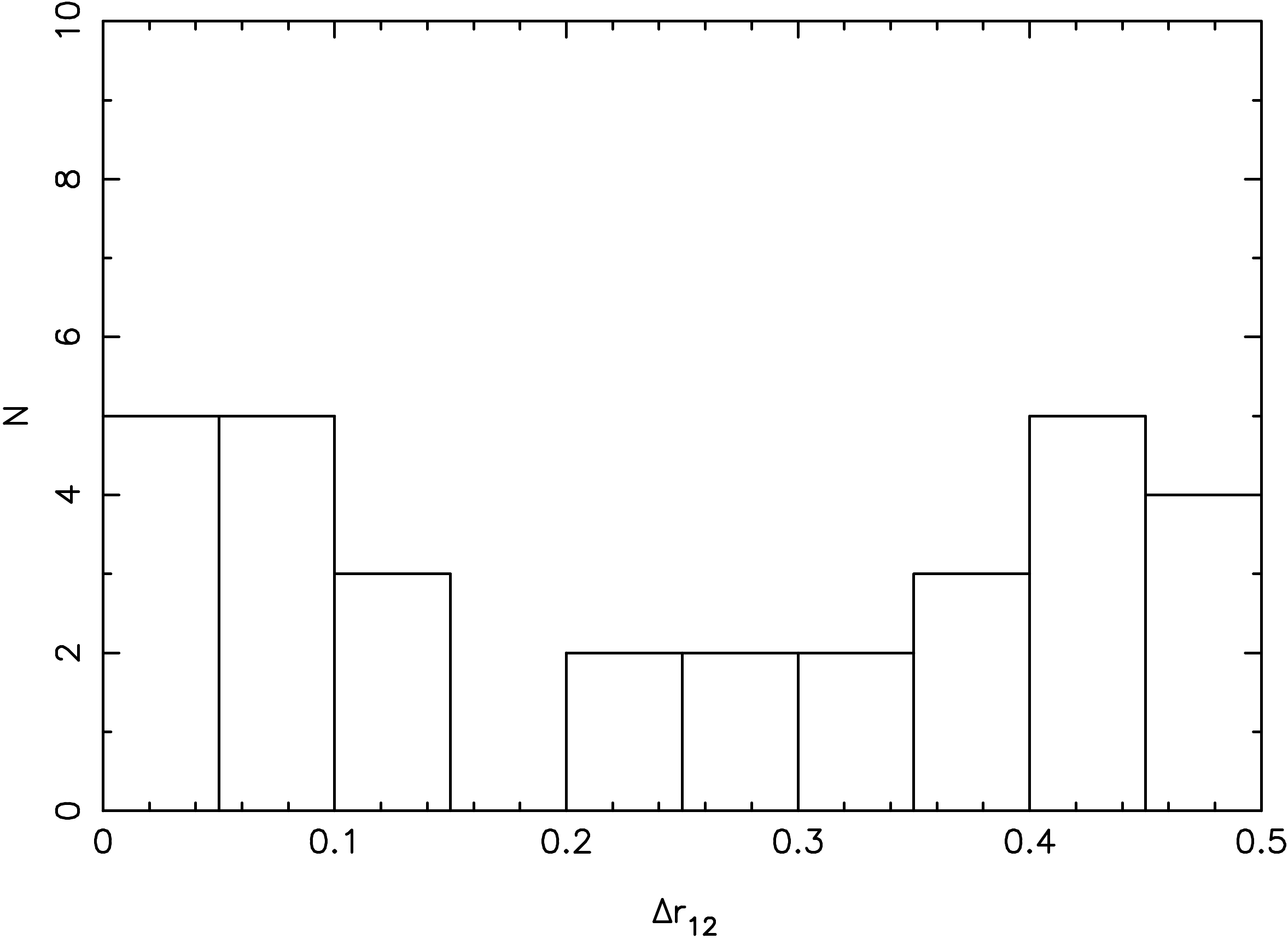}
\caption{Distribution of the asymmetry parameter $\beta$ (top-left), Fourier elongation $FE$ (top-middle), and residuals indicator $\Delta_\Sigma$ (top-right). Bottom-left: magnitude gap $\Delta m_{12}$ in bins of 0.5 mag. The vertical bars show the Poisson error in each bin. The solid line is the best fit obtained by \cite{smith10} for the LoCuSS sample. Bottom-centre: distribution of $\Delta r_{\mathrm{BCG-X}}$, the projected distance separating the X-ray emission peak and the central BCG, normalised by $R_{500}$. Bottom-right: distribution of $\Delta r_{12}$, the offset (normalised by $R_{500}$) between the central BCG and the second brightest galaxy within $0.5R_{500}$ from the peak of the galaxy surface density.}
\label{fig:histo_all} 
\end{figure*}

\subsection{The $\beta$ statistic}

The $\beta$ test probes the mirror symmetry of the galaxy distribution with respect to the cluster centre, defined here as the galaxy surface density peak. It is insensitive to elliptical galaxy distributions so we expect this indicator to be mostly correlated with the X-ray statistics involving the odd moments of the multipole decomposition of the X-ray surface brightness.

The $\beta$ statistic is computed as follows. Given a galaxy $i$ and its diametrically opposite point $o$, one measures the average distances, $d_i$ and $d_o$, to their respective $n$ closest members. These two distances are then combined to define
\begin{equation}
\beta_i=\log_{10}(d_o/d_i).
\end{equation}
The $\beta$ statistic is finally obtained by averaging the $\beta_i$ over all galaxies, in our case within $R_{500}$. While positive $\beta$ can be associated to a clumpy galaxy distribution, negative values are more difficult to interpret. However, a cluster cannot be more symmetrical than $\beta\simeq0$, so negative values can also be seen as a departure from symmetry due to substructures. We are mostly interested in rank correlations between the different indicators, therefore we used the absolute value of $\beta$ hereafter. \cite{west88} used $n=5$ to measure the average distances $d_i$ and $d_o$, regardless of the cluster richness. Here we followed the suggestion of \cite{Pinkney96} to use $n=N^{1/2}$, where $N$ is the number of galaxies within the area of interest. We ran the $\beta$ test on 500 azimuthal randomisations of the original galaxy catalogue, using the galaxy surface density peak as the cluster centre. This randomisation procedure allows to quantify the bias, a residual value to be subtracted from the measured one, and the uncertainties in the $\beta$ statistic due to Poisson fluctuations while keeping constant the radial distribution of galaxies.

The distribution of $\beta$ is presented in Figure \ref{fig:histo_all} (top-left panel). We find 13 clusters ($\sim42\%$ of the sample) with a $1\sigma$-significant asymmetry parameter; only four of them ($\sim13\%$) have $\beta$ significant at the $3\sigma$ level. The sample is divided in two halves around $\beta=0.02$, with nearly one third of the clusters having $\beta\lesssim0.01$. Two clusters are characterised by a large asymmetry $\beta>0.06$: RXCJ2129.8-5048 and RXCJ2234.5-3744.

\subsection{The $\Delta_\Sigma$ indicator}
\label{sec:O2}

The $\Delta_\Sigma$ substructure indicator requires first to build maps of the galaxy surface density, $\Sigma(\bm{r})$. To do so, we used the kernel density estimator described in \cite{pisani96}. At any position vector $\bm{r}$, the density is estimated by
\begin{equation}
\Sigma(\bm{r})\propto\sum_{i=1}^{N}K(\bm{r_i},\sigma_i;\bm{r}),
\end{equation}
where the $\bm{r_i}$ are the position vectors of the $N$ galaxies in the catalogue; the proportionality constant is obtained by requiring the integral of $\Sigma$ over the considered field of view to give $N$. For a symmetrical two-dimensional Gaussian kernel, we have
\begin{equation}
K(\bm{r_i},\sigma_i;\bm{r})=\frac{1}{2\pi\sigma_i^2}\exp\left(-\frac{1}{2}\frac{|\bm{r_i}-\bm{r}|^2}{\sigma_i^2}\right).
\end{equation}
The kernel widths are given by $\sigma_i=\lambda_i\sigma_g$, where $\sigma_g$ is a global smoothing scale. The local bandwidths, $\lambda_i$, depend on the current estimate of the density at positions $\bm{r_i}$ (typically proportional to $\Sigma^{-1/2}$). Pisani's approach consists in refining iteratively the value of $\sigma_g$ by least-squares cross-validation, making this density estimator both adaptive, through the $\lambda_i$'s, and optimal with the global parameter $\sigma_g$. 

\begin{figure}
\center
\includegraphics[width=6.5cm, angle=-90]{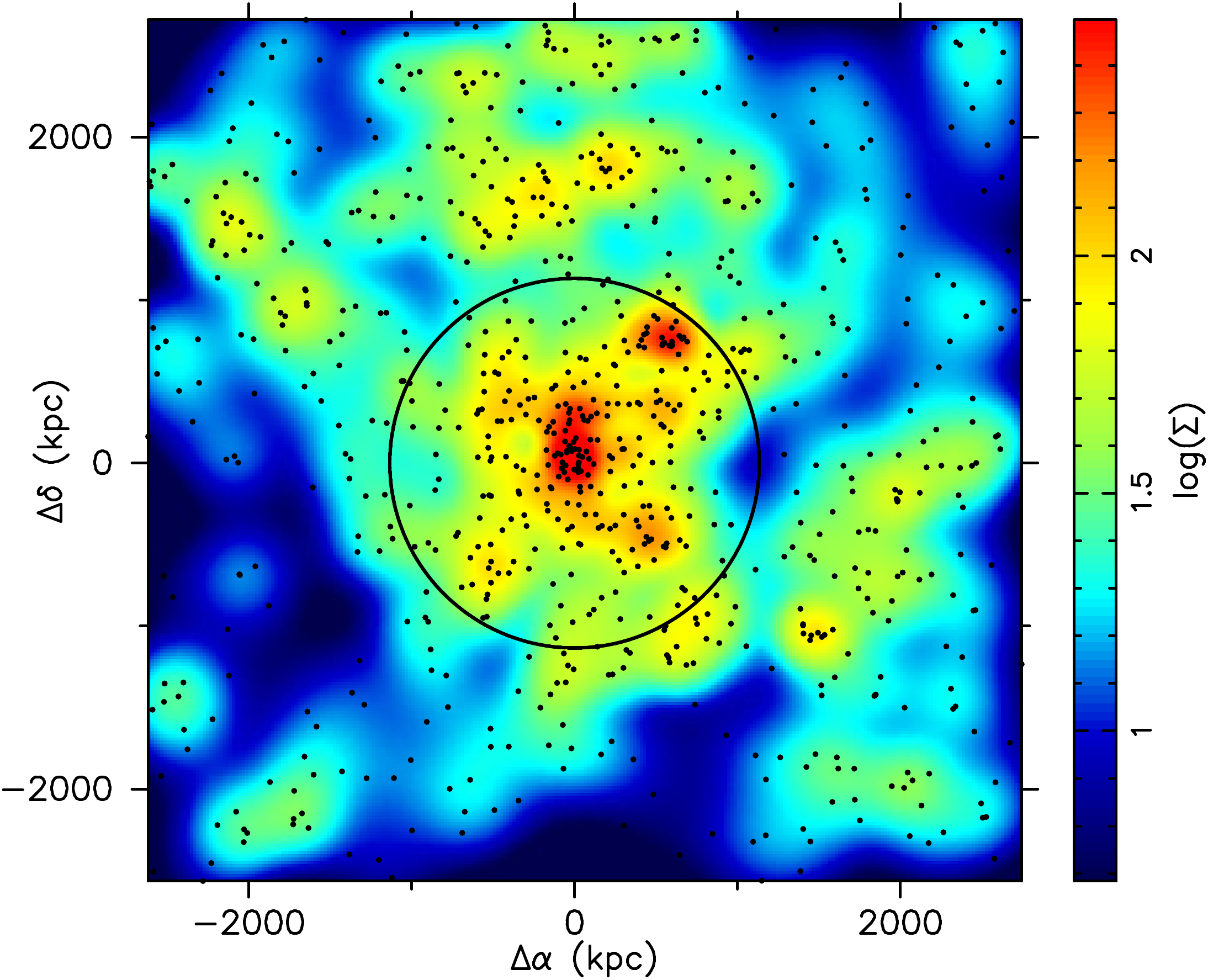}\\[4pt]
\includegraphics[width=6.5cm, angle=-90]{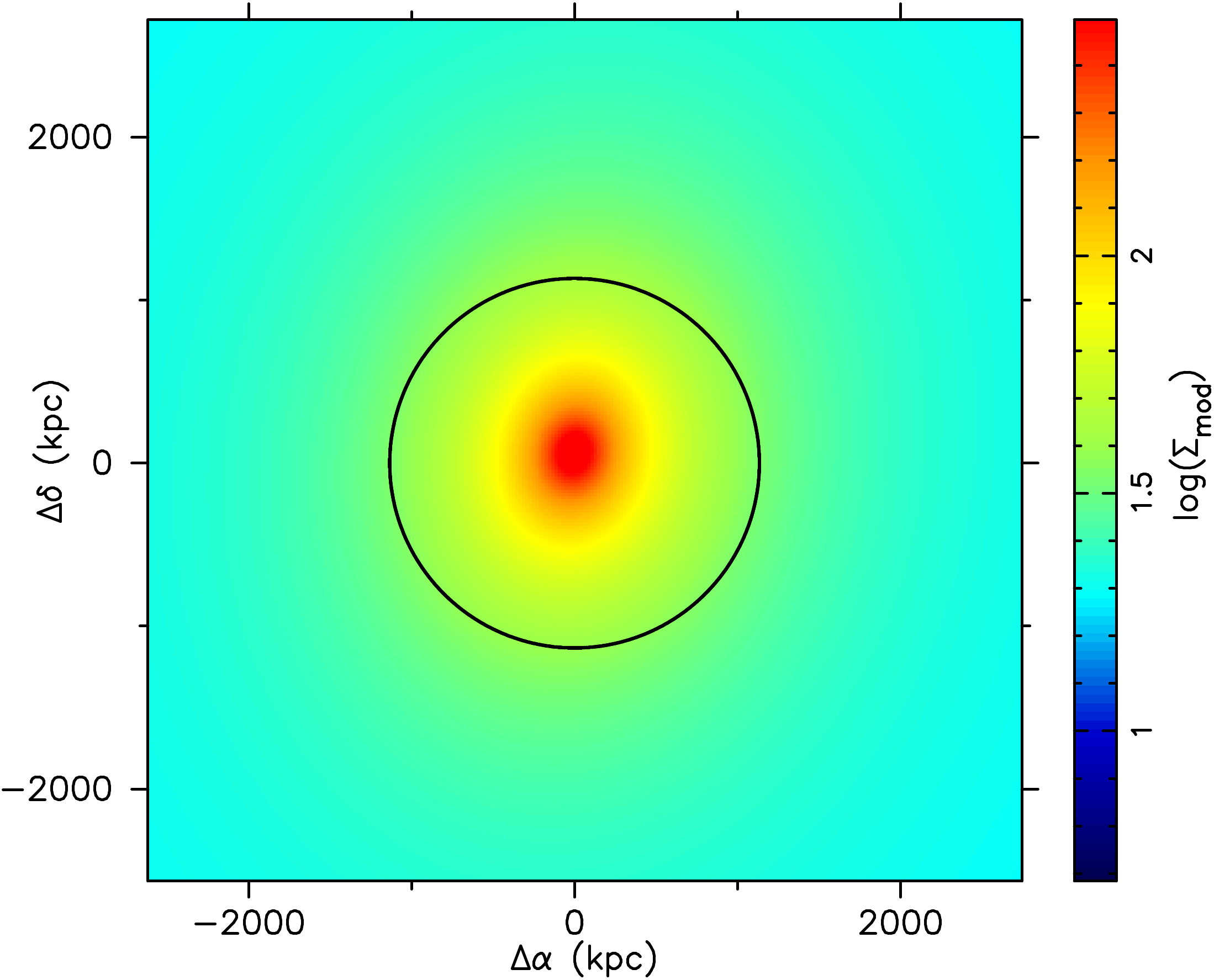}\\[4pt]
\includegraphics[width=6.5cm, angle=-90]{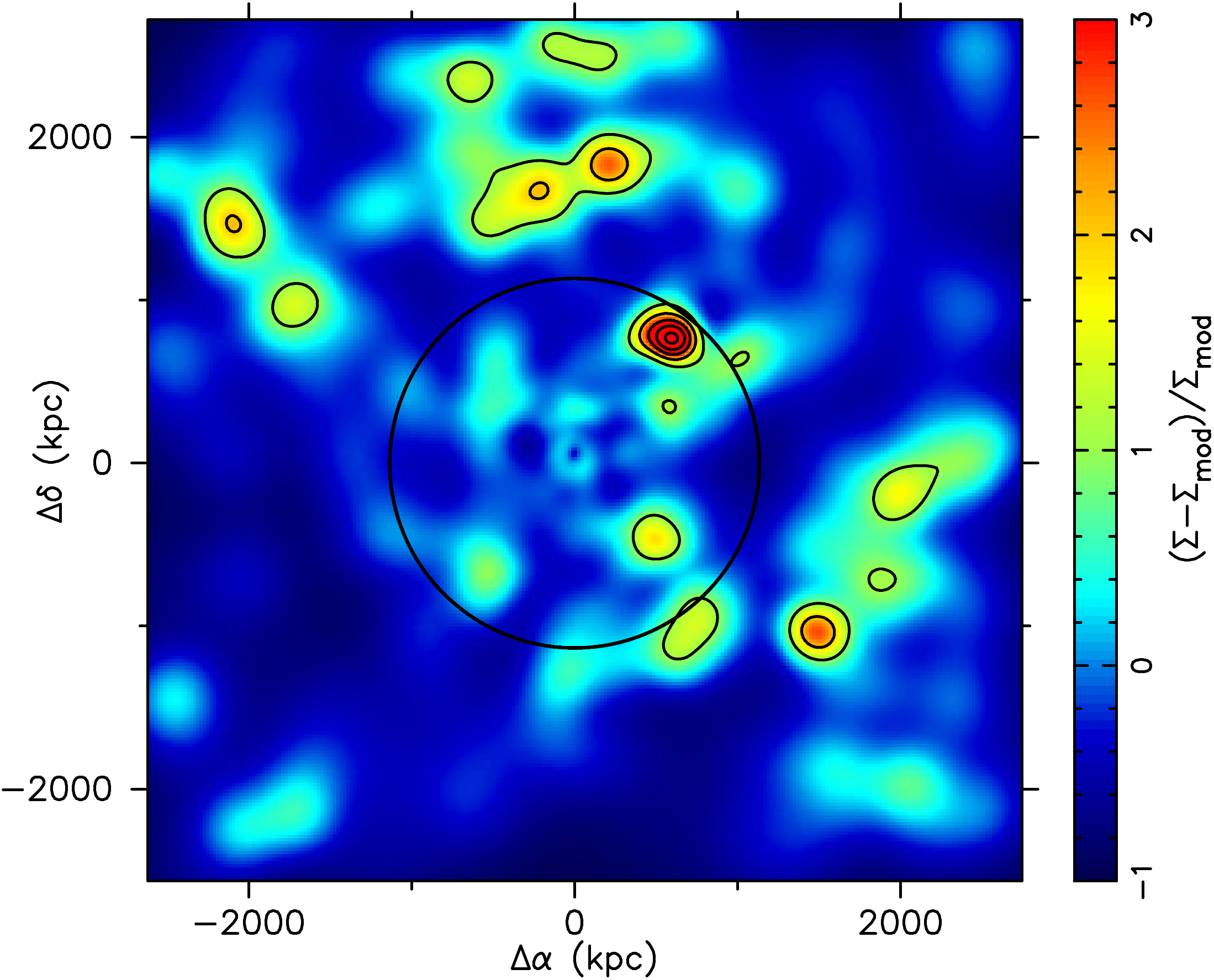}
\caption{Top panel: galaxy surface density map of RXCJ0547.6-3152, in units of $\mathrm{Mpc^{-2}}$. The red-sequence galaxies, down to $m*+3$, are marked by dots. Middle panel: best-fit model, given by an NFW component of scale radius $r_s=423$ kpc and ellipticity $e=0.20$, plus a constant background $\Sigma_{\mathrm{bckg}}=17.5\,\mathrm{Mpc^{-2}}$. Bottom panel: residuals, expressed in density contrast with respect to the best-fit model. The contours start at a density contrast of one and increase by one unit. In each panel, the circle of radius $R_{500}$ is centred on the X-ray emission peak.}
\label{fig:delta_0547} 
\end{figure}

The surface density maps were fitted by the sum of a constant background plus a projected King or NFW \citep{navarro97} model. We allowed for elliptical distributions by expressing the cluster-centric distances as $r^2=(x\cos\phi+y\sin\phi)^2+(y\cos\phi-x\sin\phi)^2/(1-e)^2$, with $e$ the ellipticity and $\phi$ the position angle of the galaxy distribution. The best-fit parameters were obtained with a standard $\chi^2$ minimisation using all pixels within $1.5R_{500}$; we selected either the King or the NFW model based on their respective $\chi^2$ value. With the galaxy surface density and its best-fit model, $\Sigma_{\mathrm{mod}}$, we computed the second substructure indicator as
\begin{equation}
\Delta_\Sigma=\frac{\sum_{i,j}[\Sigma(x_i,y_j)-\Sigma_{\mathrm{mod}}(x_i,y_j)]^2}{\sum_{i,j}\Sigma(x_i,y_j)^2},
\end{equation}
where the sum runs over the pixels $(x_i,y_j)$ within $R_{500}$. To estimate the uncertainties on $\Delta_\Sigma$ due to Poisson fluctuations, we computed its value for 500 surface density maps generated from azimuthal randomisations of the galaxy catalogue (a new best-fit model was obtained for each randomised map). We provide in Figure \ref{fig:delta_0547} an example of a galaxy surface density map, its best-fit model, and the corresponding residuals (see the Appendix for the other clusters).

\begin{figure}
\center
\includegraphics[width=6.cm, angle=-90]{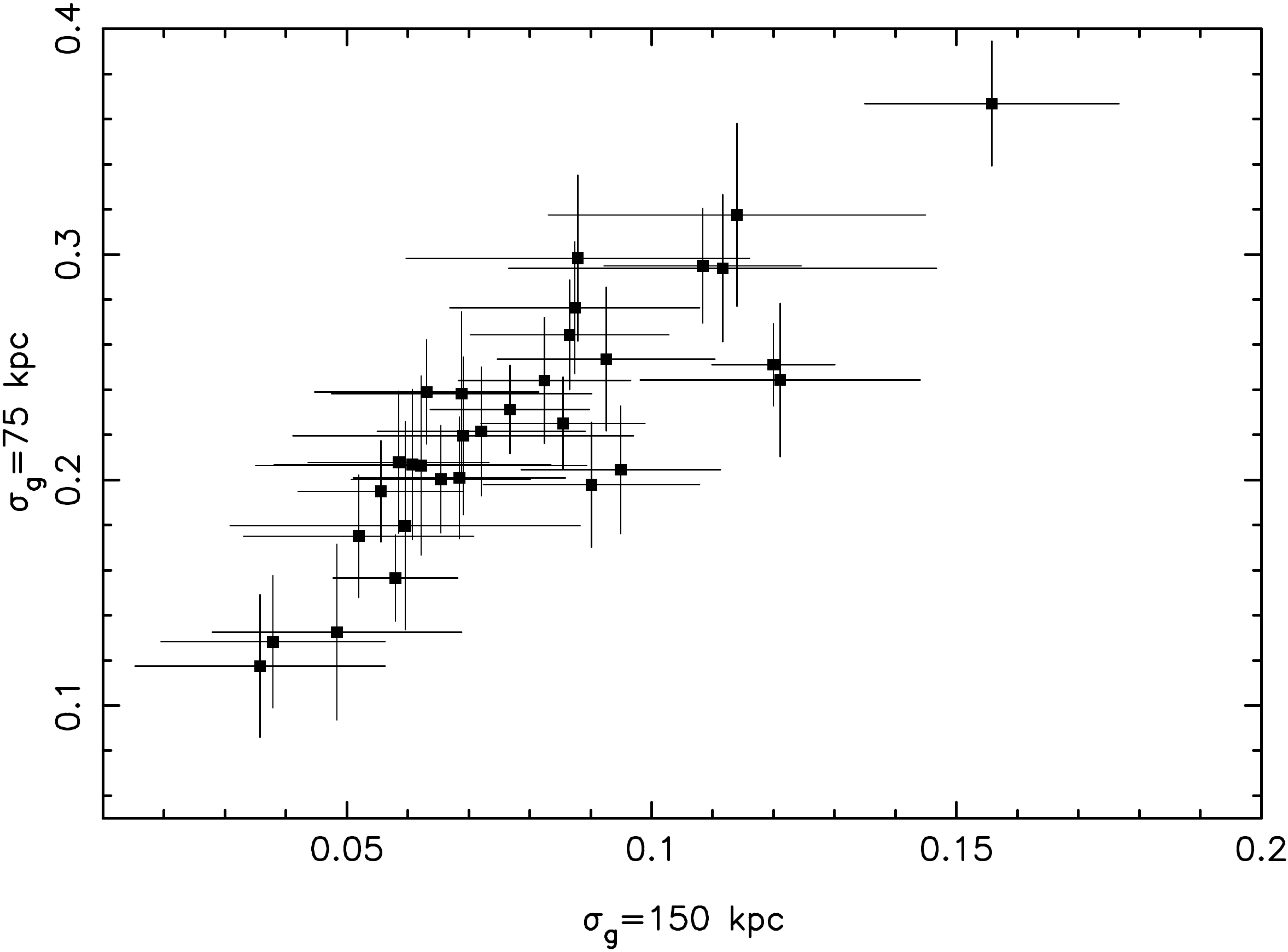}\\
\includegraphics[width=6.cm, angle=-90]{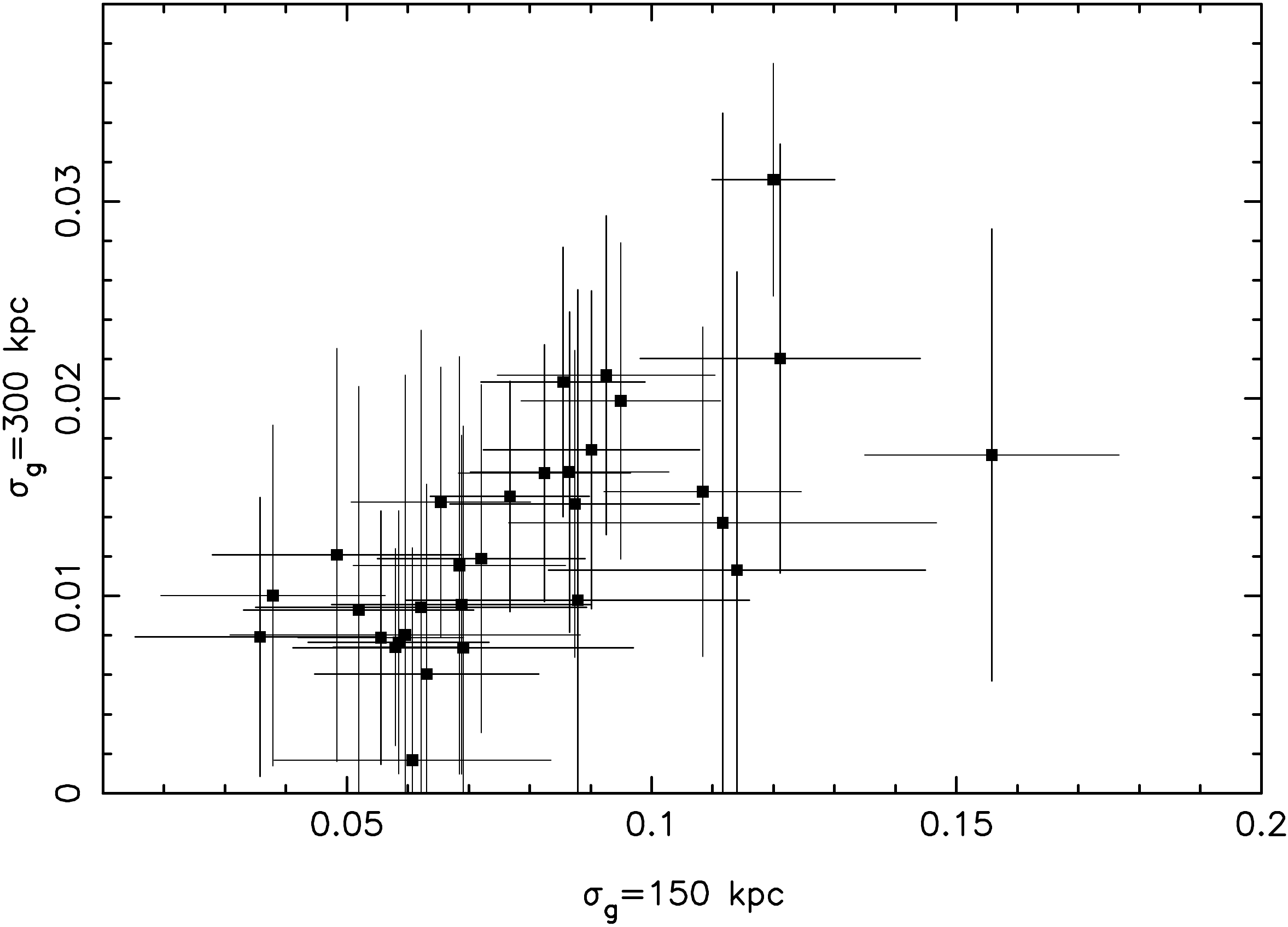}
\caption{Impact of the global smoothing scale, $\sigma_g$, on the value of the residual indicator, $\Delta_\Sigma$. We compare here the results obtained with a fixed $\sigma_g=150$ kpc to those obtained with $\sigma_g=75$ kpc (top panel) and $\sigma_g=300$ kpc (bottom panel).}
\label{fig:smooth} 
\end{figure}

For the \rexcess\ sample, we obtained smoothing scales in the range $\sigma_g\sim75-300$ kpc, with an average value of $\sim150$ kpc. The method to produce the surface density maps provides an objective way of determining the overall smoothing scale. However, the large range of $\sigma_g$ may introduce some variations in $\Delta_\Sigma$ regardless of the actual level of substructures. To investigate this, we computed $\Delta_\Sigma$ fixing $\sigma_g$ to 75, 150, and 300 kpc for all clusters. The results are presented in Figure \ref{fig:smooth}. As expected, we observe variations of the absolute scale, but the values of $\Delta_\Sigma$ are correlated nonetheless. The Spearman rank correlation coefficients are $\rho=0.83$ for $\sigma_g=(150-75)$kpc, $\rho=0.73$ for $\sigma_g=(150-300)$kpc, and $\rho=0.42$ for $\sigma_g=(300-75)$kpc. The corresponding probabilities for no correlation are $P<10^{-5}$, $P<10^{-5}$, and $P=0.02$, respectively. Since we are mostly interested in ranks, rather than precise thresholds to distinguish regular from disturbed objects, the size of the global smoothing scale should not impact significantly our final results. For consistency, we decided to compute $\Delta_\Sigma$ from the galaxy surface density maps constructed with $\sigma_g=150$ kpc for all clusters.

The distribution of $\Delta_\Sigma$ is shown in Figure \ref{fig:histo_all} (top-middle panel). According to this test, all clusters present a certain degree of substructures. The distribution peaks at $\Delta_\Sigma\sim0.06$ and 17 clusters ($\sim55\%$ of the sample) are below the dip observed at $\Delta_\Sigma\sim0.08$. The most disturbed cluster appears to be RXCJ2129.8-5048, as with the $\beta$ test.

\subsection{The $FE$ statistic}

The $FE$ test is based on the second Fourier components, $S$ and $C$, of the azimuthal galaxy distribution, $N(\phi)$. This indicator is thus sensitive to elongated galaxy distributions. It is equivalent to the quadrupole moment, $P_2$, of the multipole decomposition of the X-ray surface brightness. We evaluated the Fourier components as
\begin{equation}
S=\sum N(\phi)\sin(2\phi), \quad   C=\sum N(\phi)\cos(2\phi),
\end{equation}
with bins of $20\degree$. If one assumes that the azimuthal distribution can be described by
\begin{equation}
N(\phi)=\left(\frac{N_0}{2\pi}\right)[1+(N_1/N_0)\cos(2\phi-2\phi_0)],
\end{equation}
where $N_0$ is the total number of galaxies and $\phi_0$ the position angle of the cluster; the elongation amplitude, $N_1$, is then estimated by $N_1=2(S^2+C^2)^{1/2}$. For small deviations from circular symmetry, $N_1/N_0\ll1$, we have $\sigma_{N_1}=\sqrt{2N_0}$. Therefore, one can define a statistic based on the elongation strength as
\begin{equation}
FE=\frac{N_1}{\sigma_{N_1}}=\sqrt{\frac{2(S^2+C^2)}{N_0}}.
\end{equation}
We can note that, by construction, $FE$ measures the significance of the ellipticity rather than its absolute scale. As for the $\beta$ test, we estimated the bias and uncertainties in $FE$ due to Poisson fluctuations on 500 azimuthal randomisation of the galaxy catalogue.

The distribution of $FE$ for the \rexcess\ sample is given in Figure \ref{fig:histo_all} (top-right panel). We find that 26 clusters have a $1\sigma$-significant $FE$, that is, $\sim84\%$ of the sample. This fraction drops to $\sim42\%$ (13 clusters) when considering a $3\sigma$ significance threshold. The distribution has a minimum at $FE\sim2$, 18 clusters having an elongation smaller than this value. Two clusters stick out of the sample with $FE>4$: RXCJ0145.0-5300 and RXCJ2218.6-3853. We also note that, after the bias subtraction, three clusters have a negative $FE$: RXCJ0003.8+0203, RXCJ1236.7-3354, and RXCJ2234.5-3744. Since $FE$ cannot be negative, we will consider these clusters as censored data points and use their $1\sigma$ upper limit to compute the correlation coefficients.

\subsection{Secondary indicators}
\label{sec:indic2}

The three statistics introduced above rely on the global distribution of galaxies. So, in principle, they can be affected by the residual background or by masked area. However, we note that the surface masked by bright stars is on average less than one percent of that covered by $R_{500}$. In contrast, indicators relying on local properties of the clusters should be more robust to these sources of uncertainties, in addition of being easier to implement. For this work, we used the three following quantities:
\begin{itemize}
\item $\Delta r_{\mathrm{BCG-X}}$, the projected distance separating the X-ray emission peak and the central BCG, normalised by $R_{500}$.
\item $\Delta m_{12}$, the magnitude gap between the central BCG and the second brightest galaxy within a radius of $0.5R_{500}$ from the peak of the galaxy surface density map.
\item $\Delta r_{12}$, the projected distance separating these two galaxies, normalised by $R_{500}$.
\end{itemize}

The BCG-X-ray peak offset has been shown to be correlated to the dynamical state of a cluster (e.g. \citealt{katayama03,hudson10,mann12}), thus we can expect it to reflect a cluster substructure content. It is worth noting that a large offset is most definitely the sign of a disturbed dynamical state (ongoing merger), whereas small separations can occur for relaxed systems, due to projection effects, large impact parameters, large mass ratio, or when the BCG and surviving cool core align again due to the ram pressure slingshot effect. \cite{rossetti16} analysed the distribution of $\Delta r_{\mathrm{BCG-X}}$ for \rexcess\ based on the results of \cite{haarsma10}. Our identification of BCGs is identical to that of \cite{haarsma10} and thus leads to the same conclusion as \cite{rossetti16}, that is a proportion $23/31\sim74\%$ of relaxed clusters according the criterion $\Delta r_{\mathrm{BCG-X}}<0.02R_{500}$. We can also mention that these authors used this statistic to highlight a "cool core bias" in various X-ray selected samples with respect to a {\it Planck} SZ-selected cluster catalogue, the latter being characterised by a significantly smaller fraction of relaxed objects. 

The luminosity gap between the central and second BCGs offers an alternative way to explore the dynamical state of a cluster. Large luminosity gaps are expected for systems formed early, leaving enough time for dynamical friction to make $L^*$ galaxies spiral towards the centre of the potential and merge with the central galaxy. This process can eventually lead to the formation of fossil groups, characterised by gaps $\Delta m_{12}\ge2$ (e.g. \citealt{ponman94,jones03}). As shown for instance by \cite{smith10} with the LoCuSS cluster sample, the luminosity gap also traces the recent infall history of a system, since a bright galaxy entering a cluster at latter times can significantly decrease a previously large gap. Such an infalling bright galaxy can be isolated or part of a group formed at earlier times. In the second case, the level of substructure of the main system will also increase, therefore we can expect to observe an anti-correlation between $\Delta m_{12}$ and other substructure indicators. The distribution of $\Delta m_{12}$ is shown if Figure \ref{fig:histo_all} (bottom-left panel). Our results agree well with those of \cite{smith10}, even though they measured gaps with infrared passbands and used a different aperture to search for the second BCG, $r=640$ kpc for all clusters, corresponding to $\sim0.4R_{200}$ for a $10^{15}\,\mathrm{M_\odot}$ cluster at a redshift $z=0.2$. In particular, we find a fraction $11/31\sim35\%$ of clusters with $\Delta m_{12}\ge1$, similar to the $38\%$ quoted by these authors.

A small galaxy group falling onto a cluster and leading to a small $\Delta m_{12}$ can be observed at different stages: pre-, ongoing, or post-merger. In the former case, the infalling substructure has not yet affected the central gas distribution of the main system, which thus will appear relaxed in X-rays. Similarly, large impact parameters or small mass ratios would not affect the gas distribution in a measurable way. Therefore, the statistic $\Delta m_{12}$ alone does not constitute an entirely reliable indicator of a cluster's dynamical state as inferred from X-ray data. Combining this statistic with the offset between the central and second BCGs, $\Delta r_{12}$, can provide additional information to distinguish between these different scenarios. In particular, a small $\Delta m_{12}$ with a large $\Delta r_{12}$ is likely to probe a pre-merger configuration, whereas a small $\Delta m_{12}$ and $\Delta r_{12}$ are expected for ongoing mergers. We note that the statistic $\Delta r_{12}$ will not be sensitive to substructure with a large impact parameter, since we limited the search for the second BCG to within $0.5R_{500}$. The distribution of $\Delta r_{12}$ for \rexcess\ is presented in Figure \ref{fig:histo_all} (bottom-right panel); it is characterised by a gap at $\Delta r_{12}=0.2R_{500}$, separating a group of $13/31\sim42\%$ clusters with small offsets, thus potentially dynamically young systems.

\subsection{Correlations between the different indicators}
\label{sec:OO}

To compare the results of the different indicators, we used Spearman's rank correlation coefficient, since we do not expect the various statistics to be linear response variables of the clusters substructure content or dynamical state. The coefficients were estimated with the ASURV package (Astronomical Survival Statistics, \citealt{isobe86}), which accounts for censored data; the results are listed in Table \ref{table:O_O}.

\begin{figure}
\center
\includegraphics[width=6cm, angle=-90]{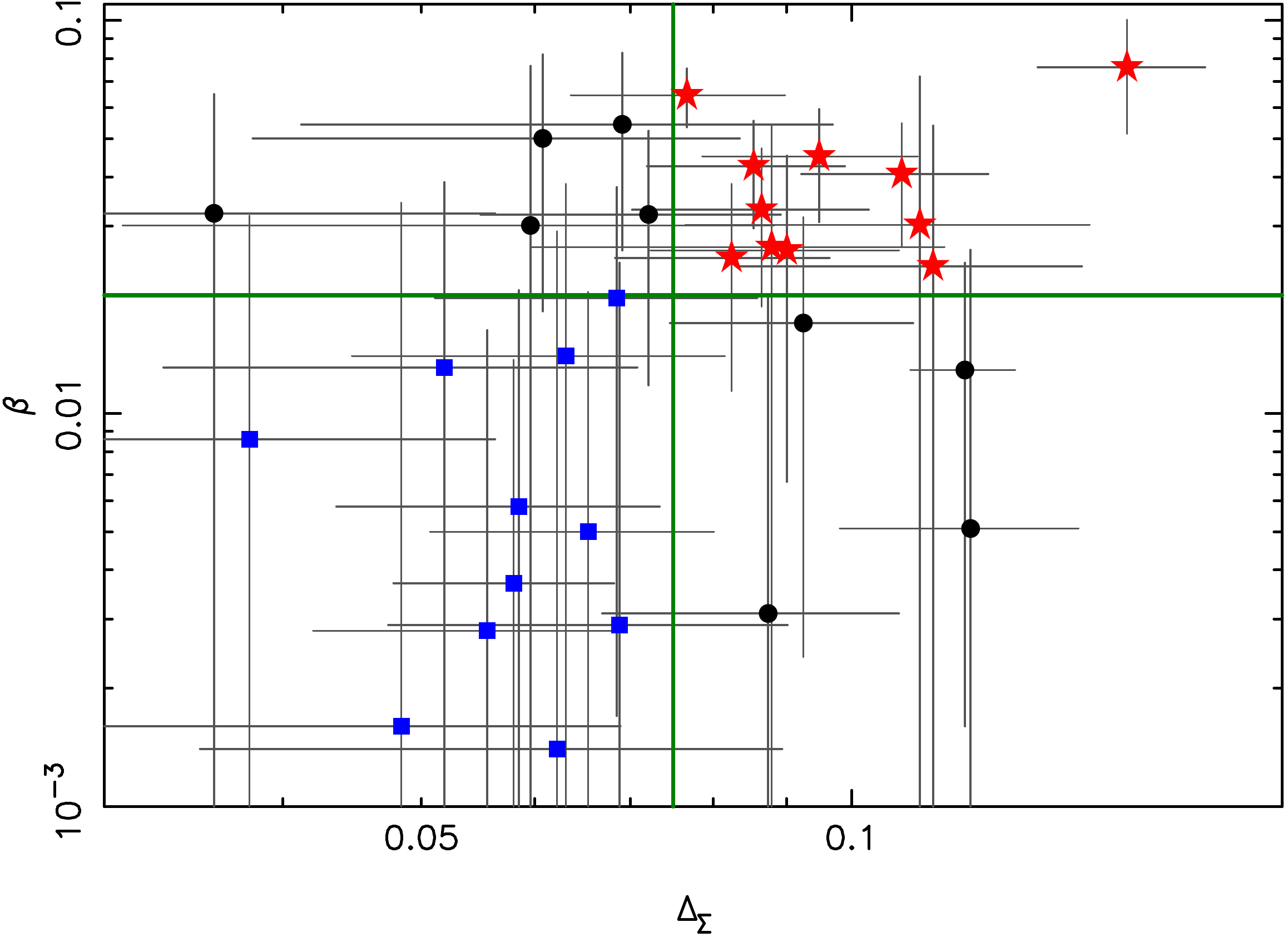}
\caption{Comparison of the asymmetry and residuals substructure indicators. Our tentative classification of morphologically regular (blue squares) and disturbed (red stars) clusters, based on the thresholds $\beta=0.02$ and $\Delta_\Sigma=0.075$, is shown by the two green lines. Clusters falling outside these two sub-classes are marked with black dots. Error bars indicate the statistical uncertainties estimated from the randomisation procedures outlined in the text.}
\label{fig:OO} 
\end{figure}

The correlation between $\beta$ and $\Delta_\Sigma$ is the strongest one among the primary indicators, with $\rho=0.36$ and a probability for the null hypothesis (no correlation) $P=0.06$. Therefore, we can attempt at classifying morphologically regular and disturbed cluster based on these two statistics. Their distributions (Fig. \ref{fig:histo_all}) suggest using the thresholds $\beta>0.02$ and $\Delta_\Sigma>0.075$ to qualify a cluster as disturbed. The Figure \ref{fig:OO} presents the partitioning of the sample based on these two selection criteria. Only nine systems fall outside the two sub-classes, thus $\sim71\%$ of the clusters are classified in the same way according to the asymmetry and residuals indicators. If we require a cluster to satisfy both criteria to be safely classified as disturbed, we find that \rexcess\ contains $11/31\sim35\%$ of systems with a significant level of substructure; this proportion increases to $\sim65\%$ when considering clusters flagged by either $\beta$ or $\Delta_\Sigma$.

\begin{figure}
\center
\includegraphics[width=6cm, angle=-90]{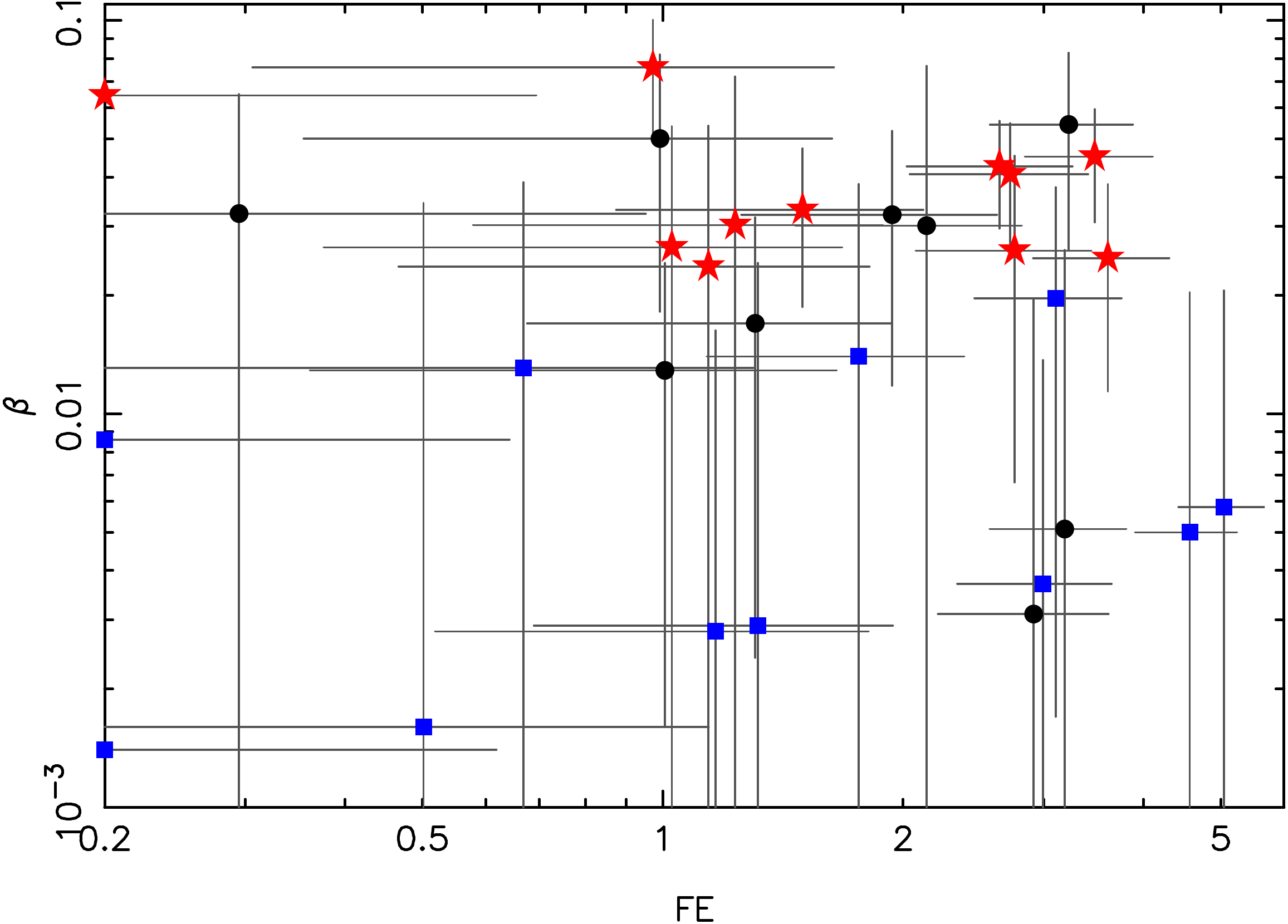}\\[4pt]
\includegraphics[width=6cm, angle=-90]{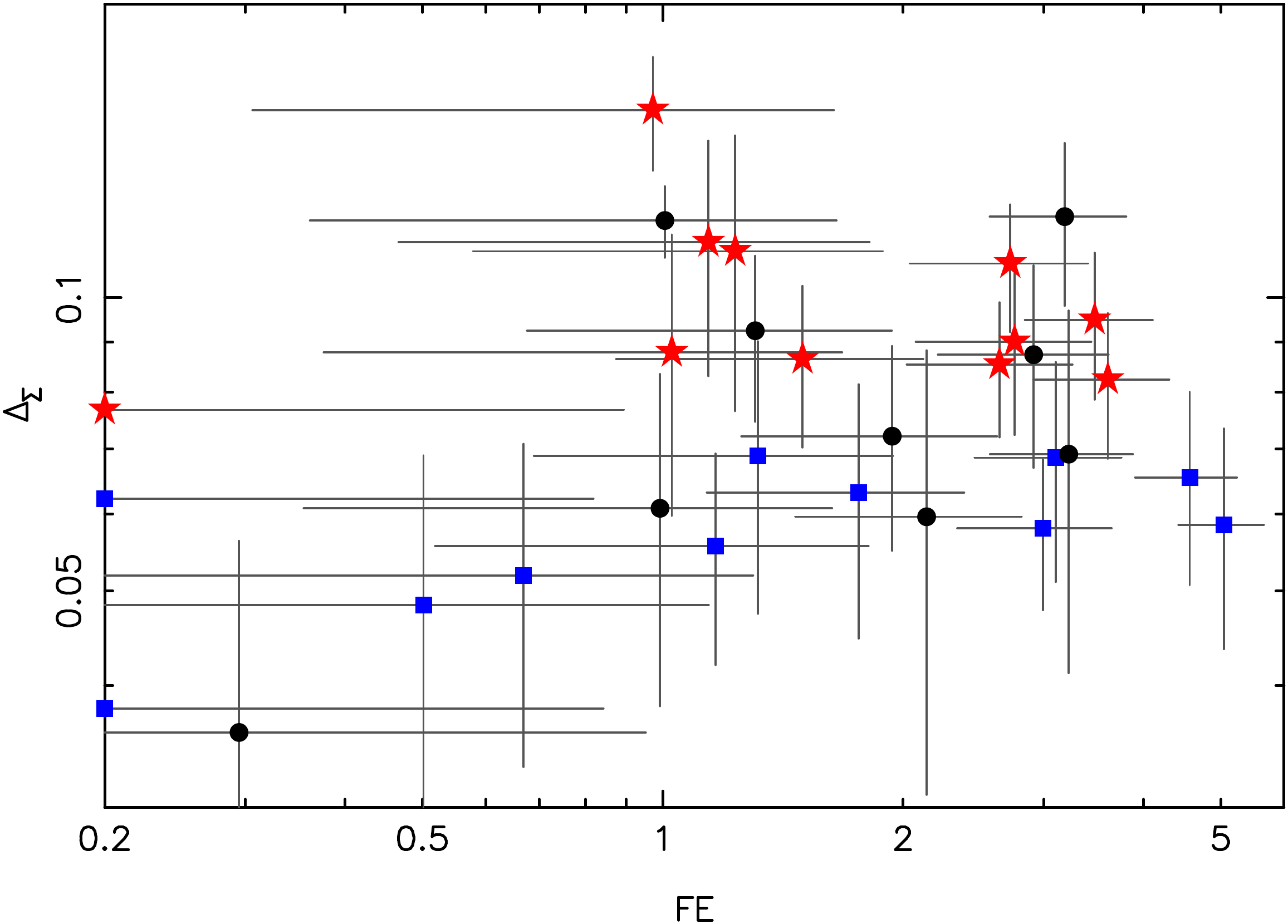}
\caption{Comparison of the Fourier elongation and the two other main substructure indicators. The colour code follows the selection presented in Fig. \ref{fig:OO}.}
\label{fig:O_FE} 
\end{figure}

The Fourier elongation is sensitive to flattened galaxy distributions, which can occur for elliptical but otherwise relaxed clusters, or for systems having substructures distributed along an axis passing through their centre. Therefore, a significant $FE$ does not constitute an indisputable proof of substructures. Conversely, a highly asymmetric distribution of substructures would suppress the $FE$ signal, so that small values of this statistic do not occur only for relaxed spherical systems; this is the case for RXCJ2234.5-3744, which has $FE=0$ and the second largest $\beta$. The lack of correlation of $FE$ with the asymmetry ($\rho=0.00$, $P=1.00$) and residuals ($\rho=0.21$, $P=0.24$) indicators agrees with these remarks. Furthermore, we can see in Figure \ref{fig:O_FE} that the clusters falling into the regular sub-class cover the full range of elongation, which confirms that regular clusters are not necessarily spherical. It is interesting to note, however, that most of the morphologically complex systems have $FE\gtrsim1$ and that nearly half of them are characterised by a large $FE>2$. 

\begin{figure}
\center
\includegraphics[width=6cm, angle=-90]{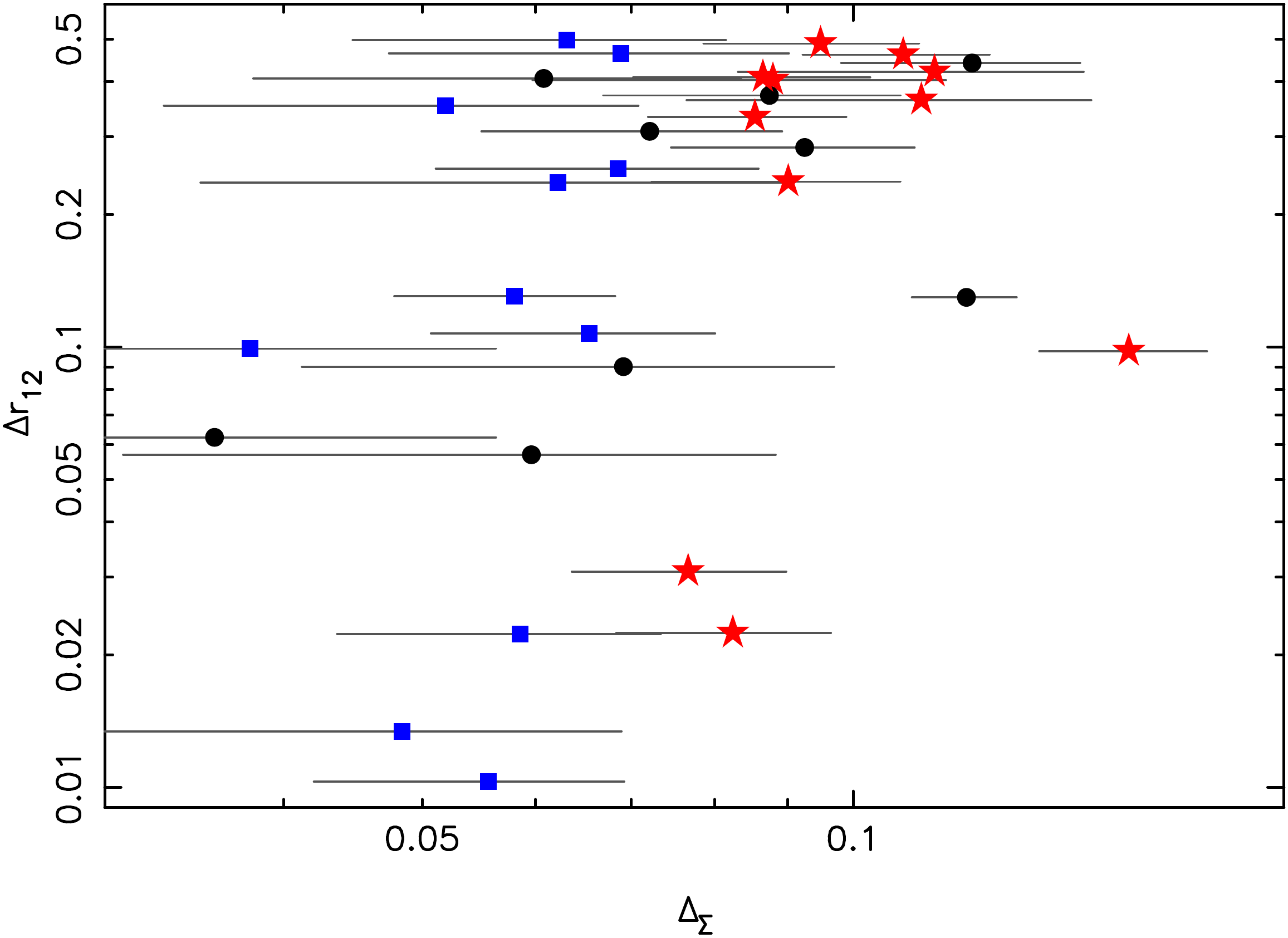}
\caption{Correlation of the residuals, $\Delta_\Sigma$, with the offset between the central and second BCGs, $\Delta r_{12}$; error bars are smaller than the symbols. See Fig. \ref{fig:OO} for the colour code.}
\label{fig:delta_dr12} 
\end{figure}

The secondary indicators are local quantities, probing mostly the clusters core, since they only involve the central BCG (position or luminosity) and a second data point (second BCG or X-ray emission peak), whereas the primary statistics rely on the global distribution of cluster members. Therefore, the lack of strong correlation between them is not surprising. For instance, an off-axis merger of a substructure large enough to produce a significant asymmetry $\beta$ would not lead to a large offset $\Delta r_{\mathrm{BCG-X}}$. The only significant correlation between a primary and a secondary indicator is obtained for the pair $\Delta_\Sigma-\Delta r_{12}$, which has a coefficient $\rho=0.46$ and a probability $P=0.01$ for the null hypothesis. This correlation is better explained in terms of the regular and disturbed sub-classes, since $\Delta r_{12}$ is not a continuous response variable of a cluster's substructure content. As we can see in Figure \ref{fig:delta_dr12}, $8/11\sim73\%$ of the morphologically disturbed clusters have large offsets $\Delta r_{12}>0.2R_{500}$, whereas the regular ones cover the full range of values; the correlation is thus driven by the disturbed population. On the one hand, for a cluster without significant substructures, the second BCG is most likely a satellite observed at a random stage of its orbit, hence at no particular position. On the other hand, a significant substructure could host a red galaxy that is bright enough to be selected as the second BCG. Since $\Delta_\Sigma$ is more sensitive to high density contrasts produced by substructures located in regions of low density, away from the cluster core, it is natural to observe most of the disturbed clusters in the top-right quadrant of the $\Delta_\Sigma-\Delta r_{12}$ plan.

\begin{figure}
\center
\includegraphics[width=6cm, angle=-90]{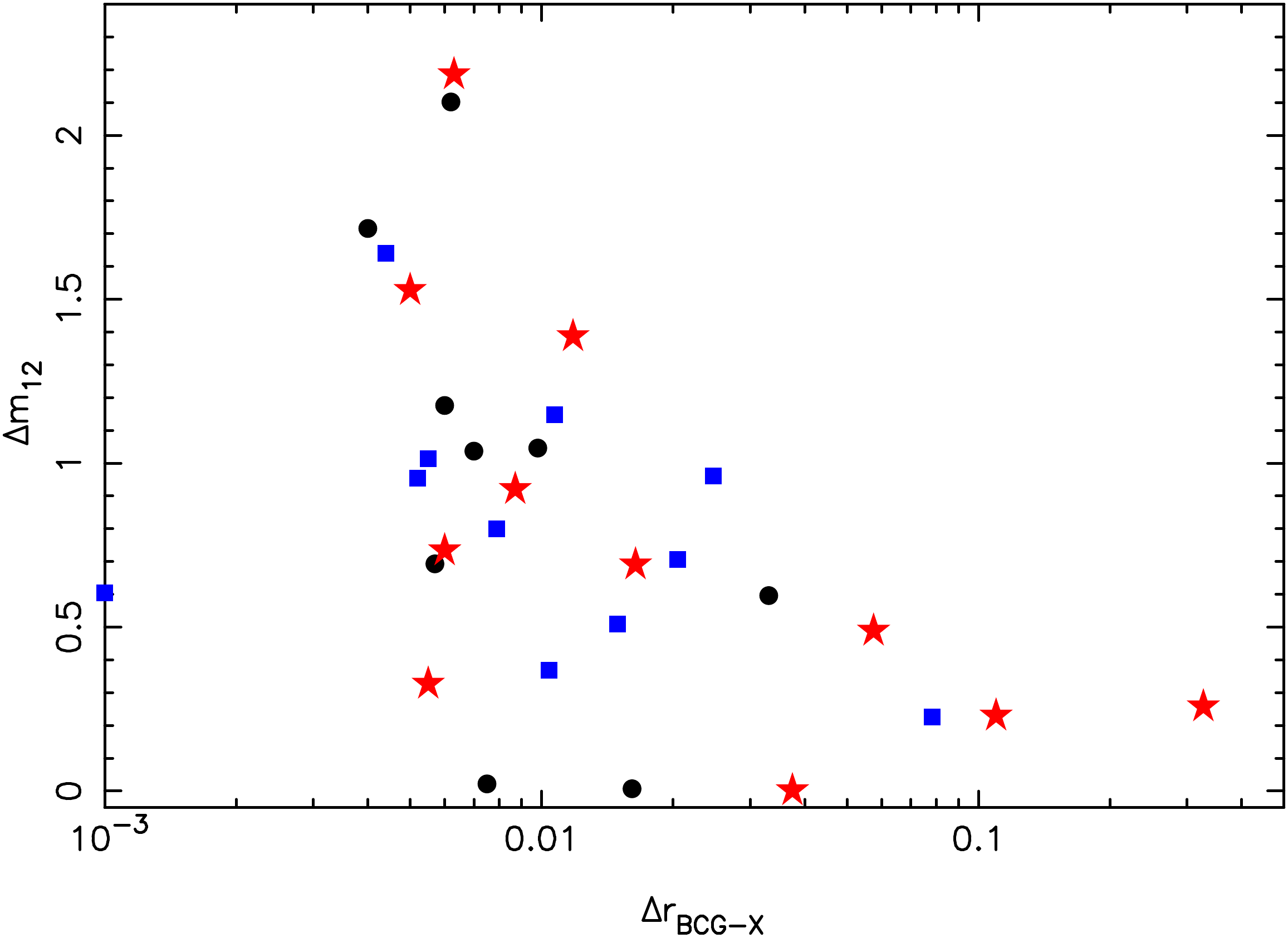}
\caption{Correlation between the X-ray peak-central BCG offset, $\Delta r_{\mathrm{BCG-X}}$, and the magnitude gap between the central and second brightest BCGs, $\Delta m_{12}$. See Fig. \ref{fig:OO} for the colour code.}
\label{fig:rXBCG_m12} 
\end{figure}

Proceeding with the secondary indicators, we find that their cross correlations are all significant, as expected since these tests are mainly sensitive to the clusters core and their recent mass assembly. In particular, the pair $\Delta r_{\mathrm{BCG-X}}-\Delta m_{12}$ has the strongest correlation with $\rho=-0.56$ and $P=0.002$ (see Fig. \ref{fig:rXBCG_m12}). \cite{smith10} argued that small magnitude offsets can be attributed to dynamically young systems characterised by a recent or ongoing merger. The observed anti-correlation with $\Delta r_{\mathrm{BCG-X}}$ supports their findings, since a large separation between the central BCG and the X-ray emission peak is a clear signature of a recent major merger. We can also note on Figure \ref{fig:rXBCG_m12} that there is no distinct separation between the morphologically regular and disturbed clusters. This indicates that our optical classification, derived from the overall spatial distribution of cluster galaxies, is likely to reflect the substructure content of a cluster regardless of the recent dynamical history of its core.

A substructure, whose own central BCG is brighter than the brightest satellite of the main component, may be observed shortly after its core crossing, and thus be associated with a small $\Delta r_{12}$. In that case, it should also imprint the central ICM distribution. The anti-correlation between $\Delta r_{12}$ and $\Delta r_{\mathrm{BCG-X}}$ ($\rho=-0.31$, $P=0.10$) seems to favour this scenario, where the recent merger of a substructure massive enough to host a bright galaxy creates a displacement of the X-ray emission peak with respect to the central BCG. This is also supported by the correlation between $\Delta r_{12}$ and $\Delta m_{12}$ ($\rho=0.33$, $P=0.08$), which shows that small offsets between the central and second BCGs occur in dynamically young systems. Clearly, the interpretation of $\Delta r_{12}$ is not straightforward: we argued previously that disturbed systems can present large offsets due to a substructure at intermediate distance, whereas we find here that small offsets could be the sign of a recent merger, hence also characteristic of disturbed systems. We can interpret this apparent contradiction as the consequence of a lack of correlation between the central properties of a cluster, probed by the secondary indicators, and its substructure content at larger scales, probed by $\Delta_\Sigma$. In other words, the recent mass assembly of a cluster, which is responsible for the current dynamical state of its core, appears to not be correlated to a possible future accretion of substructures, which is responsible for its overall morphology. 

\begin{table}
\centering 
\begin{threeparttable}
\caption{Spearman correlation coefficients between the optical substructure indicators.}
\label{table:O_O}
\begin{tabularx}{7.cm}{l Y Y}
\hline\hline\noalign{\smallskip}
 & $\rho$ & $P$\\
\noalign{\smallskip}\hline\noalign{\smallskip}
$FE-\beta$ & 0.00 & 1.00\\[3pt]
$FE-\Delta_\Sigma$ & 0.21 & 0.24\\[3pt]
$FE-\Delta r_{\mathrm{BCG-X}}$ & 0.17 & 0.35\\[3pt]
$FE-\Delta m_{12}$ & -0.20 & 0.17\\[3pt]
$FE-\Delta r_{12}$ & 0.11 & 0.54\\[3pt]
\hline\noalign{\smallskip}
$\beta-\Delta_\Sigma$ & 0.36 & 0.06\\[3pt]
$\beta-\Delta r_{\mathrm{BCG-X}}$ & 0.27 & 0.15\\[3pt]
$\beta-\Delta m_{12}$ & -0.12 & 0.52\\[3pt]
$\beta-\Delta r_{12}$ & 0.12 & 0.52\\[3pt]
\hline\noalign{\smallskip}
$\Delta_\Sigma-\Delta r_{\mathrm{BCG-X}}$ & 0.17 & 0.38\\[3pt]
$\Delta_\Sigma-\Delta m_{12}$ & -0.03 & 0.90\\[3pt]
$\Delta_\Sigma-\Delta r_{12}$ & 0.46 & 0.01\\[3pt]
\hline\noalign{\smallskip}
$\Delta r_{\mathrm{BCG-X}}-\Delta m_{12}$ & -0.56 & 0.002\\[3pt]
$\Delta r_{\mathrm{BCG-X}}-\Delta r_{12}$ & -0.31 & 0.10\\[3pt]
\hline\noalign{\smallskip}
$\Delta m_{12}-\Delta r_{12}$ & 0.33 & 0.08\\
\noalign{\smallskip}\hline
\end{tabularx}
    \begin{tablenotes}
      \small
      \item Columns: (1) Pair of indicators. (2) Spearman rank correlation coefficient. (3) Probability of the null hypothesis (no correlation).
    \end{tablenotes}
  \end{threeparttable}
\end{table}

\section{Discussion}

\subsection{Comparison of the ICM and cluster galaxies distributions}
\label{sec:XO}

The morphological analysis of \rexcess\ based on X-ray data is presented in \cite{pratt09,boehringer10}. To sort the clusters into the regular and disturbed categories, a classification reminded in Table \ref{table:sub}, they used the X-ray centroid shift, $\omega$, which is defined as the standard deviation of the projected separation between the X-ray emission peak and the centroid of ten apertures of radius in the$[0.1-1]R_{500}$ range (see e.g. \citealt{poole06}). According to their threshold $\omega>0.01$, \rexcess\ contains $12/31\sim39\%$ of disturbed clusters. \cite{boehringer10} also included the power ratios approach of \cite{buote95} to characterise the overall morphology of the ICM. It is based on the moments of the X-ray surface brightness, measured within a circular aperture of radius $R_{500}$ and normalised by the total intensity of the cluster X-ray emission. By combining the $\omega$ criterion with their threshold on the hexapole power ratio, $P_3/P_0>1.5\times10^{-7}$, \cite{boehringer10} found a fraction $7/31\sim23\%$ of disturbed clusters, whereas a selection based solely on $P_3/P_0$ gives $11/31\sim35\%$. We can also mention that the classifications based on either $\omega$ or $P_3/P_0$ agree for $22/31\sim71\%$ of the clusters. Therefore, it leads to the same number of outliers as the optical $(\beta,\Delta_\Sigma)$ selection presented in the previous section (see Fig. \ref{fig:OO}).

\begin{figure}
\center
\includegraphics[width=6cm, angle=-90]{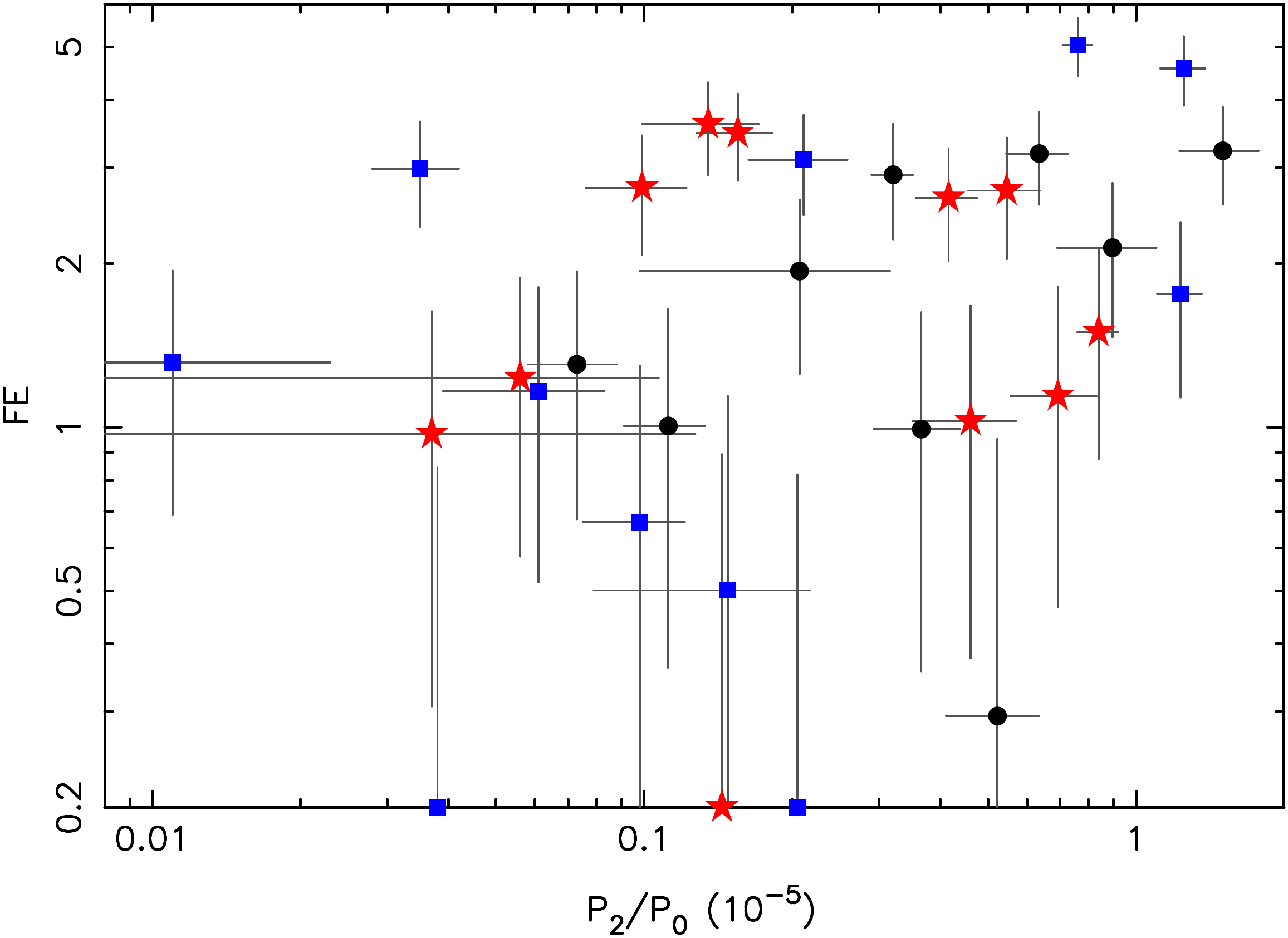}
\caption{Comparison of the X-ray power ratio $P_2/P_0$ and the Fourier elongation. See Fig. \ref{fig:OO} for the colour code.}
\label{fig:P2} 
\end{figure}

To compare the substructure content of the ICM and cluster galaxies, we estimated the Spearman correlation coefficients between the optical indicators and the X-ray centroid shift, the quadrupole, $P_2/P_0$, and hexapole power ratios; the results are summarised in Table \ref{table:O_X}. We start with the quadrupole moment, which probes the elongation of the X-ray surface brightness. In contrast to the collisionless cluster galaxies, the ICM particles experience strong interactions that lead to a local isotropic velocity distribution. Therefore, the ICM spatial distribution in a relaxed cluster should be closer to spherical compared to the shape of the underlying gravitational potential, making $P_2/P_0$ a good indicator of a disturbed cluster. As we can see in Figure \ref{fig:P2}, we detect a fairly strong correlation between this indicator and $FE$, characterised by $\rho=0.39$ and a probability $P=0.03$ for the null hypothesis: clusters presenting an elongated ICM distribution also tend have a flattened galaxy distribution. By construction, both indicators are sensitive to axis-symmetric distributions of substructures. Clusters with large $FE$ and $P_2/P_0$ are thus likely to be systems experiencing a mass growth along a preferential axis.

\begin{figure}
\center
\includegraphics[width=6cm, angle=-90]{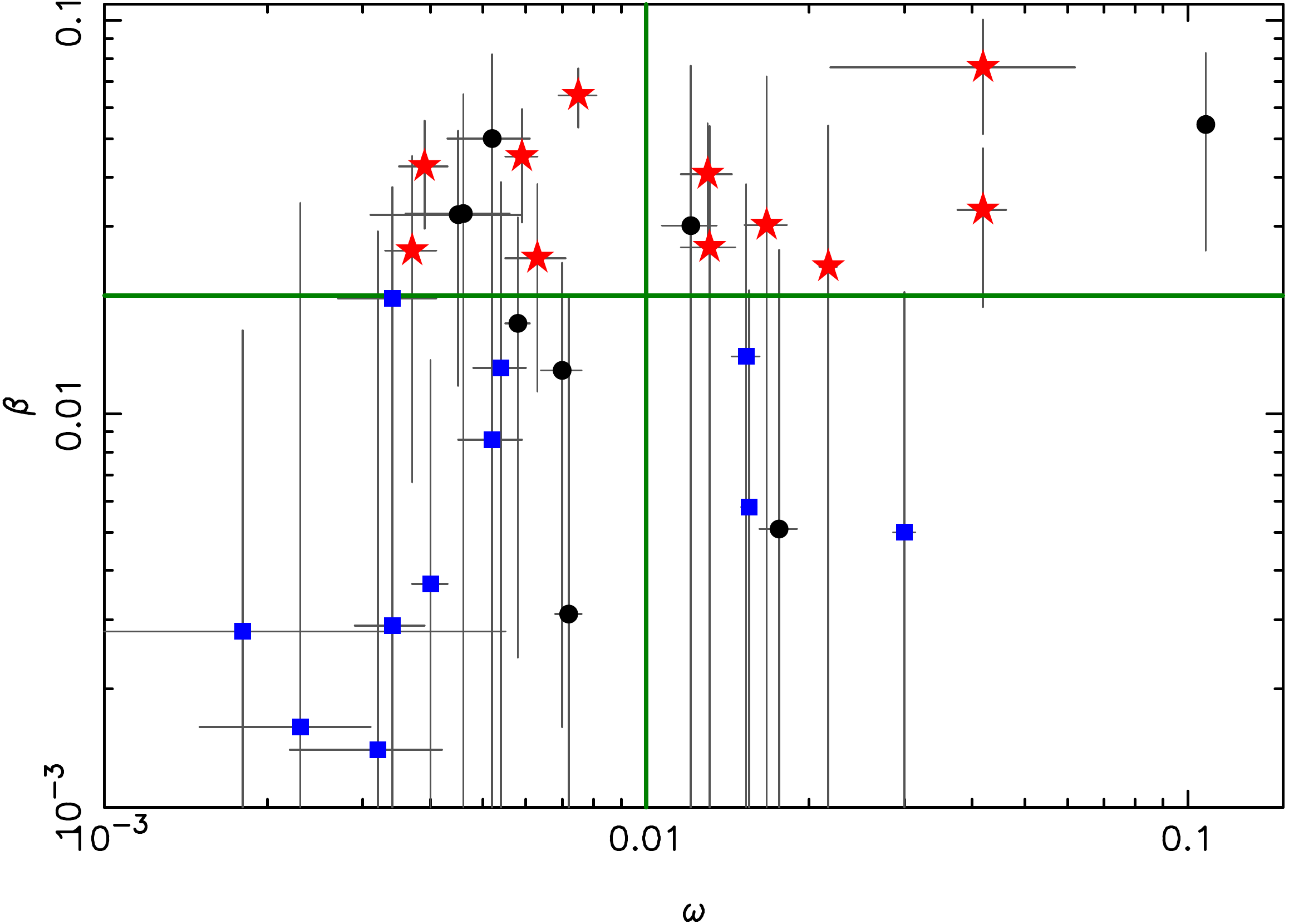}\\[4pt]
\includegraphics[width=6cm, angle=-90]{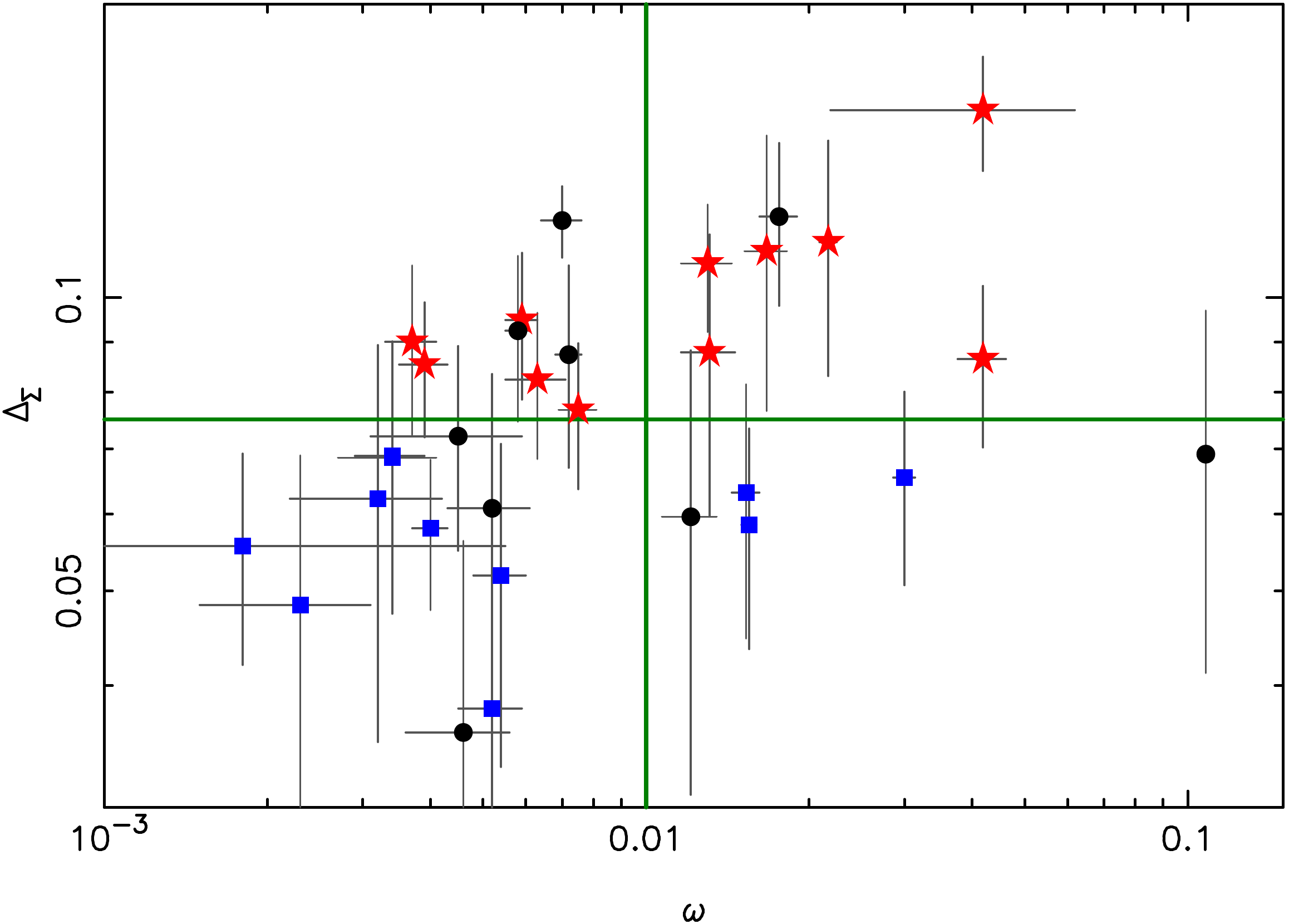}\\[4pt]
\includegraphics[width=6cm, angle=-90]{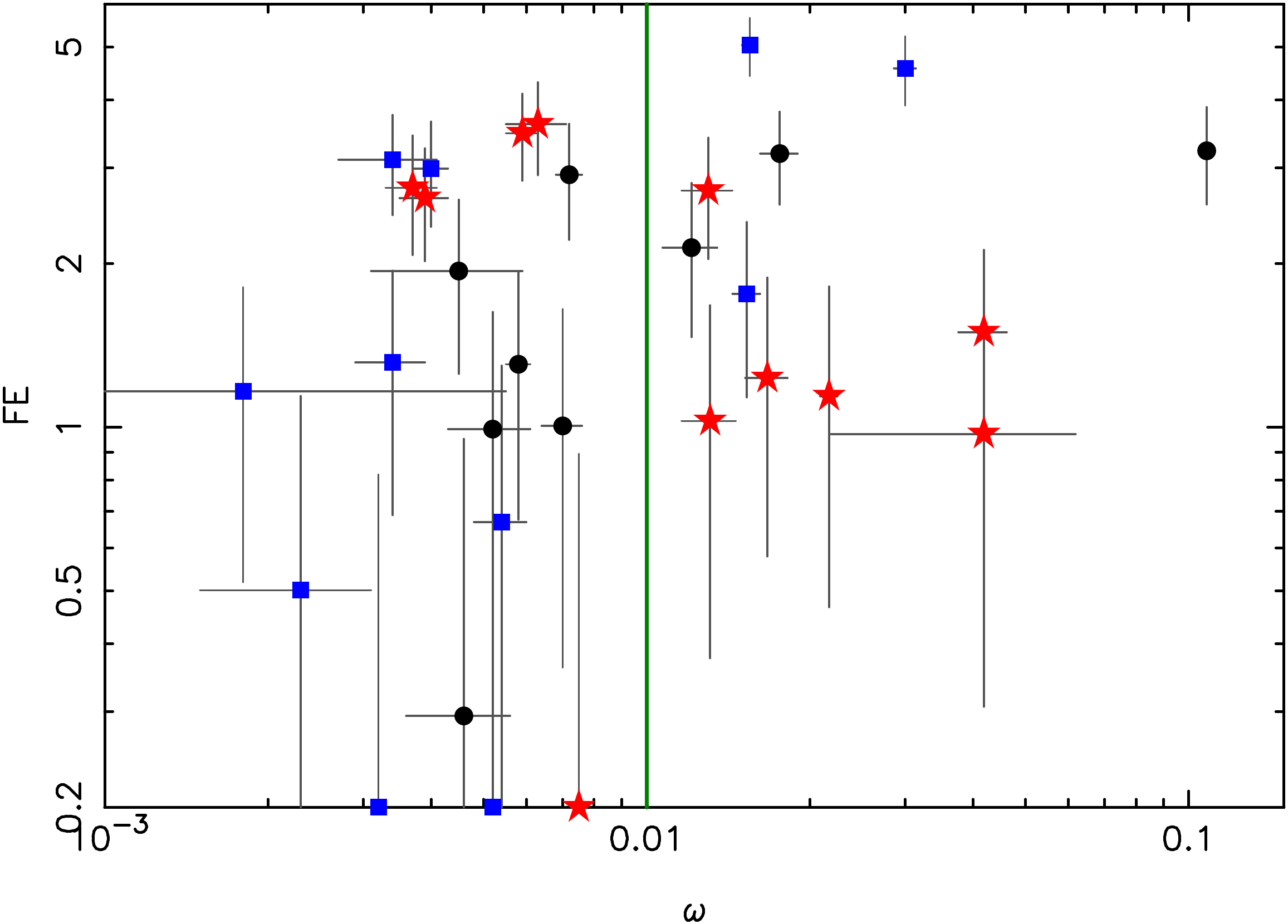}
\caption{Comparison of the X-ray centroid shift parameter, $\omega$, and the three main optical substructure indicators. In each panel, the threshold $\omega=0.01$, used by \cite{pratt09} to distinguish the morphologically disturbed clusters, is shown by the green vertical line. The green horizontal lines in the top and middle panels mark the thresholds used to classify the clusters according to the $\beta$ and $\Delta_\Sigma$ indicators. The colour code is the same as in Fig. \ref{fig:OO}.}
\label{fig:wx} 
\end{figure}

The comparison between the centroid shift and the main optical indicators is presented in Figure \ref{fig:wx}. Regarding the Fourier elongation, we observe the same trend as in Figure \ref{fig:O_FE}: the regular clusters, as classified by $\omega$, span the full range of elongation, whereas disturbed systems tend to be more elliptical with $FE>1$; the same applies for a selection based on $P_3/P_0$ (see the bottom panel of Fig. \ref{fig:P3}), which confirms that $FE$ should not be used alone as a measure of substructures. The strongest correlation between $\omega$ and the optical indicators is obtained for the residuals, $\Delta_\Sigma$. With a Spearman coefficient $\rho=0.48$ and a corresponding probability $P=0.01$ for the null hypothesis, it shows a clear relationship between the X-ray and optical morphologies. In particular, we find that $8/11\sim73\%$ of the optically regular clusters are also classified as such according to $\omega$. On the other hand, only $6/11\sim55\%$ of the optically disturbed clusters have $\omega>0.01$, suggesting that the optical indicators are sensitive to substructures that are not affecting in a sensible way the gas distribution, e.g. pre-mergers, small-mass clumps, large impact parameters. Alternatively, low statistics or projection effects from foreground/background galaxy clumps, part of a large-scale structure connected to the main clump or not, could explain these outliers; without spectroscopic information, we cannot confirm or rule out this latter possibility. Regarding the correlation between $\omega$ and the secondary indicators, we find a tight relationship with $\Delta m_{12}$ ($\rho=-0.40$, $P=0.03$) and $\Delta r_{\mathrm{BCG-X}}$ ($\rho=0.43$, $P=0.02$), which confirms that these two simple statistics trace a cluster's dynamical state.

\begin{figure}
\center
\includegraphics[width=6cm, angle=-90]{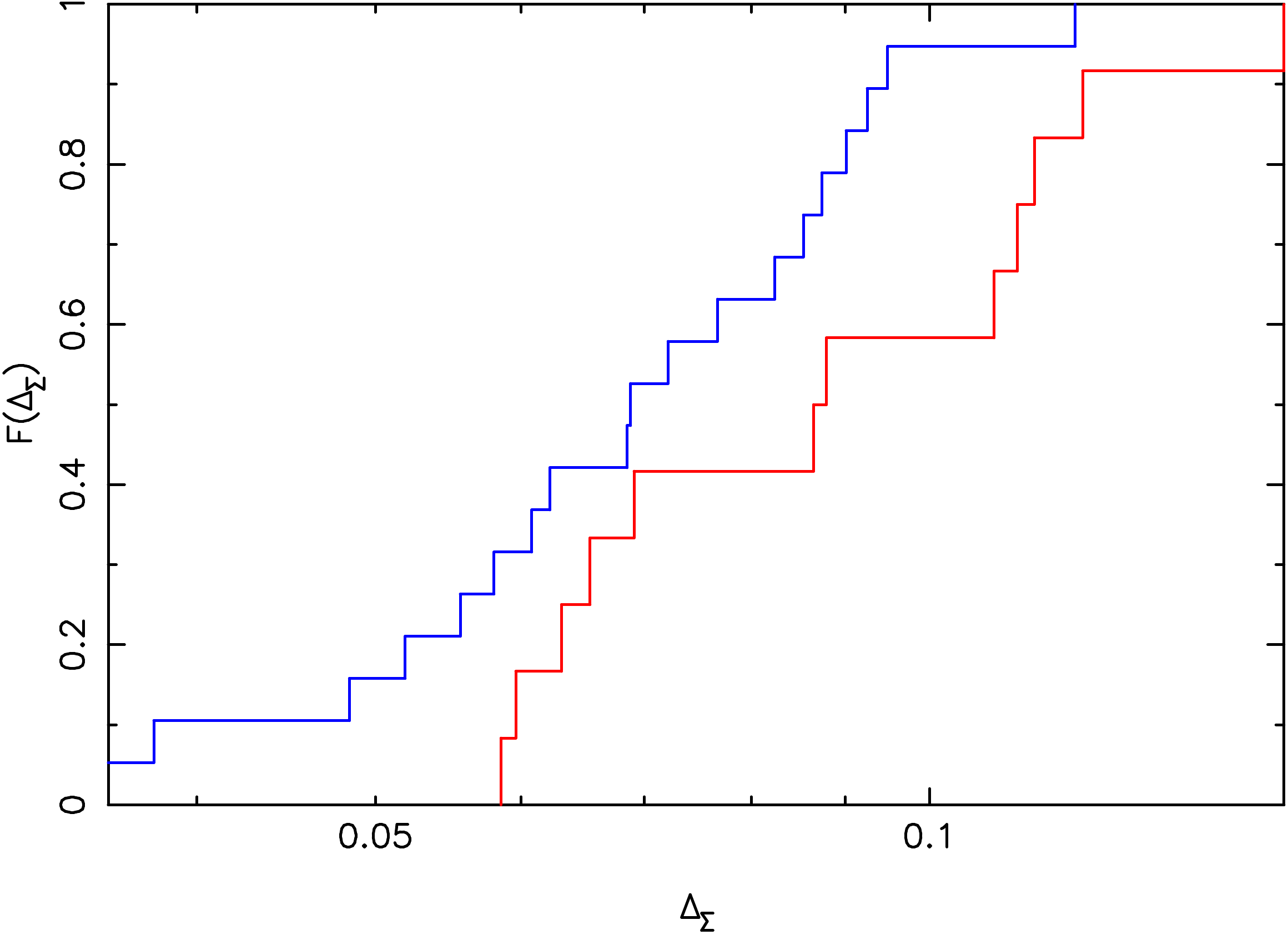}
\caption{Empirical cumulative functions of the residuals parameter, $F(\Delta_\Sigma)$, for the regular (blue) and disturbed (red) $\omega$-based populations. The maximum separation is $D=0.36$, giving a probability $P=0.22$ that they were drawn from the same parent distribution.}
\label{fig:KS_Delta} 
\end{figure}

\begin{figure}
\center
\includegraphics[width=6cm, angle=-90]{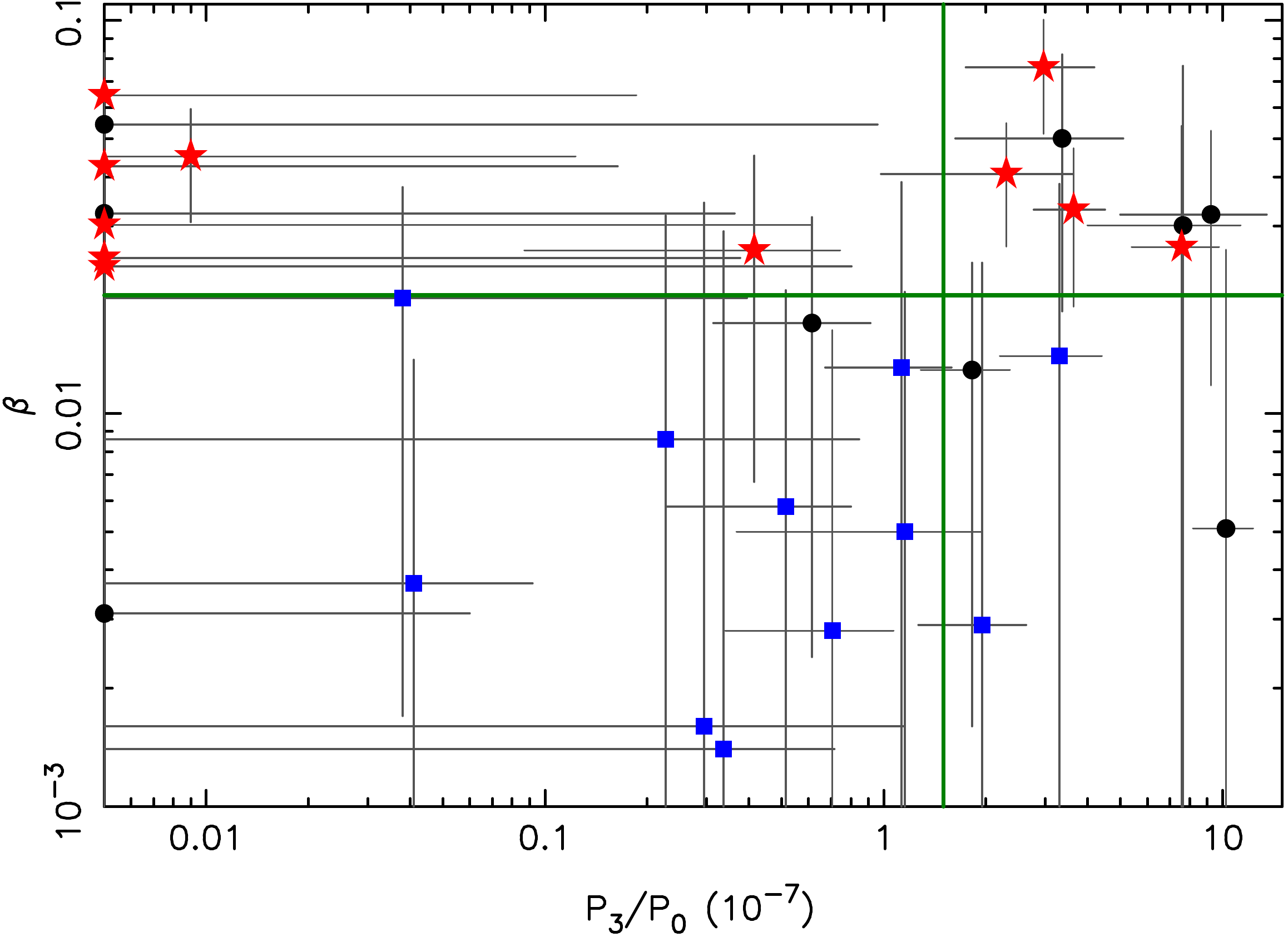}\\[4pt]
\includegraphics[width=6cm, angle=-90]{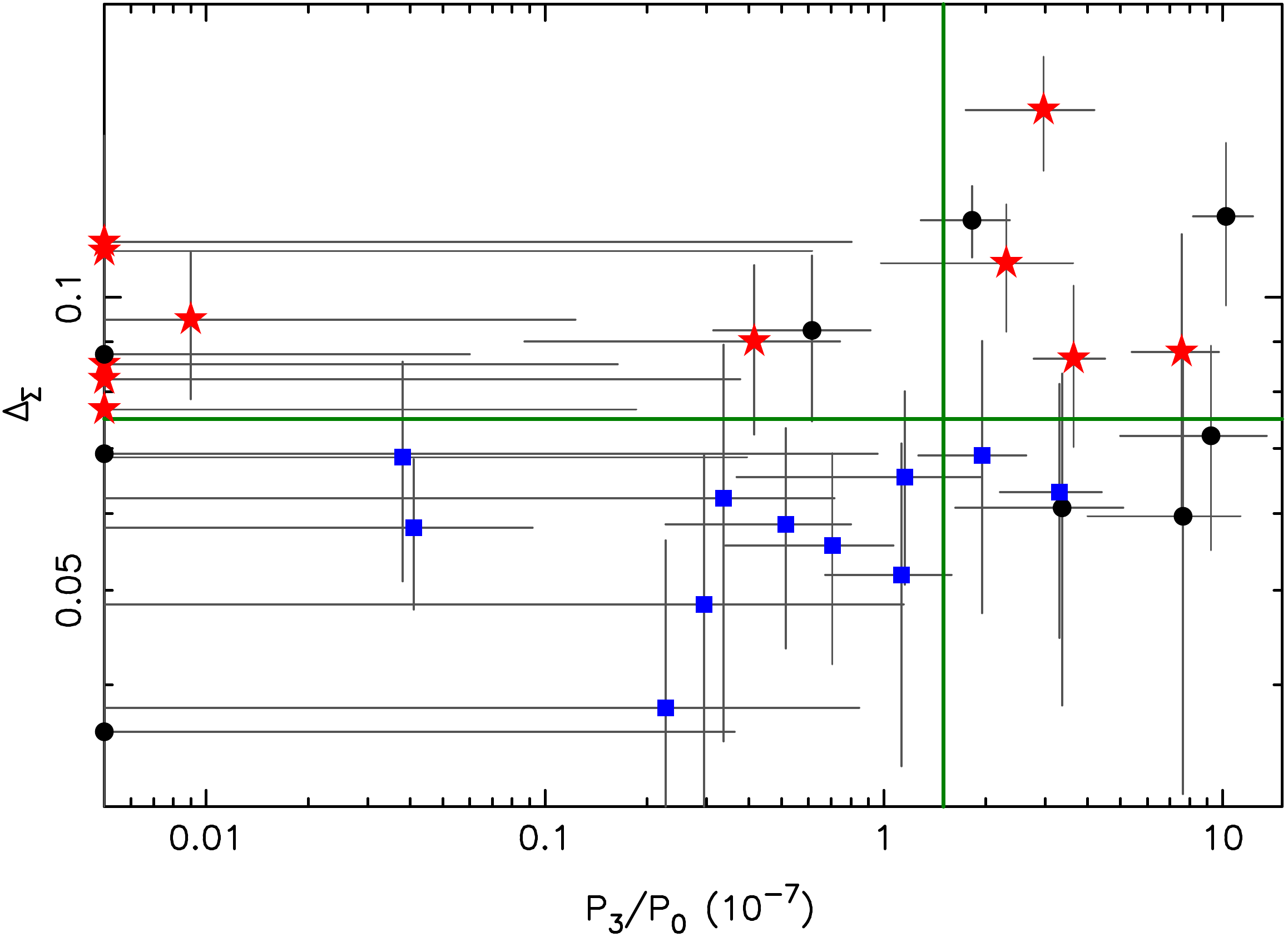}\\[4pt]
\includegraphics[width=6cm, angle=-90]{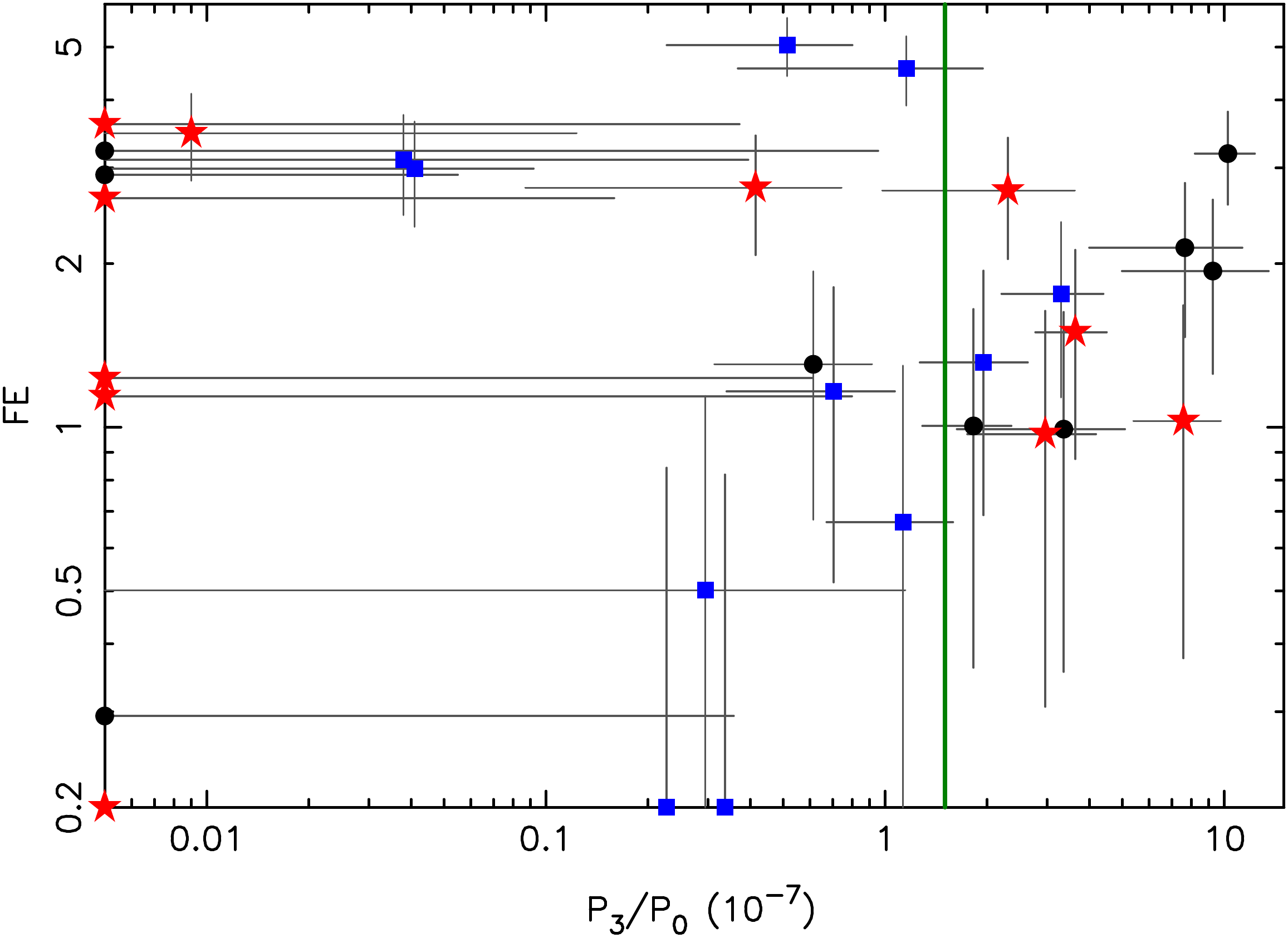}
\caption{Comparison of the X-ray power ratio $P_3/P_0$ and the three main optical substructure indicators. In each panel, the green vertical line shows the threshold $P_3/P_0=1.5\times10^{-7}$, used by \cite{boehringer10} to characterised the dynamical state of a system in terms of the power ratios $P_3/P_0$ and $P_4/P_0$. The green horizontal lines in the top and middle panels mark the thresholds used to classify the clusters according to the $\beta$ and $\Delta_\Sigma$ indicators. The colour code is the same as in Fig. \ref{fig:OO}.}
\label{fig:P3} 
\end{figure}

We also investigated the link between the properties of the ICM and the galaxy distribution by means of the two-sample Kolmogorov-Smirnov test applied to the $\omega$-based regular and disturbed sub-classes. The results, summarised in Table \ref{table:KS}, show that this classification does not separate the cluster sample into two populations with significantly different optical properties. Among the primary indicators, the residuals $\Delta_\Sigma$ provide the best way to distinguish between the two populations, however with a rather large probability $P=0.22$ that they were drawn from the same parent distribution (see Fig. \ref{fig:KS_Delta}). The photometric indicators exhibit fairly continuous distributions, in contrast to the bi-modality seen in that of $\omega$. In addition to the outliers mentioned previously, this prevents a simple mapping from the ICM morphology to that of the spatial distribution of cluster galaxies. As we will see in the next subsection, the sub-classes of cool cores and non-cool cores present much cleaner separations in terms of their optical properties. However, it is interesting to note that, among the 12 clusters with $\omega>0.01$, half are also classified as disturbed by our combined $(\beta,\Delta_\Sigma)$ criteria, and three more by either one of them. Furthermore, two of the three X-ray disturbed clusters missed by $\beta$ or $\Delta_\Sigma$ are characterised by the two largest values of $FE$ (RXCJ0145.0-5300 and RXCJ2218.6-3853). In other words, $11/12$ of the clusters with a disturbed ICM spatial distribution are also flagged as morphologically disturbed by at least one of our three main substructure indicators.

Before we detail the correlation between the hexapole power ratio $P_3/P_0$ and the other substructure indicators, it is worth recalling here that this quantity involves a weighting of the X-ray surface brightness by the third power of the cluster-centric distance. Therefore, $P_3/P_0$, whose value is mainly driven by the shape of the outer isophotes, should be a better tracer of future mergers rather than the recent mass assembly that affects the central ICM distribution. A representative example is provided in \cite{weissmann13} for the Bullet cluster (see their Fig. 14): this cluster, which is a known merger, appears regular when computing $P_3/P_0$ within an aperture of radius $R_{500}$, since the merging sub-halo lies at a distance $r\sim0.3R_{500}$ from the centre of the main halo. For the \rexcess\ sample, we do not find any significant correlation between the hexapole moment and the optical indicators. Due to the previous remark, this was expected for the secondary indicators. The lack of correlation with the primary statistics is a bit more problematic at first sight. For comparison, \cite{boehringer10} also found a rather weak correlation between $P_3/P_0$ and $\omega$, characterised by $\rho=0.21$ and $P=0.25$. The same authors investigated the influence of the aperture used to compute the power ratios on this correlation and found that it becomes significant for apertures of radii $r=0.7-0.8R_{500}$. This dependence of $P_3/P_0$ on the aperture radius makes it a rather poor indicator when applied blindly on a sample of clusters; a better approach was proposed by \cite{weissmann13} to account for disturbed clusters with substructures located close to their core.

The distribution of the clusters in the $(P_3/P_0,\beta)$ and $(P_3/P_0,\Delta_\Sigma)$ plans is presented in Figure \ref{fig:P3}. As for the $\omega$ parameter, we see that most ($9/11\sim82\%$) of the optically regular clusters are also classified as such according to the criterion $P_3/P_0<1.5\times10^{-7}$, whereas only $4/11\sim36\%$ of the disturbed systems have a $P_3/P_0$ larger than this threshold. Eight clusters have $P_3/P_0<0$, sign of a very regular X-ray morphology at large radii. It is interesting to see that five of these clusters are classified as disturbed by our optical analysis, two of which being also classified as disturbed according to $\omega$. These outliers are the reason why we do not observe any significant correlations between $P_3/P_0$ and the asymmetry $\beta$ or the residuals $\Delta_\Sigma$; removing these eight objects, we find Spearman coefficients of $\rho=0.36$ ($P=0.11$) for the former and $\rho=0.35$ ($P=0.12$) for the latter. These results confirm that the implementation of the power ratio used by \cite{boehringer10} can underestimate the substructure content of some clusters.

\begin{table}
\centering 
\begin{threeparttable}
\caption{Spearman correlation coefficients between the X-ray and optical substructure indicators.}
\label{table:O_X}
\begin{tabular}{l c c c c c c c c c}
\hline\hline\noalign{\smallskip}
 & \multicolumn{2}{c}{$\omega$} & \multicolumn{2}{c}{$P_2/P_0$} & \multicolumn{2}{c}{$P_3/P_0$}\\
  & $\rho$ & $P$ & $\rho$ & $P$ & $\rho$ & $P$\\
\noalign{\smallskip}\hline\noalign{\smallskip}
$FE$ & 0.28 & 0.13 & 0.39 & 0.03 & -0.07 & 0.76\\[3pt]
$\beta$ & 0.39 & 0.04 & 0.21 & 0.28 & -0.04 & 0.82\\[3pt]
$\Delta_\Sigma$ & 0.48 & 0.01 & -0.05 & 0.81 & 0.04 & 0.82\\[3pt]
\hline\noalign{\smallskip}
$\Delta r_{\mathrm{BCG-X}}$ & 0.43 & 0.02 & 0.21 & 0.28 & 0.15 & 0.42\\[3pt]
$\Delta m_{12}$ & -0.40 & 0.03 & -0.12 & 0.52 & 0.23 & 0.22\\[3pt]
$\Delta r_{12}$ & 0.18 & 0.35 & 0.15 & 0.45 & 0.26 & 0.16\\
\noalign{\smallskip}\hline
\end{tabular}
    \begin{tablenotes}
      \small
      \item For each pair of indicators, we give the rank correlation coefficient, $\rho$, and the probability $P$ of the null hypothesis.
    \end{tablenotes}
  \end{threeparttable}
\end{table}

\subsection{Cool cores and disturbed clusters}
\label{sec:CC}

Besides a morphological classification of the \rexcess\ clusters, \cite{pratt09} also provided a cool core classification based on the criterion $h(z)^{-2}n_{e,0}>4\times10^{-2}\,\mathrm{cm^{-3}}$, where $n_{e,0}$ is the central gas density. Ten clusters, that is $\sim32\%$ of the sample, are classified accordingly as cool cores (see Table \ref{table:sub}), for which the central cooling time is $t_{\mathrm{cool},0}<10^9$ years. Now the question that we want to address here is whether the population of cool cores differs from that of the non-cool cores in regard of their galaxy population. To investigate this, we used the two-sample Kolmogorov-Smirnov test to search for differences in the distribution of the morphological indicators between the two populations. The results are summarised in Table \ref{table:KS}.

\begin{table}
\centering 
\begin{threeparttable}
\caption{Two-sample Kolmogorov-Smirnov test applied on the substructure indicators, to distinguish cool cores from non-cool cores and disturbed from relaxed clusters.}
\label{table:KS}
\begin{tabularx}{8.cm}{l Y Y Y Y}
\hline\hline\noalign{\smallskip}
 & \multicolumn{2}{c}{CC / nCC} & \multicolumn{2}{c}{R / D}\\
 & $D$ & $P$ & $D$ & $P$\\
\noalign{\smallskip}\hline\noalign{\smallskip}
$FE$ & 0.24 & 0.80 & 0.34 & 0.33\\[3pt]
$\beta$ & 0.42 & 0.14 & 0.32 & 0.38\\[3pt]
$\Delta_\Sigma$ & 0.33 & 0.36 & 0.36 & 0.22\\[3pt]
\hline\noalign{\smallskip}
$\Delta r_{\mathrm{BCG-X}}$ & 0.51 & 0.04 & 0.32 & 0.37\\[3pt]
$\Delta m_{12}$ & 0.29 & 0.56 & 0.34 & 0.29\\[3pt]
$\Delta r_{12}$ & 0.57 & 0.01 & 0.37 & 0.20\\[3pt]
$c_{500}$ & 0.46 & 0.08 & 0.28 & 0.53\\[3pt]
\hline\noalign{\smallskip}
$\omega$ & 0.32 & 0.45 & - & -\\[3pt]
$P_2/P_0$ & 0.28 & 0.60 & 0.78 & $<10^{-4}$\\[3pt]
$P_3/P_0$ & 0.24 & 0.85 & 0.48 & 0.09\\
\noalign{\smallskip}\hline
\end{tabularx}
    \begin{tablenotes}
      \small
      \item Columns: (1) Substructure indicator. (2,3) Two-sample Kolmogorov-Smirnov statistic, $D$, and probability of the null hypothesis that cool cores and non-cool core clusters are drawn from the same parent distribution. (4,5) Same as columns (2,3) but for regular and disturbed clusters, as classified according to the X-ray centroid shift parameter $\omega$.
    \end{tablenotes}
  \end{threeparttable}
\end{table}

For the main optical indicators, we find that the asymmetry parameter $\beta$ provides the best separation, with a maximum distance between the two cumulative distributions $D=0.42$, giving a probability $P=0.14$ that cool cores and non-cool cores were drawn from the same parent distribution. The former are also characterised by smaller residuals $\Delta_\Sigma$, the probability for the null hypothesis being larger in this case ($P=0.36$). Furthermore, we find that only $2/10=20\%$ of the cool cores are morphologically disturbed according to our optical classification based on ($\beta,\Delta_\Sigma$). Therefore, we can conclude that cool cores are more likely to be found in regular systems. In terms of the X-ray substructure indicators, we find that the two populations are better separated by $\omega$, but with a large probability $P=0.45$ for the null hypothesis. So according to the Kolmogorov-Smirnov test, cool cores and non-cool cores have a similar distribution of substructures. This result is rather surprising and is certainly the consequence of the test being more sensitive to the median of the cumulative distributions rather than their wings. To highlight this point, we can remark that, according to the $\omega$ classification, only two cool cores are morphologically disturbed, one being also classified as such by our optical analysis (RXCJ2319.6-7313).

Looking at the results of the Kolmogorov-Smirnov test for the secondary indicators, we see that they separate better the two populations than the primary indicators. In particular, the offset between the central BCG and the X-ray emission peak, $\Delta r_{\mathrm{BCG-X}}$, gives $D=0.51$ and $P=0.04$. This clearly indicates that cool cores are hosted by clusters for which no significant mergers occurred in their recent past. This result also confirms that the different $\Delta r_{\mathrm{BCG-X}}$ distributions of X-ray and SZ selected cluster samples are the consequence of a cool-core bias, that is, an over-representation of cool cores in X-ray samples depending on the selection method. The offset between the central and second BCGs, $\Delta r_{12}$, provides the best way to distinguish cool cores from non-cool cores ($D=0.57$, $P=0.01$), the latter being characterised by smaller offsets. This is consistent with cool cores forming in systems with a quiet recent formation history: the absence of major mergers leaves enough time for the ICM to condensate via radiative cooling and for the massive satellites to infall on the central BCG via dynamical friction, thus increasing the probability for the second BCG to be found at a large distance from the cluster centre. In that case, the second BCG may be a bright galaxy observed at an early stage of its first orbit, isolated or part of a substructure (as argued in Sec. \ref{sec:OO} given the correlation between $\Delta r_{12}$ and $\Delta_\Sigma$), leading to a small magnitude gap $\Delta m_{12}$. This explains why the $\Delta m_{12}$ distributions of cool cores and non-cool cores are not significantly different according to the Kolmogorov-Smirnov test ($D=0.29$, $P=0.56$).

\begin{figure}
\center
\includegraphics[width=6.5cm, angle=-90]{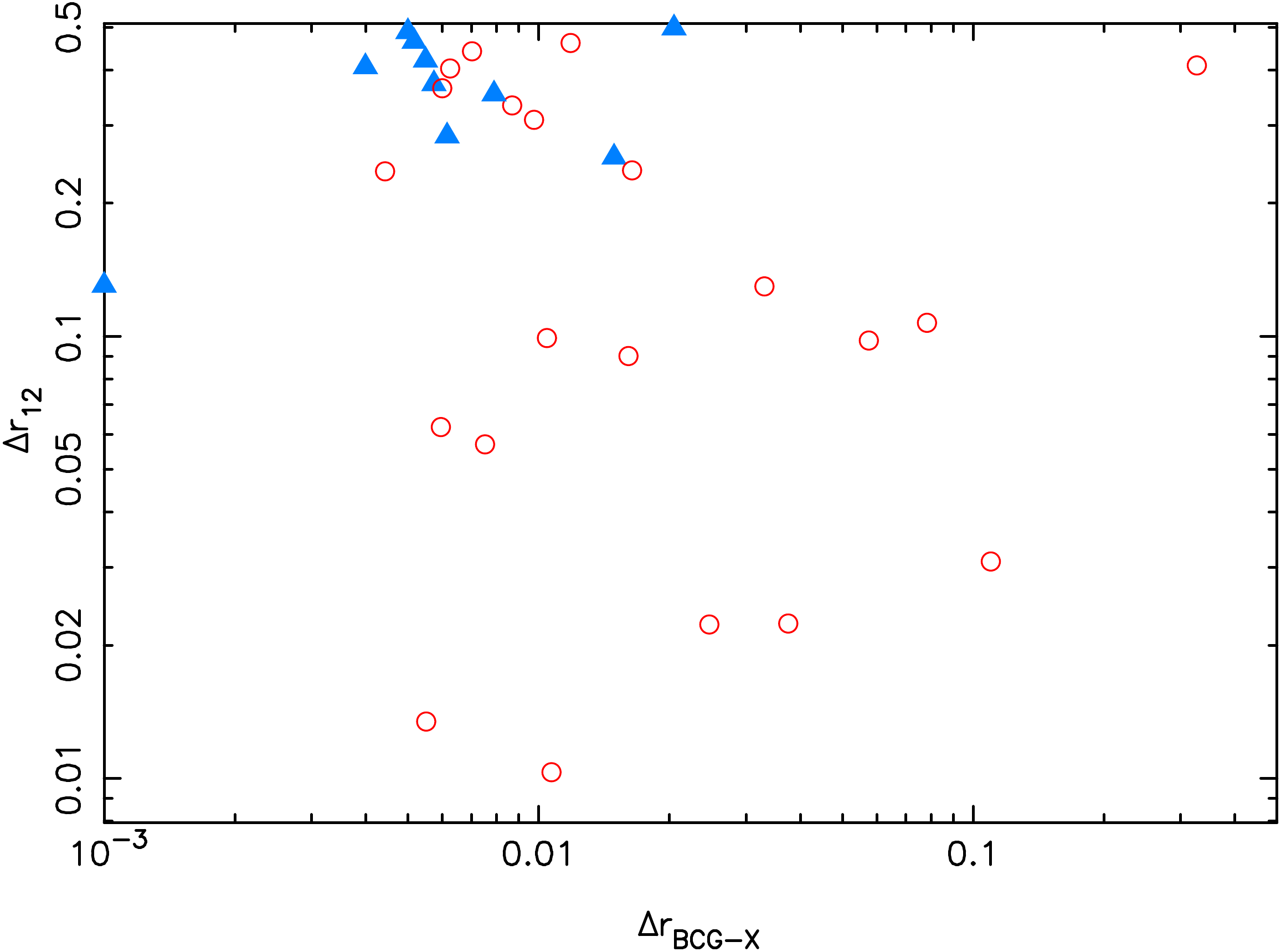}
\caption{Distribution of the cool cores (blue-filled triangles) and non-cool cores (empty red circles) in the $(\Delta r_{\mathrm{BCG-X}},\Delta r_{12})$ diagram.}
\label{fig:rXBCG_r12} 
\end{figure}

Given the results obtained for $\Delta r_{\mathrm{BCG-X}}$ and $\Delta r_{12}$, we can look whether the combination of these two simple parameters could provide a way of selecting cool cores without relying on the more complex X-ray analysis. The distribution of the \rexcess\ clusters in the $(\Delta r_{\mathrm{BCG-X}},\Delta r_{12})$ plan is presented in Figure \ref{fig:rXBCG_r12}. All the cool cores are located in a region defined by $\Delta r_{\mathrm{BCG-X}}\lesssim0.02$ and $\Delta r_{\mathrm{12}}>0.1$. Conversely, $8/18\sim44\%$ of the clusters located in this region are non-cool cores, four of which being also classified as disturbed according to $\omega$. Therefore, these criteria on $\Delta r_{\mathrm{BCG-X}}$ and $\Delta r_{12}$ do not provide a secure, albeit complete, selection of cool-core systems. However, it could be used as an interesting diagnosis to investigate possible selection biases between different cluster samples in a more precise way than what was done by \cite{rossetti16} with $\Delta r_{\mathrm{BCG-X}}$ only. It is also worth noting that the cut $\Delta r_{\mathrm{BCG-X}}<0.02$ that separates cool cores from non-cool cores is the same as the criterion used by these authors to classify a cluster as relaxed, thus all \rexcess\ cool cores are found in relaxed systems, as one would expect. 

\begin{figure}
\center
\includegraphics[width=6.5cm, angle=-90]{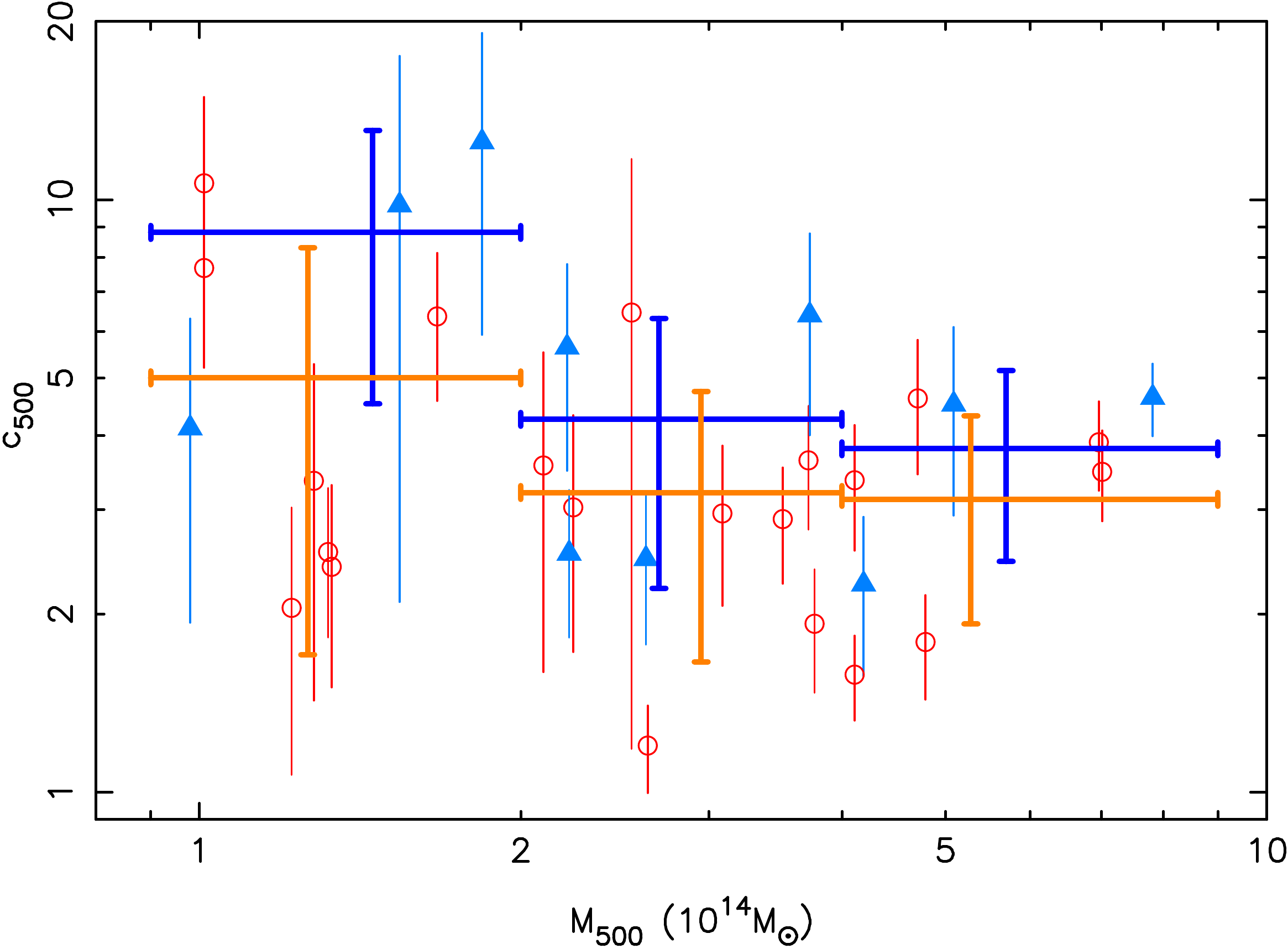}
\caption{Optical concentration as a function of mass for the cool cores (blue-filled triangles) and non-cool cores (empty red circles). The average value and scatter of the concentration in three mass bins are highlighted in dark blue and orange for the cool cores and non-cool cores, respectively.}
\label{fig:c500} 
\end{figure}

In addition to the set of primary and secondary indicators, we also estimated to the optical concentration parameter of each cluster. It is defined as $c_{500}=R_{500}/r_s$, where the scale radius, $r_s$, was obtained by fitting a spherical NFW model (plus background) to the galaxy distribution, using the maximum likelihood estimator technique \citep{sarazin80}. The results are presented in Figure \ref{fig:c500}. We can first remark an anti-correlation between mass and optical concentration, a well-known trend for the concentration parameter of the mass profile, which indicates that galaxies are a good tracer of their cluster-scale dark matter halo host. Moreover, we find that clusters with a cool core tend to have a more concentrated galaxy distribution than non-cool cores, regardless of their mass. It has been shown that relaxed clusters are, on average, characterised by higher concentrations (e.g. \citealt{deboni13}). Thus our results agree again with the scenario of cool cores forming in systems having evolved to a relaxed state due to the absence of major mergers.

\section{Summary}

In this work we have studied the substructure content of the \rexcess\ sample of galaxy clusters, based on the spatial distribution of their red-sequence galaxies, up to a radius $R_{500}$. We used two sets of substructure indicators. Our primary tools were the asymmetry test, $\beta$, the residuals between the galaxy surface density map and its best-fit model, $\Delta_\Sigma$, and the Fourier elongation, $FE$. We probed the cluster cores with secondary indicators, namely the offset between the central BCG and the X-ray emission peak, $\Delta r_{\mathrm{BCG-X}}$, the magnitude offset between the first and second BCGs, $\Delta m_{12}$, and their radial offset, $\Delta r_{12}$. We searched for rank correlations between these statistics and with the X-ray results derived by \cite{pratt09} and \cite{boehringer10}, based on the centroid shift parameter, $\omega$, and the power ratios $P_2/P_0$ and $P_3/P_0$. Finally, we compared the optical properties of cool cores and non-cool cores. Our main results are as follows.

\begin{itemize}
\item The distribution of the main indicators are rather continuous, preventing a simple partition of the clusters into regular and disturbed systems. This contrasts with the bimodal distribution of $\omega$, used by \cite{pratt09} to classify the clusters.
\item The strongest correlation among the main statistics is obtained for $\Delta_\Sigma-\beta$. Based on these two parameters, we proposed a tentative classification of the clusters leading to a fraction $\sim35\%$ of clusters with a significant level of substructures, a value similar to the $\sim39\%$ obtained by \cite{pratt09} using $\omega$.
\item $FE$ does not correlate with the other substructure indicators. Specifically, we found that the regular clusters cover the full range of $FE$, whereas the systems with substructures tend to have larger $FE$.
\item The secondary indicators are, in most cases, uncorrelated with the main statistics, suggesting that the dynamical state of the clusters core does not need to be representative of their overall substructure content.
\item The correlation $\Delta r_{\mathrm{BCG-X}}-\Delta m_{12}$ is the strongest one among all correlations, confirming previous claims that $\Delta m_{12}$ is a fair tracer of a cluster's dynamical state. However, we did not find a clear separation between the optically regular and disturbed sub-classes in the $\Delta r_{\mathrm{BCG-X}}-\Delta m_{12}$ plan.
\item $\Delta r_{12}$ has a significant anti-correlation with $\Delta r_{\mathrm{BCG-X}}$ and correlation with $\Delta m_{12}$: small offsets $\Delta r_{12}$ are expected to be more frequent in systems going through the merger of a substructure massive enough to produce a large $\Delta r_{\mathrm{BCG-X}}$ and to host a massive elliptical galaxy.
\item $\omega$ presents an interesting correlation with $\Delta_\Sigma$ and $\beta$. In particular, $\sim73\%$ of the optically regular systems are also classified as such according to $\omega$. However, only $\sim55\%$ of the disturbed ones have large $\omega$, possibly due to the presence of pre-mergers, small-mass substructures or large impact parameters that would not affect the ICM in a sensible way; alternatively, projection effects could increase the observed substructure level of the galaxy distribution.
\item 11 of the 12 $\omega$-disturbed clusters are also flagged as such by at least one of our main optical indicators: nine by either $\Delta_\Sigma$ or $\beta$ (six by both), two of the remaining three having the largest $FE$ values in the entire sample.
\item $\omega$ correlates with the secondary indicators $\Delta r_{\mathrm{BCG-X}}$ and $\Delta m_{12}$, which also suggests that this statistic is more sensitive to the current dynamical state of a cluster as compared to its global substructure content.
\item $FE$ and $P_2/P_0$ correlate with each other. The latter is a clear sign of substructures, thus clusters with an elongated ICM and a galaxy spatial distributions characterised by a large $FE$ are likely to be systems growing via a collimated infall along a preferential axis.
\item We did not find any significant correlation between $P_3/P_0$ and the main optical indicators. We did find that $\sim82\%$ of the optically regular systems have small $P_3/P_0$, however, only $\sim36\%$ of the disturbed ones are also characterised by large $P_3/P_0$. Moreover, we found that among the eight systems with $P_3/P_0=0$, five are classified as disturbed in optical due to the fact that the hexapole ratios are biased to the cluster outskirts. Excluding the $P_3/P_0=0$ systems, we did find that this statistic correlates with the main optical substructure indicators.
\item Clusters hosting a cool core are characterised by higher optical concentrations, smaller $\Delta r_{\mathrm{BCG-X}}$, and larger $\Delta r_{12}$ compared to the non-cool core systems. We interpreted the difference in $\Delta r_{12}$ between the two populations as the consequence of galaxy cannibalism due to dynamical friction.
\item We found that cool cores are located within a small region in the $\Delta r_{\mathrm{BCG-X}}-\Delta r_{12}$ plan, making the combination of these two statistics an interesting tool to investigate a possible selection bias in X-ray selected cluster samples.
\end{itemize}

We started the photometric substructure analysis of \rexcess\ as a complement of the existing X-ray studies, with the goal of understanding possible causes for the wide range of results found in the literature regarding the frequency of morphologically-disturbed galaxy clusters. By working on the same cluster sample, we were able to focus on the impact of using different data sets and substructure indicators, putting aside intrinsic selection effects. Besides the fact that clusters cannot be readily sorted into two simple sub-classes, our results suggest that it is important to make the distinction between disturbed morphologies and disturbed dynamical states. On the one hand, a cluster can exhibit a complex distribution of its galaxy members on large scales, while having a core showing no sign of a recent merger, hence a relaxed dynamical state. On the other hand, a cluster hosting a single, massive substructure that is observed shortly after crossing its core will be easily classified as dynamically young but otherwise characterised by a regular overall shape. The secondary indicators and the X-ray centroid shift appears to be more sensitive to the recent formation history of a cluster and, consequently, its current dynamical state, whereas the global statistics can pick up substructures in its outer region, hence probing its future mass assembly.

The various statistics used in this work respond differently to the orbital stage, mass ratio, and impact parameter of a substructure. The situation is further complicated by the practical implementation of these tests. For instance, we saw that the residuals of the galaxy surface density depend on the smoothing of the maps. Another striking example was found with $P_3/P_0$, whose weighting scheme makes it too sensitive to the shape of the ICM close to $R_{500}$, thus missing substructures at intermediate radii. The choice of the aperture is another important factor. We restricted our analysis to within $R_{500}$ for consistency with the work of \cite{pratt09} and \cite{boehringer10}. However, a quick look at the galaxy surface density maps reveals the presence of substructures at larger radii, whose inclusion would certainly affect our results.

To summarise, it appears important to combine different data sets and different approaches to fully appreciate the structure of galaxy clusters. The cheap and fast secondary indicators may be sufficient to identify dynamically young systems for which the mass estimators that rely on the hypothesis of dynamical equilibrium may be biased. Statistics involving the overall spatial distribution of cluster members should provide more accurate results regarding the total amount of substructures, bearing in mind the possibility of false detection due to projection effects. This work, on a somewhat small sample, shows the utility of a multi-wavelength approach to assess the structure of a cluster. A similar analysis on a larger sample would provide more statistically reliable constraints regarding the occurence of clusters hosting a substantial amount of substructures.

%

\bibliography{../references}

\appendix

\onecolumn
\section{Galaxy surface density maps}
\begin{figure}[h]
\center
\includegraphics[width=5.cm, angle=-90]{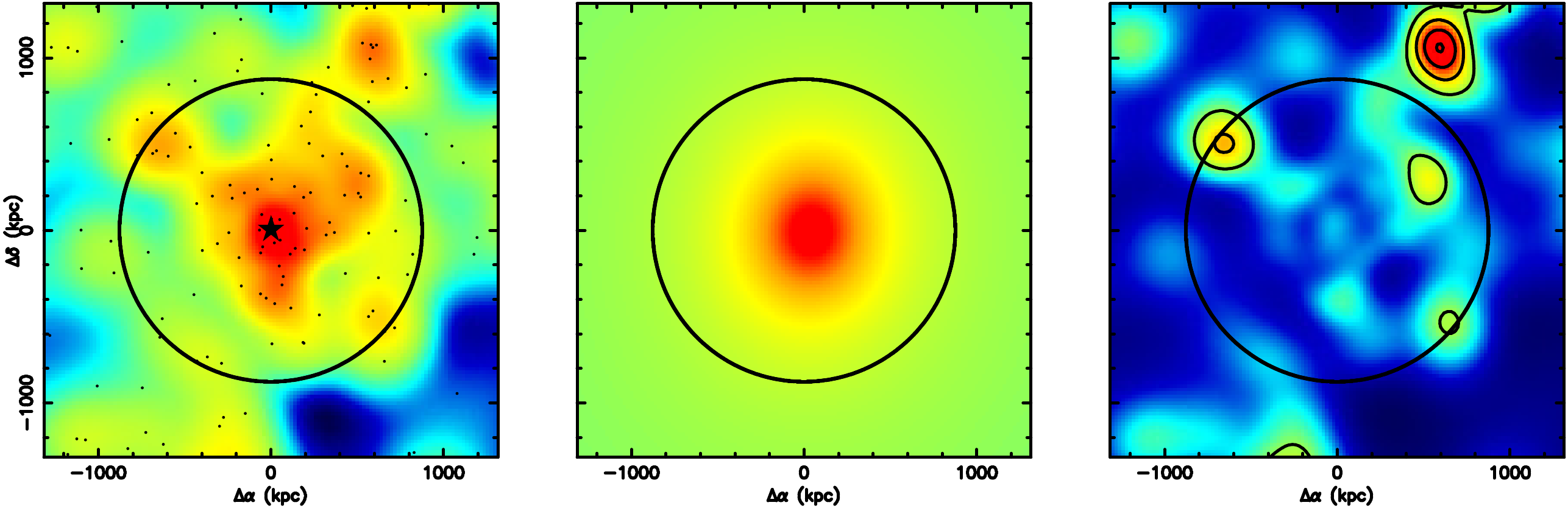}\\[3pt]
\includegraphics[width=5.cm, angle=-90]{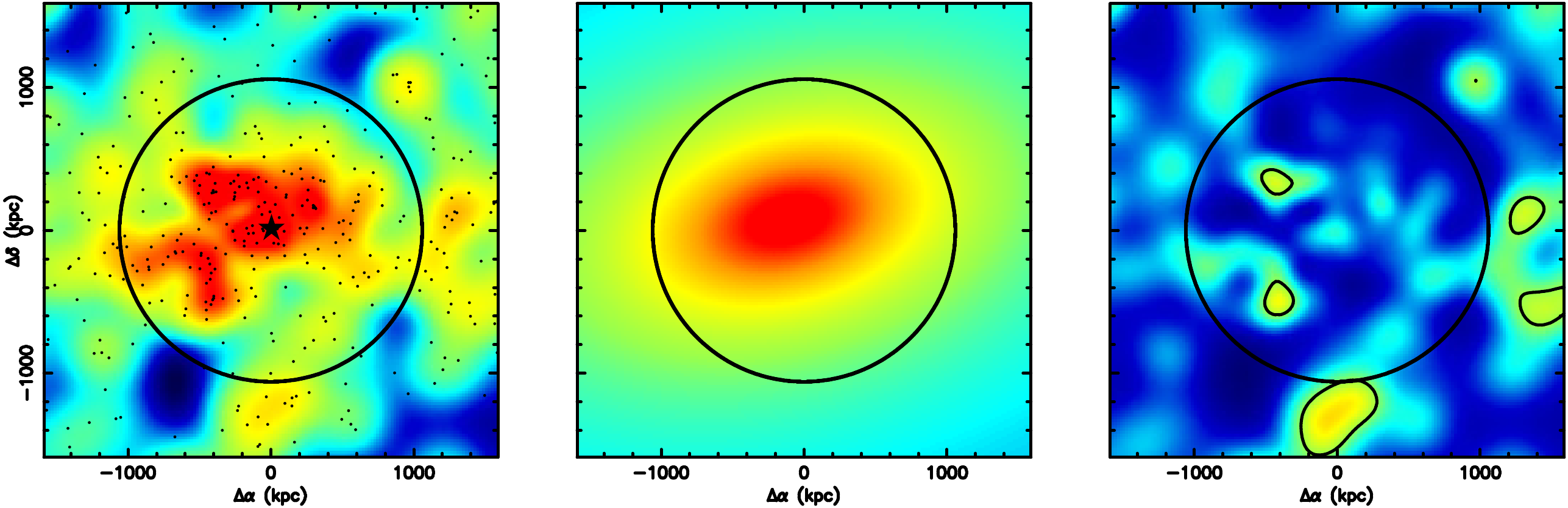}\\[3pt]
\includegraphics[width=5.cm, angle=-90]{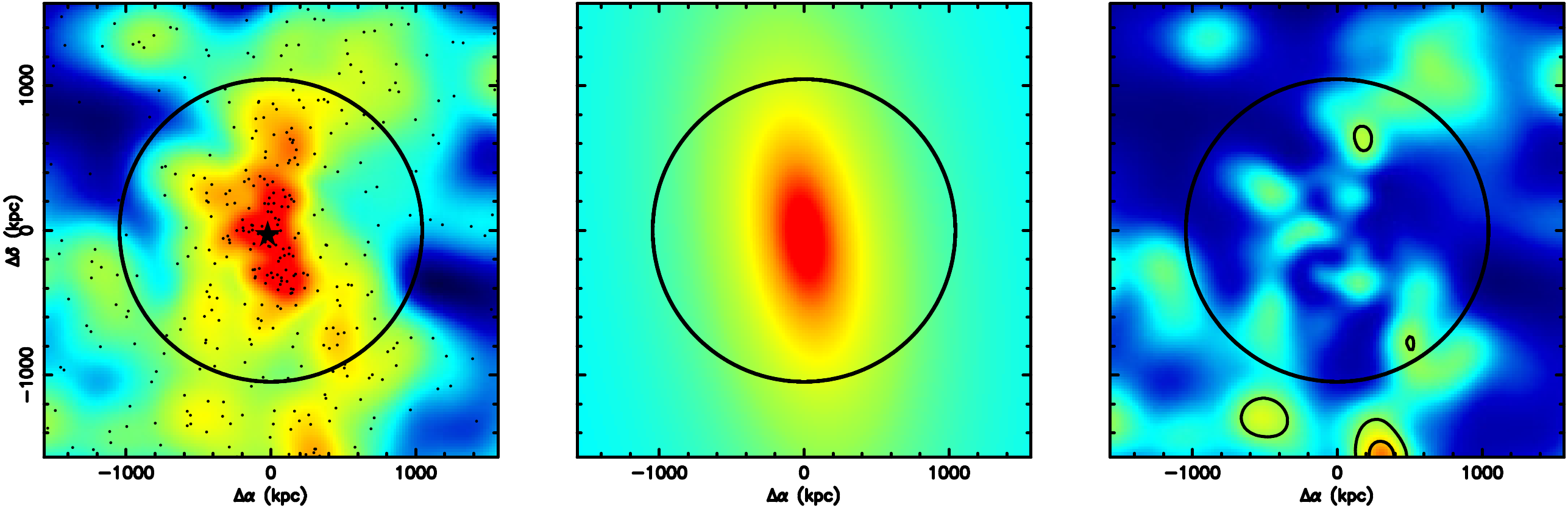}\\[3pt]
\includegraphics[width=5.cm, angle=-90]{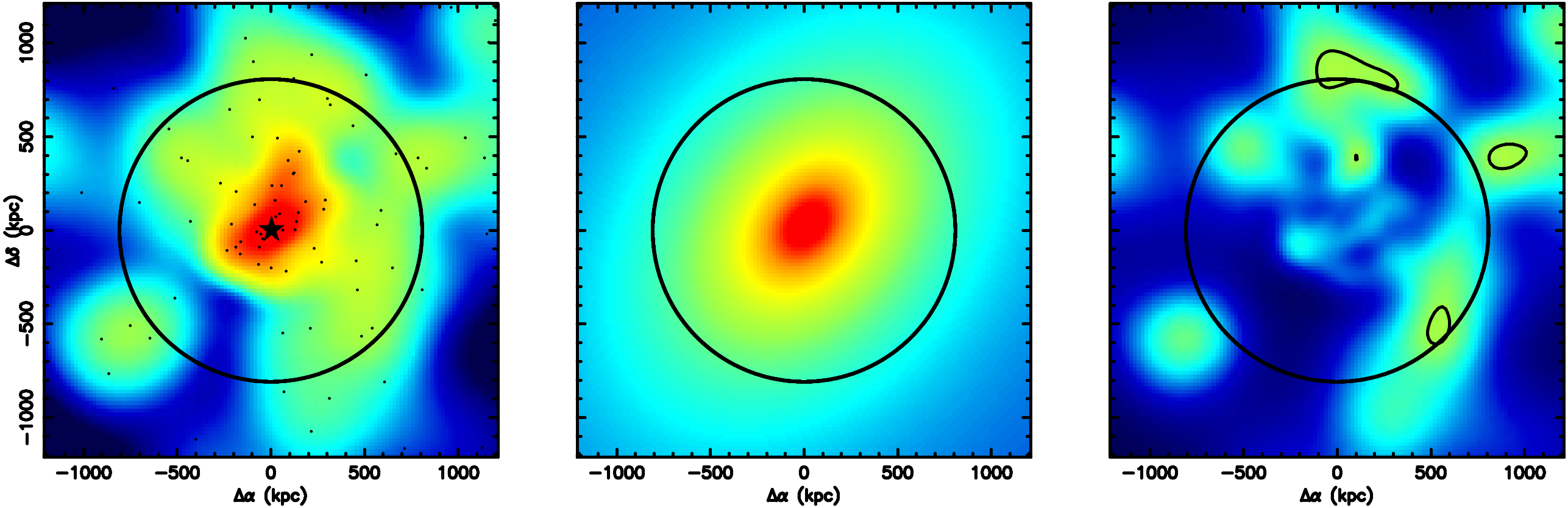}
\caption{Galaxy surface density map (left), best-fit model (middle) and residuals (right) for RXCJ0003.8+0203, RXCJ0006.0-3443, RXCJ0020.7-2542, and RXCJ0049.4-2931 from top to bottom, respectively. See Figure \ref{fig:delta_0547} for more details.}
\label{fig:Nmaps}
\end{figure}

\begin{figure}[h]
\center
\includegraphics[width=5.cm, angle=-90]{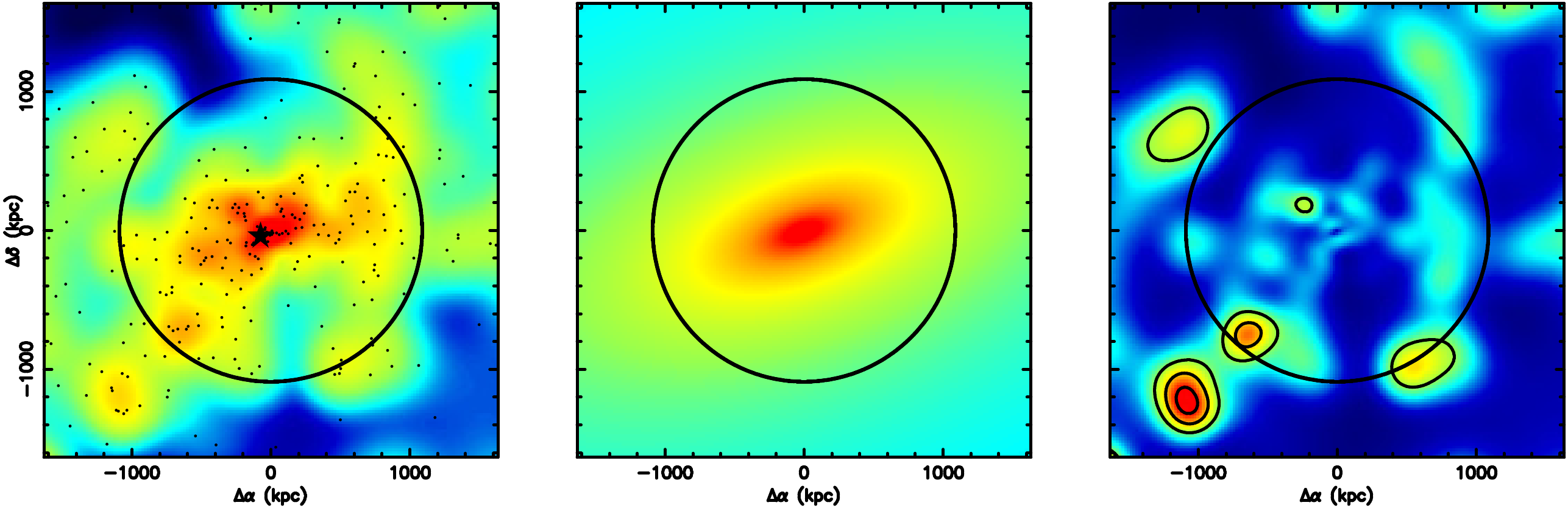}\\[3pt]
\includegraphics[width=5.cm, angle=-90]{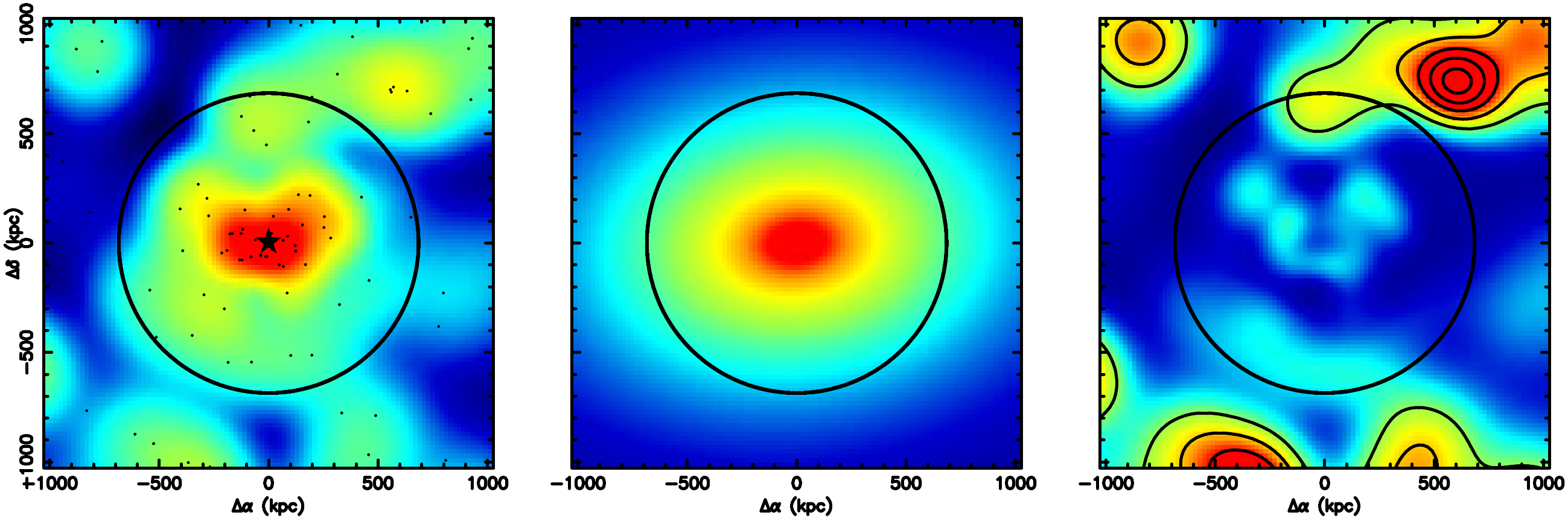}\\[3pt]
\includegraphics[width=5.cm, angle=-90]{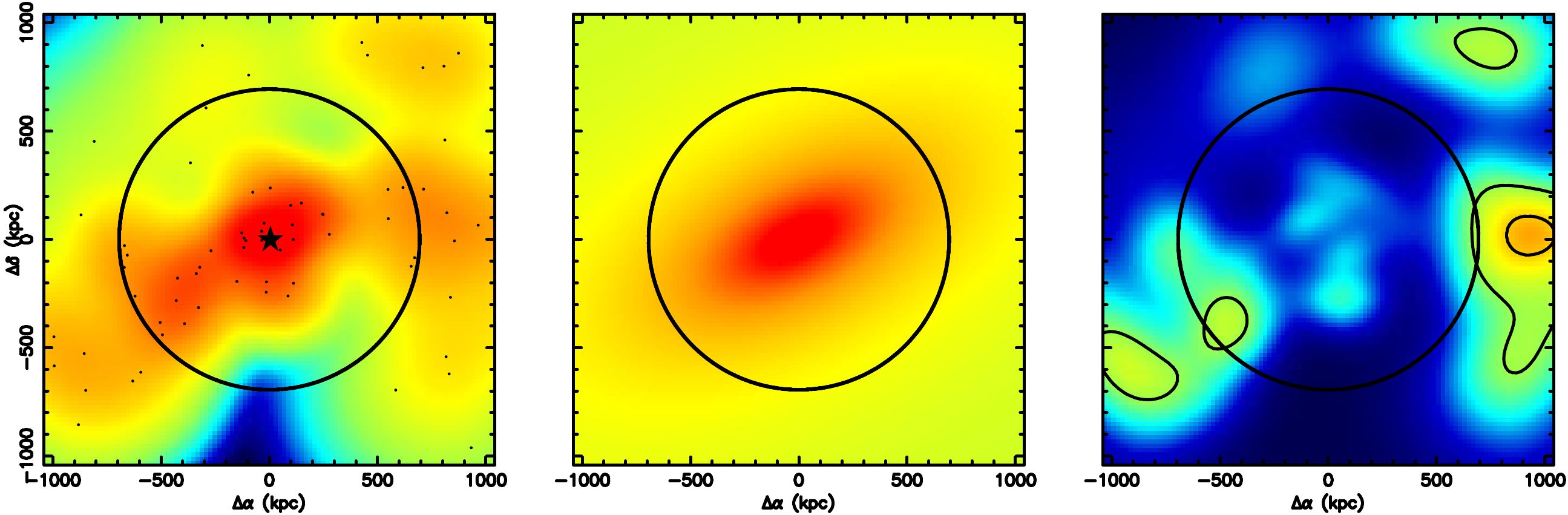}\\[3pt]
\includegraphics[width=5.cm, angle=-90]{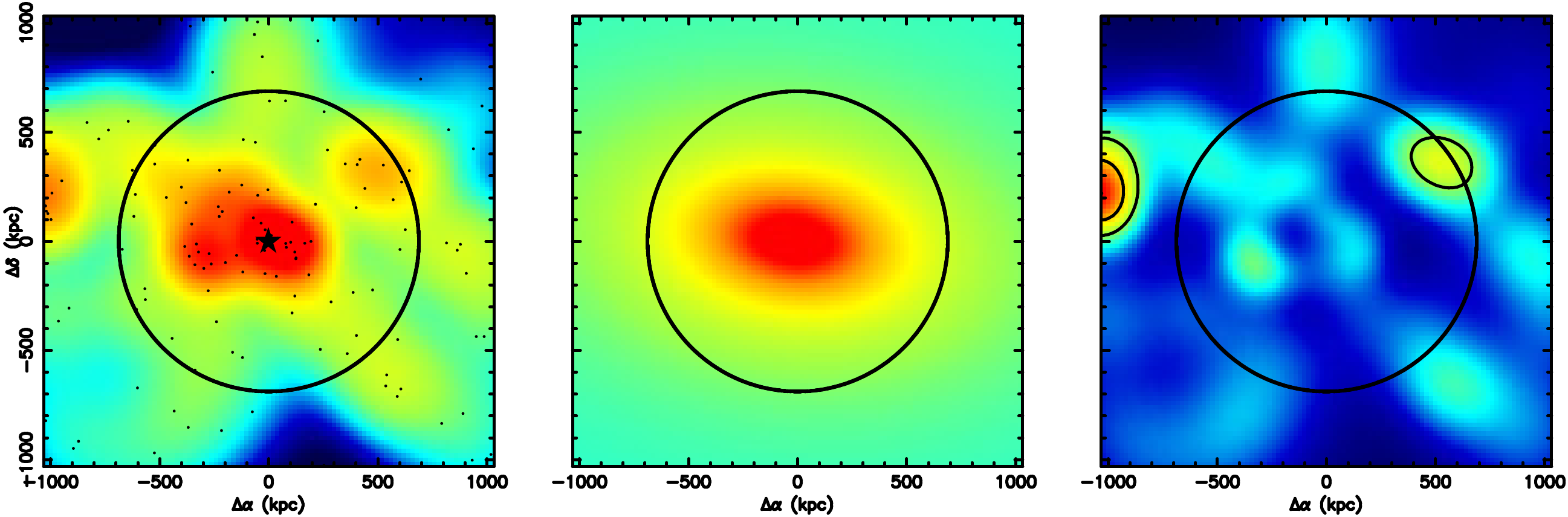}
\caption{Same as Figure \ref{fig:Nmaps}, for RXCJ0145.0-5300, RXCJ0211.4-4017, RXCJ0225.1-2928, and RXCJ0345.7-4112 from top to bottom, respectively.}
\end{figure}

\begin{figure}[h]
\center
\includegraphics[width=5.cm, angle=-90]{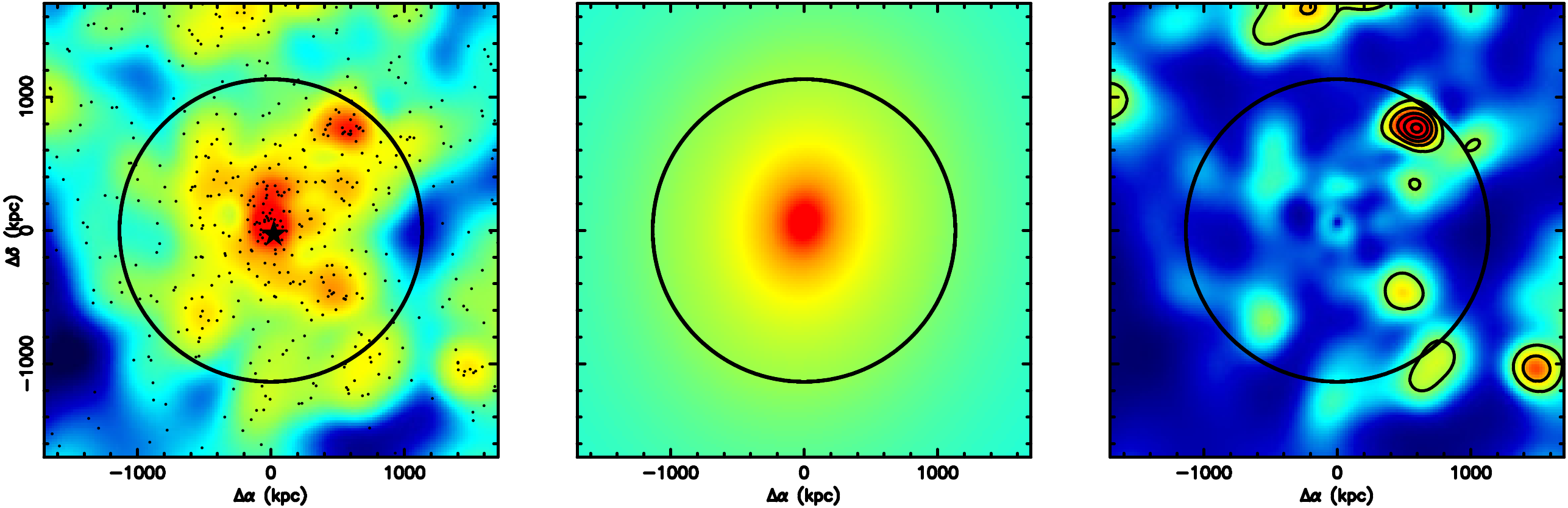}\\[3pt]
\includegraphics[width=5.cm, angle=-90]{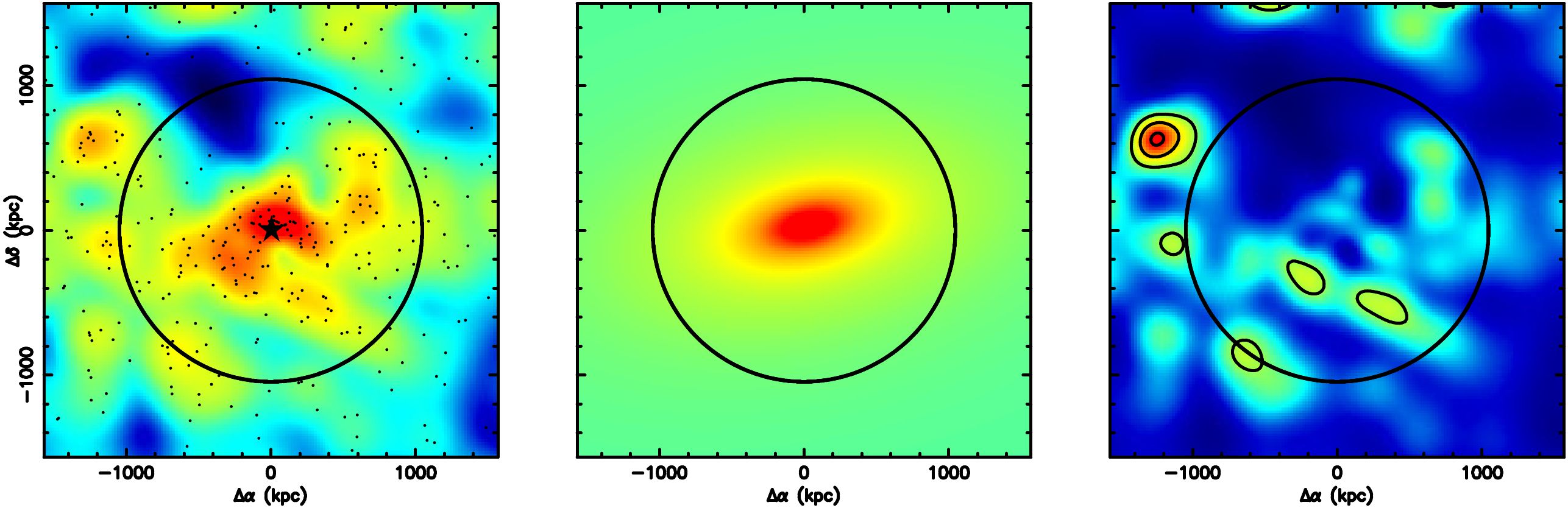}\\[3pt]
\includegraphics[width=5.cm, angle=-90]{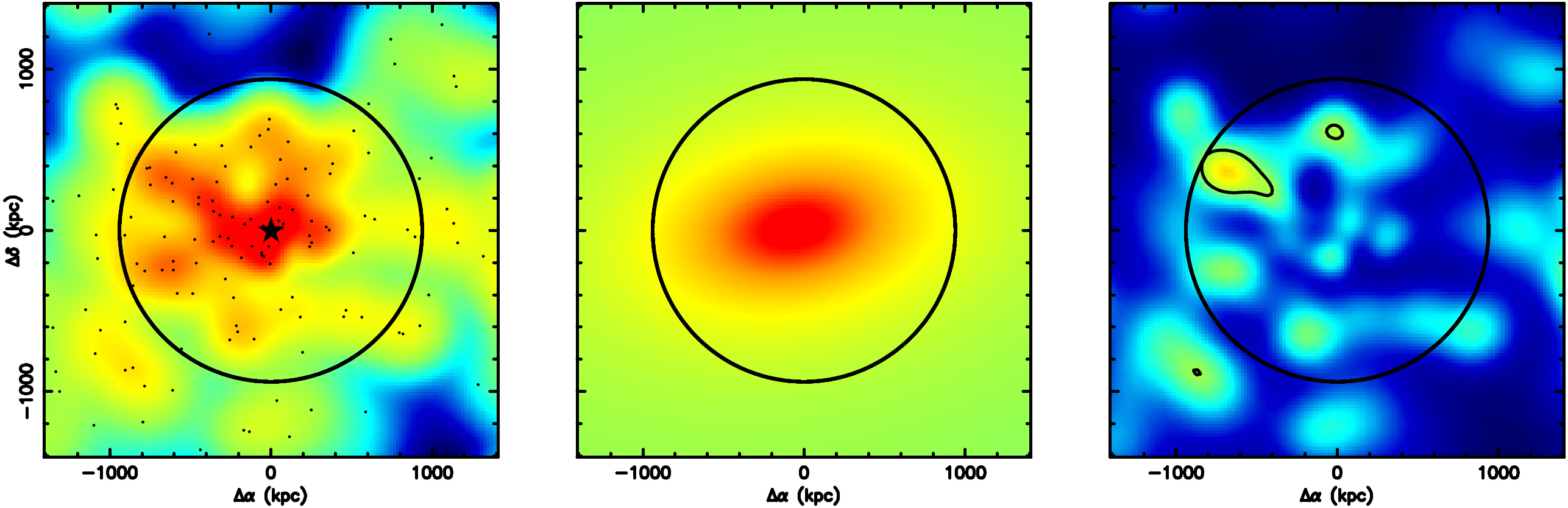}\\[3pt]
\includegraphics[width=5.cm, angle=-90]{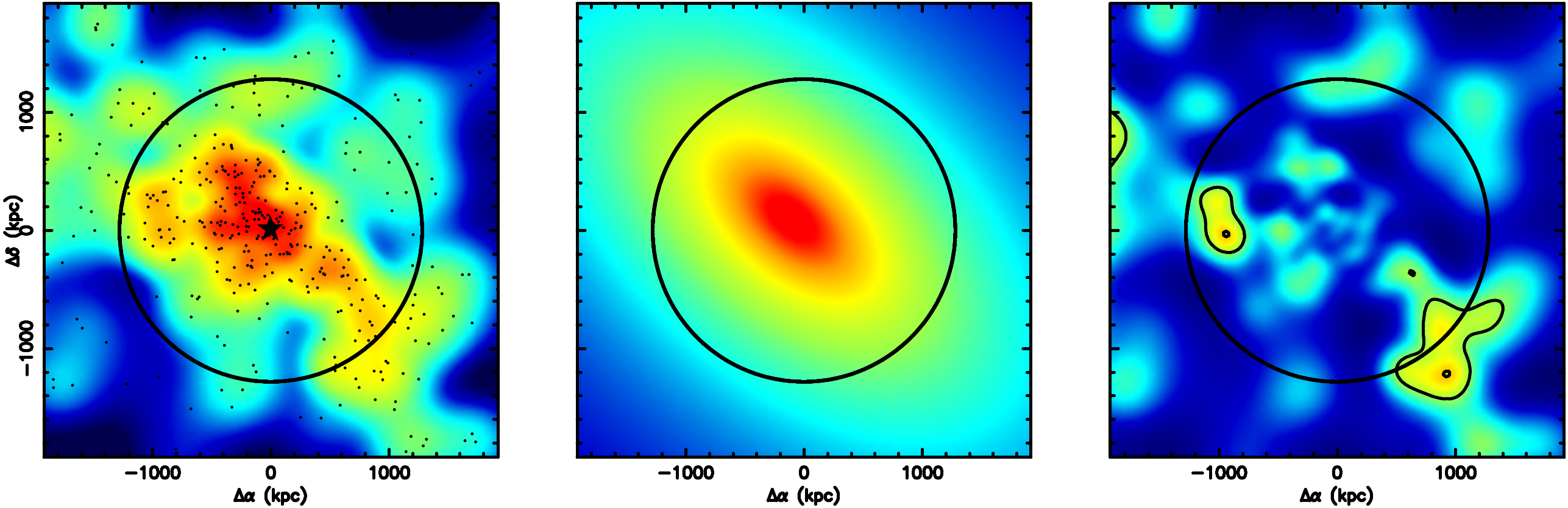}
\caption{Same as Figure \ref{fig:Nmaps}, for RXCJ0547.6-3152, RXCJ0605.8-3528, RXCJ0616.8-4748, and RXCJ0645.4-5413 from top to bottom, respectively.}
\end{figure}

\begin{figure}[h]
\center
\includegraphics[width=5.cm, angle=-90]{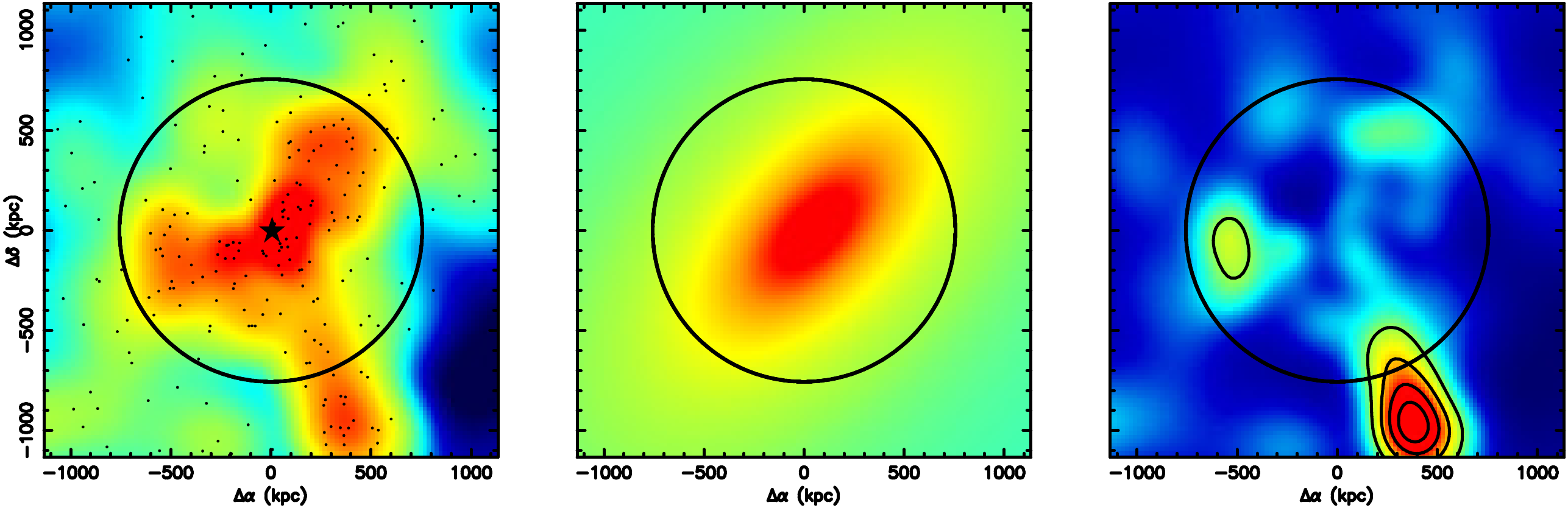}\\[3pt]
\includegraphics[width=5.cm, angle=-90]{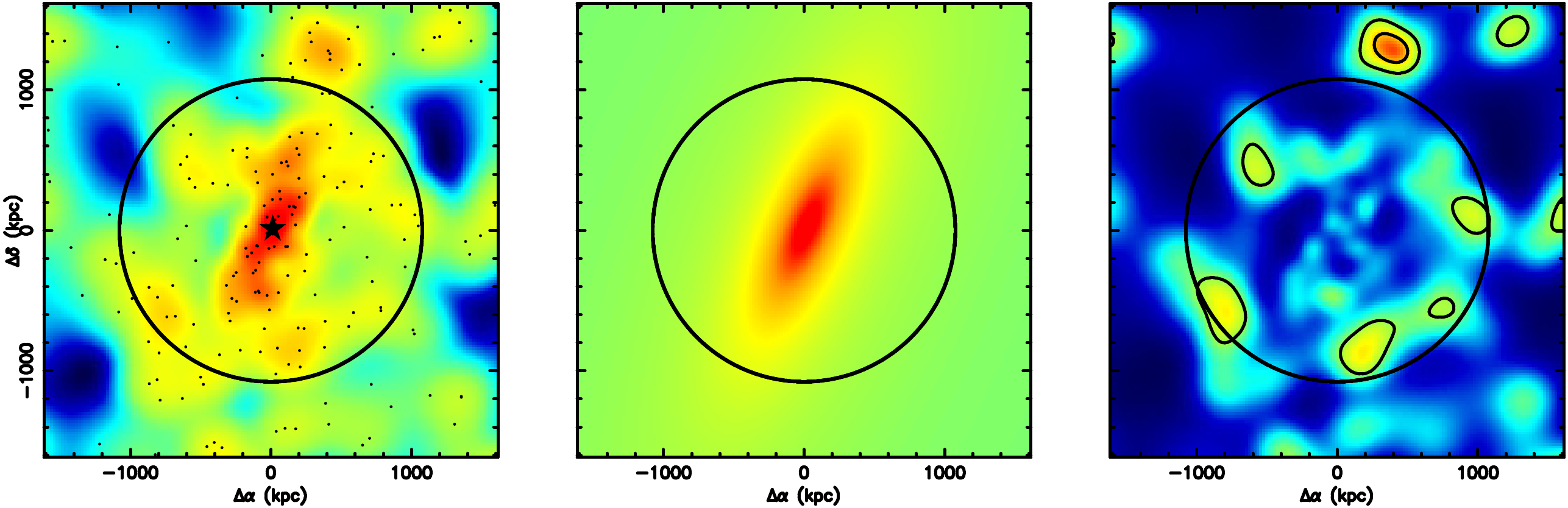}\\[3pt]
\includegraphics[width=5.cm, angle=-90]{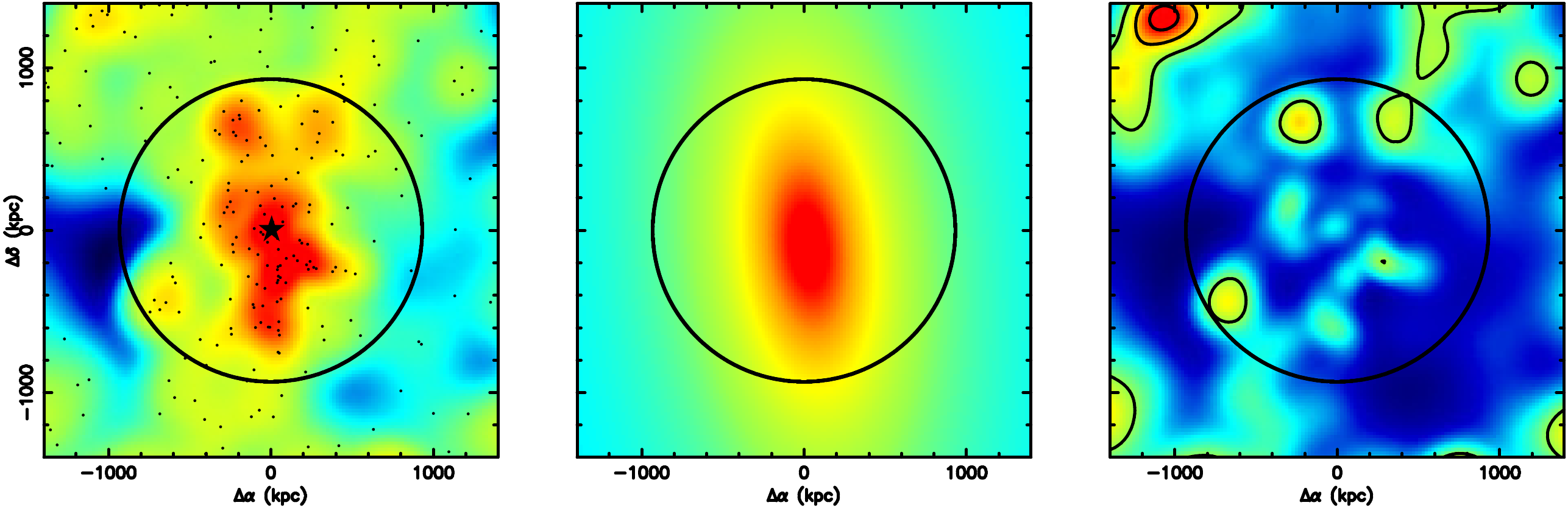}\\[3pt]
\includegraphics[width=5.cm, angle=-90]{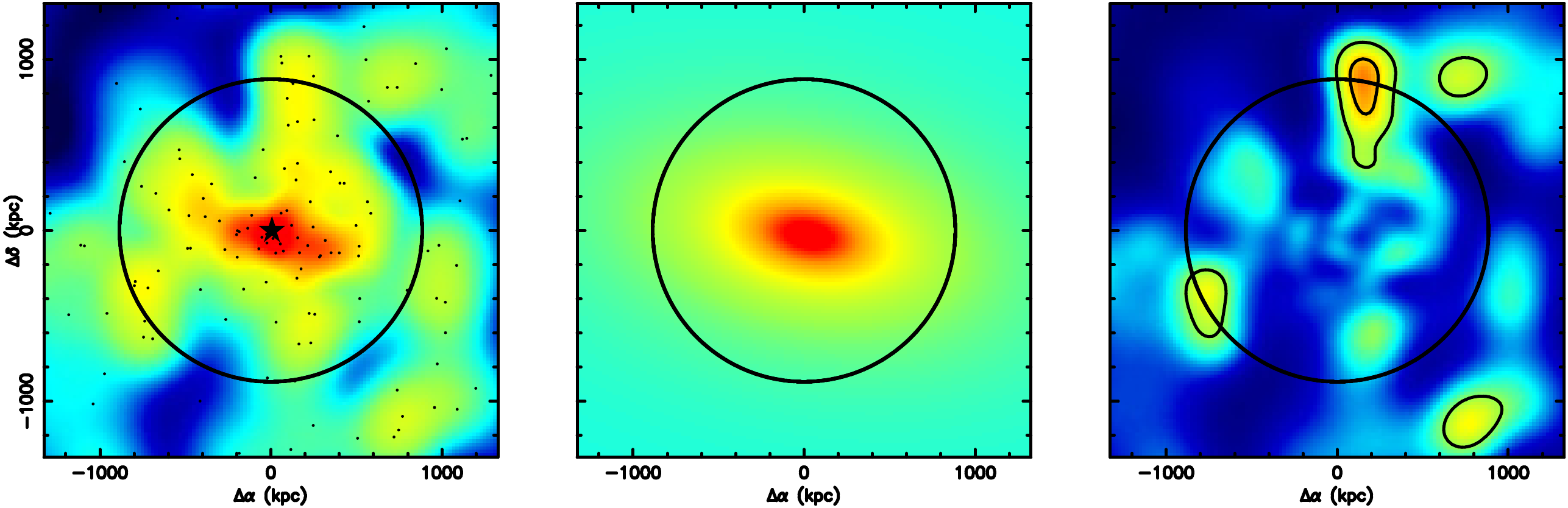}
\caption{Same as Figure \ref{fig:Nmaps}, for RXCJ0821.8+0112, RXCJ0958.3-1103, RXCJ1044.5-0704, and RXCJ1141.4-1216 from top to bottom, respectively.}
\end{figure}

\begin{figure}[h]
\center
\includegraphics[width=5.cm, angle=-90]{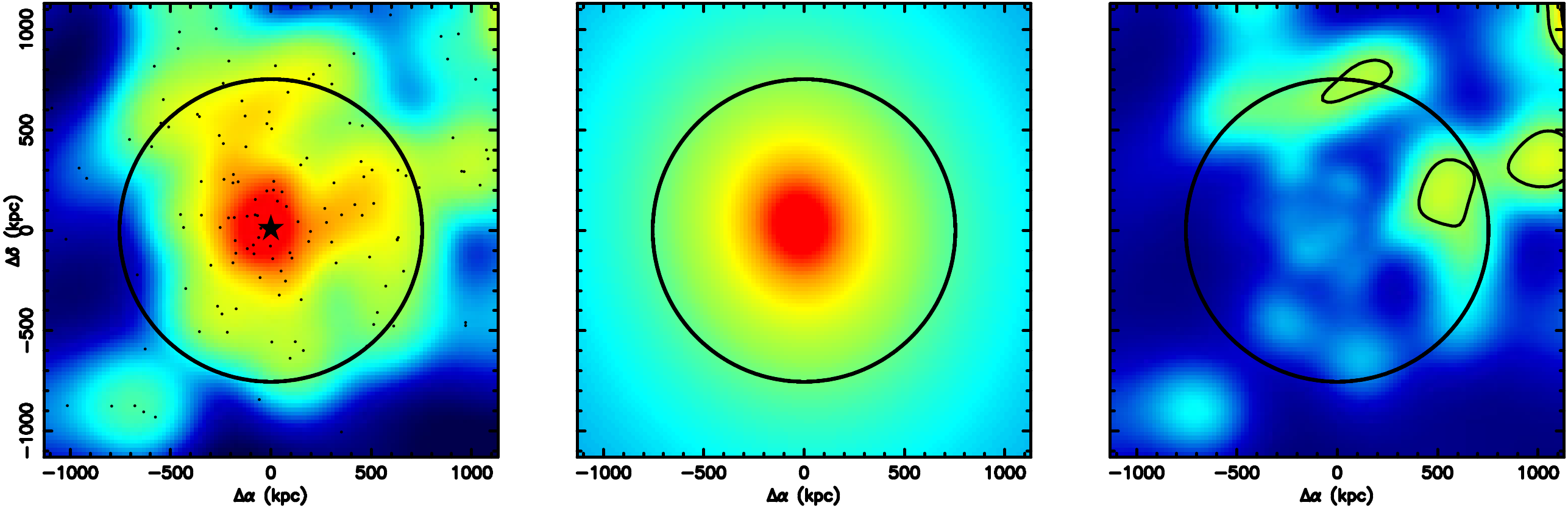}\\[3pt]
\includegraphics[width=5.cm, angle=-90]{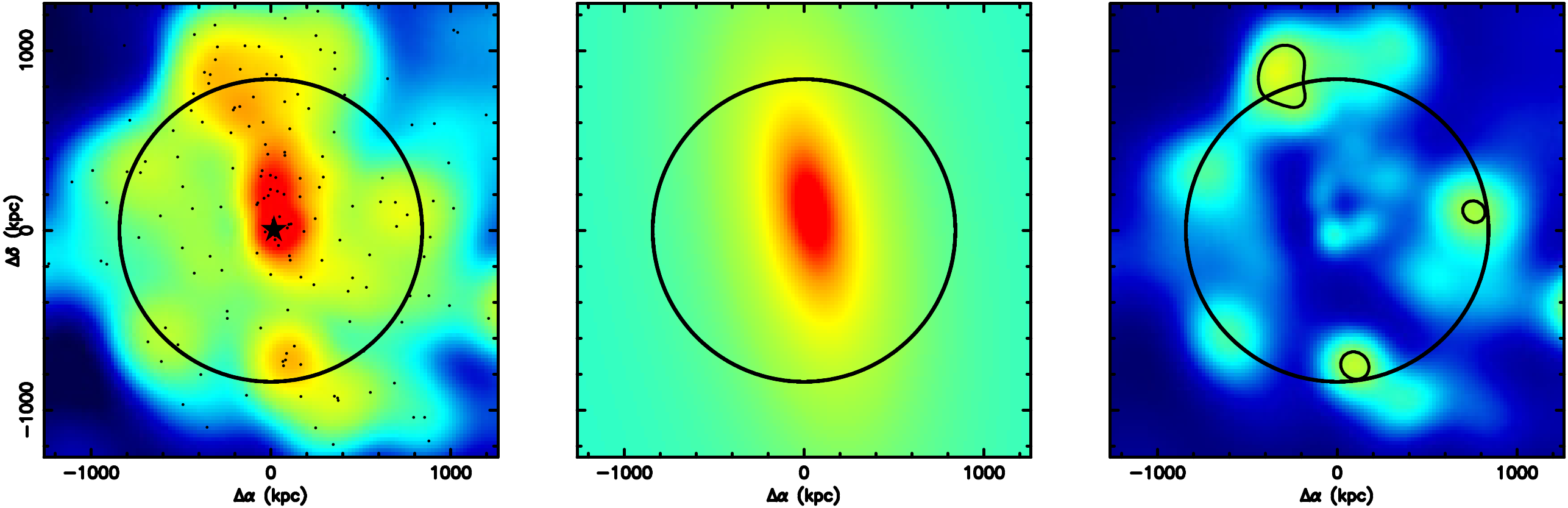}\\[3pt]
\includegraphics[width=5.cm, angle=-90]{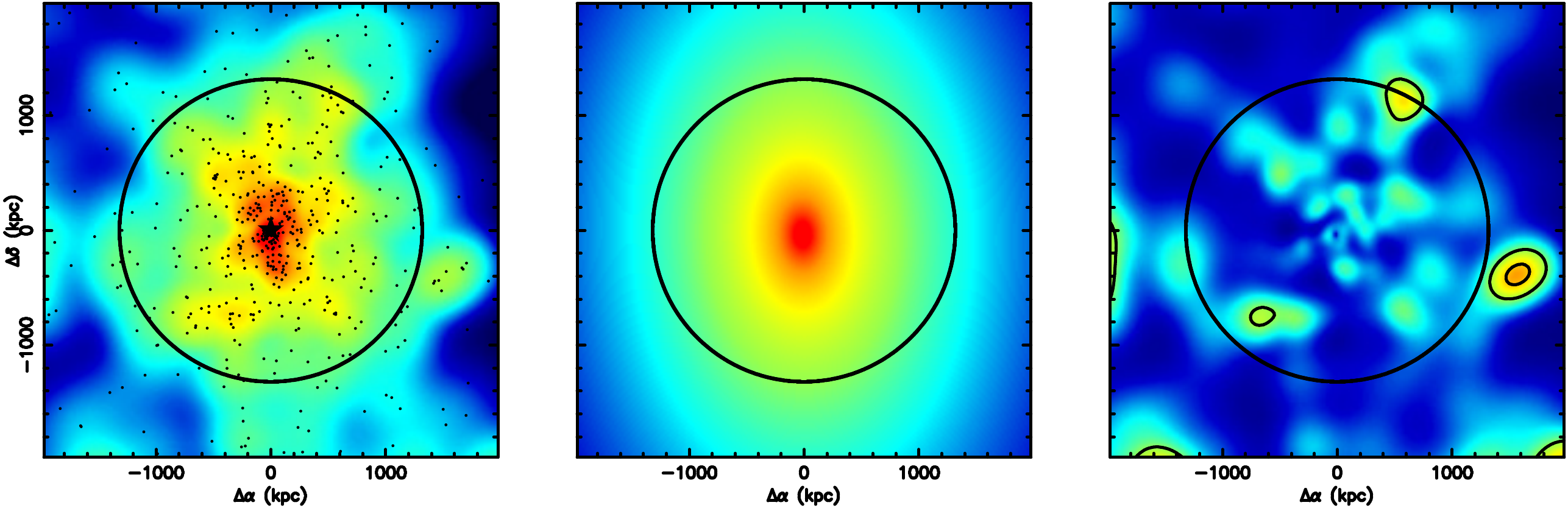}\\[3pt]
\includegraphics[width=5.cm, angle=-90]{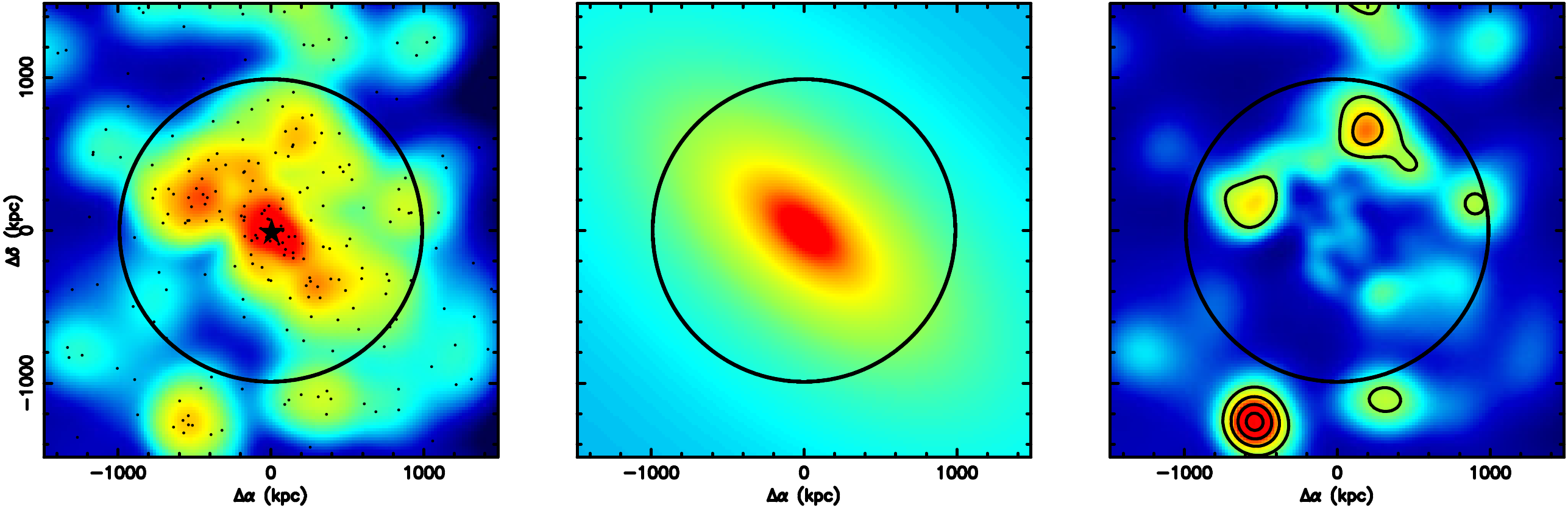}
\caption{Same as Figure \ref{fig:Nmaps}, for RXCJ1236.7-3354, RXCJ1302.8-0230, RXCJ1311.4-0120, and RXCJ1516.3+0005 from top to bottom, respectively.}
\end{figure}

\begin{figure}[h]
\center
\includegraphics[width=5.cm, angle=-90]{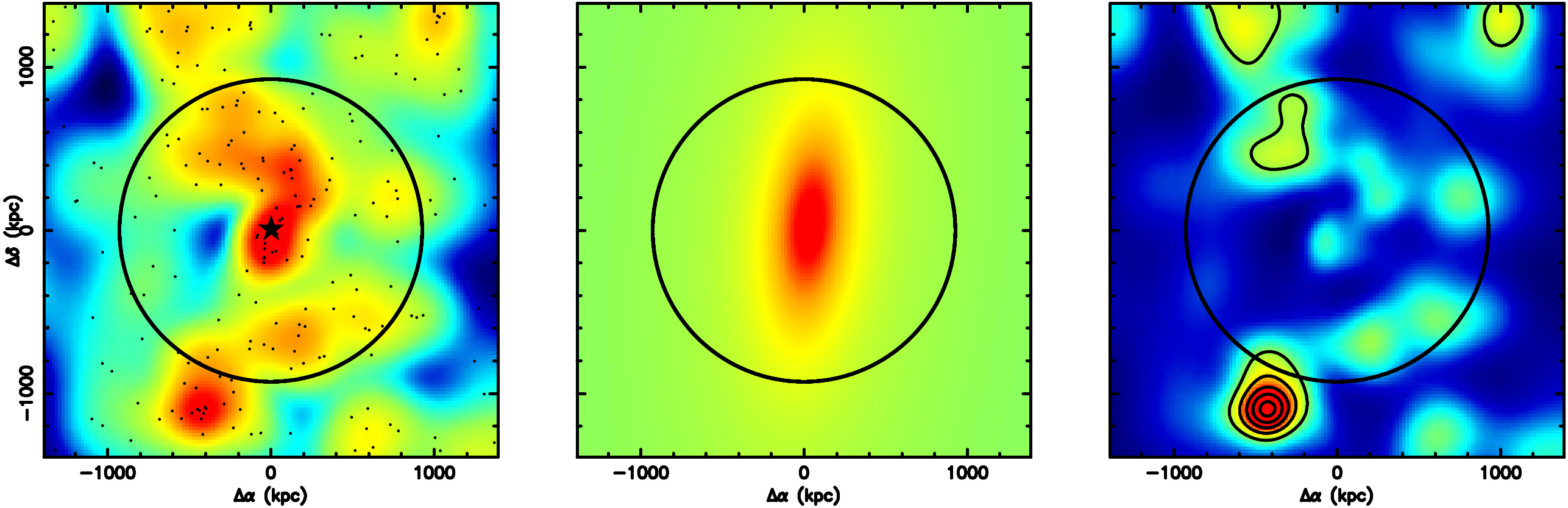}\\[3pt]
\includegraphics[width=5.cm, angle=-90]{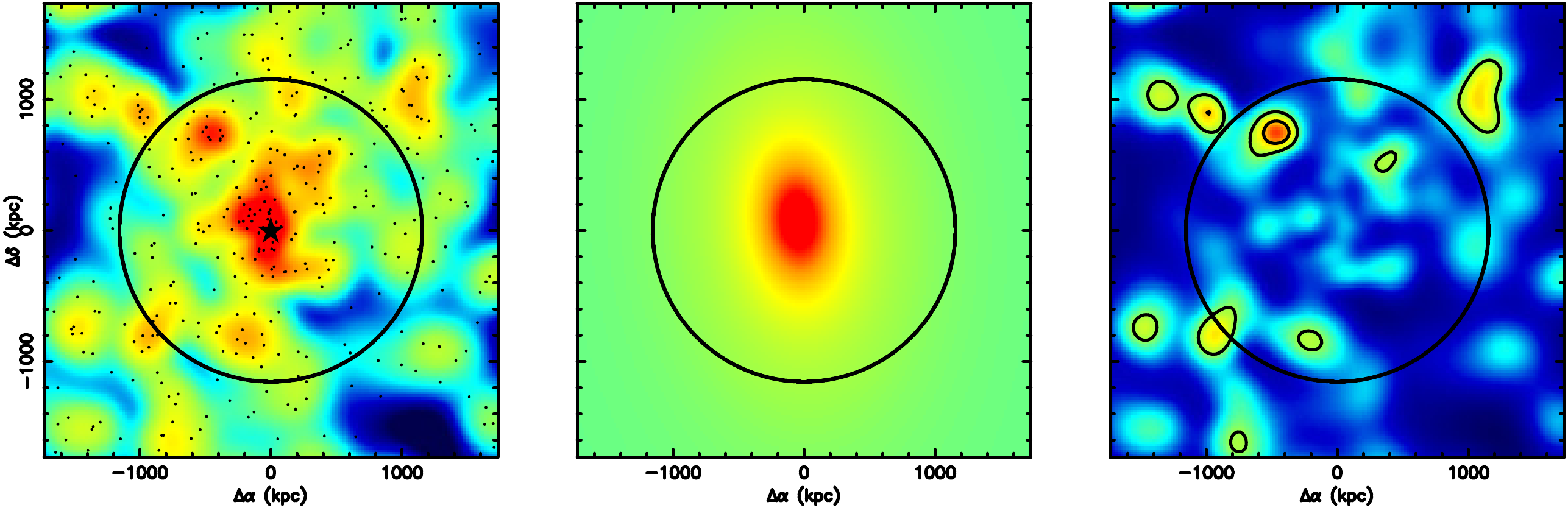}\\[3pt]
\includegraphics[width=5.cm, angle=-90]{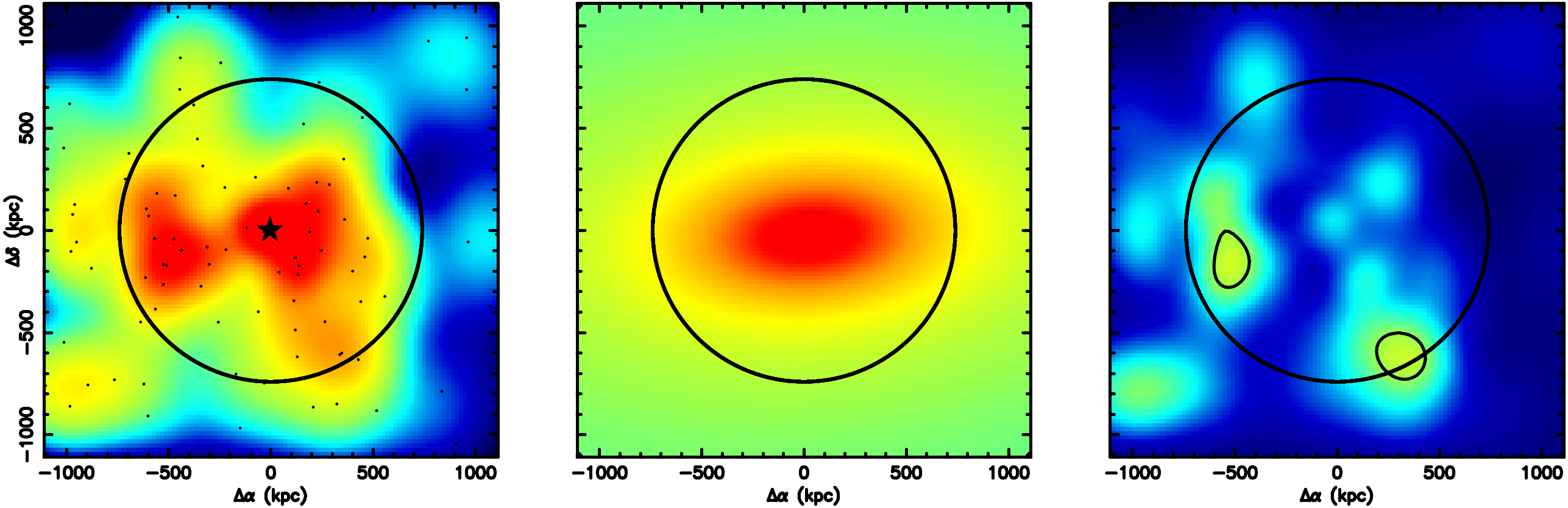}\\[3pt]
\includegraphics[width=5.cm, angle=-90]{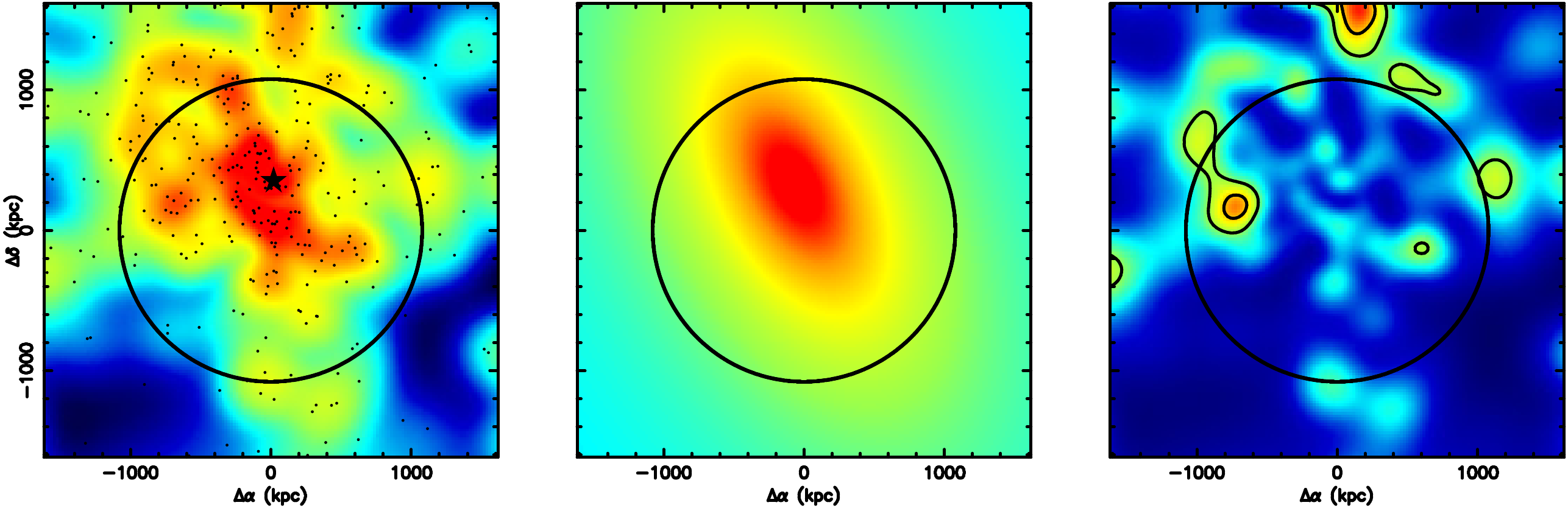}
\caption{Same as Figure \ref{fig:Nmaps}, for RXCJ1516.5-0056, RXCJ2014.8-2430, RXCJ2023.0-2056, and RXCJ2048.1-1750 from top to bottom, respectively.}
\end{figure}

\begin{figure}[h]
\center
\includegraphics[width=5.cm, angle=-90]{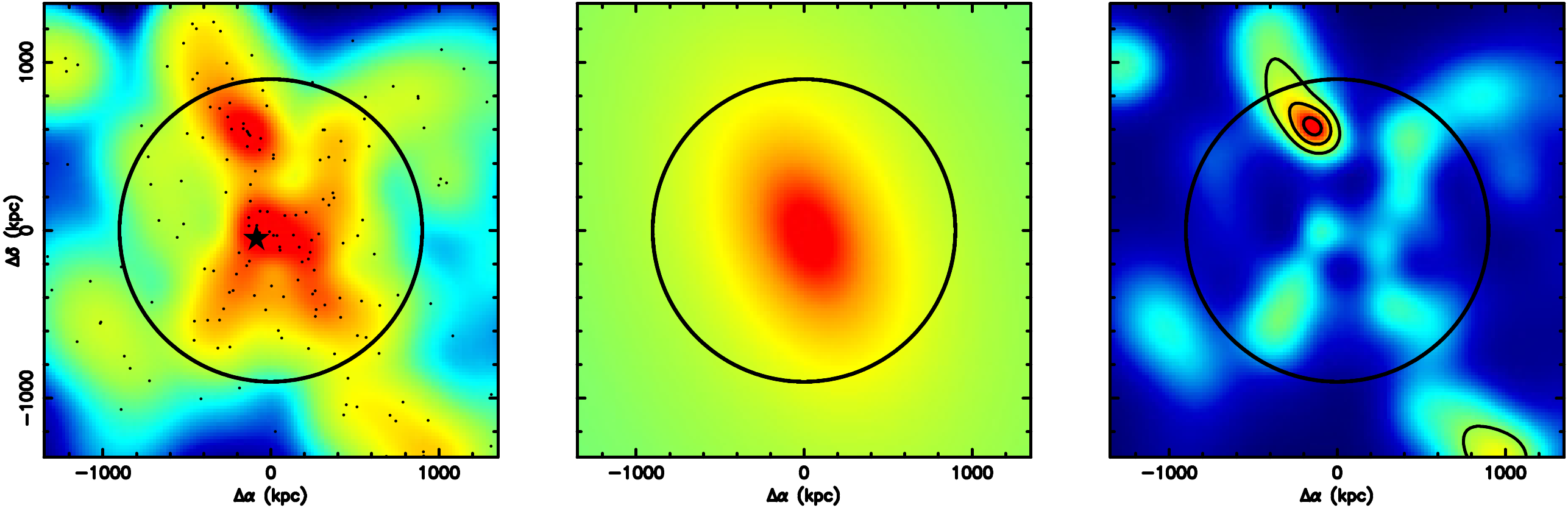}\\[3pt]
\includegraphics[width=5.cm, angle=-90]{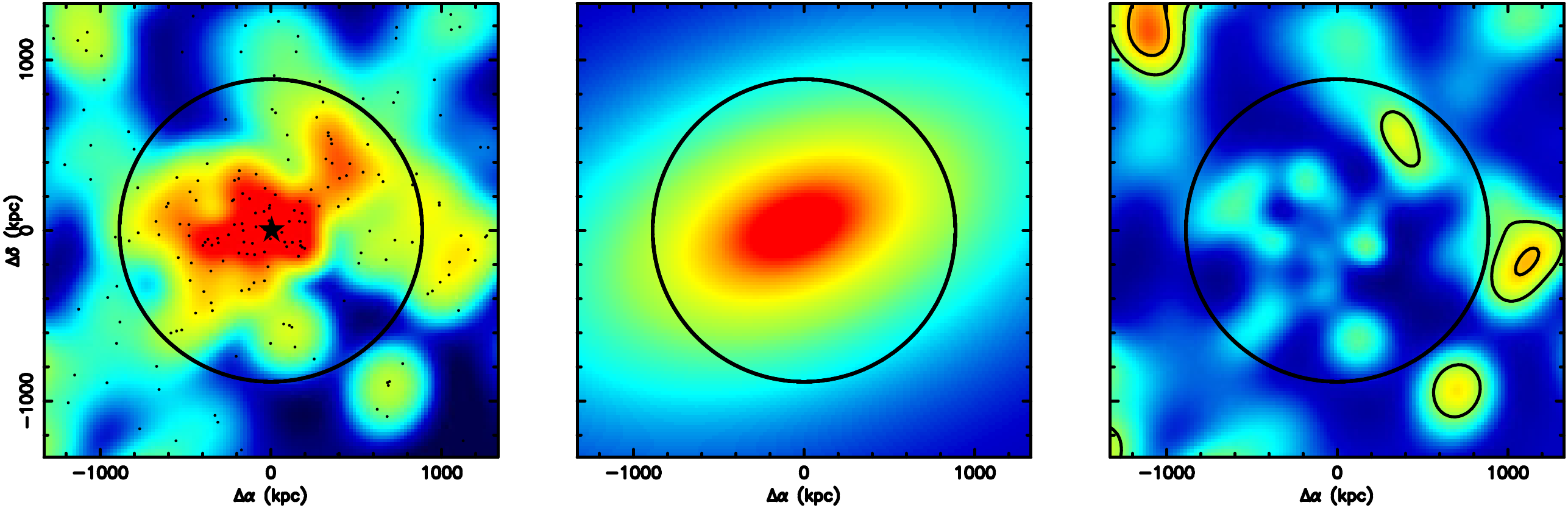}\\[3pt]
\includegraphics[width=5.cm, angle=-90]{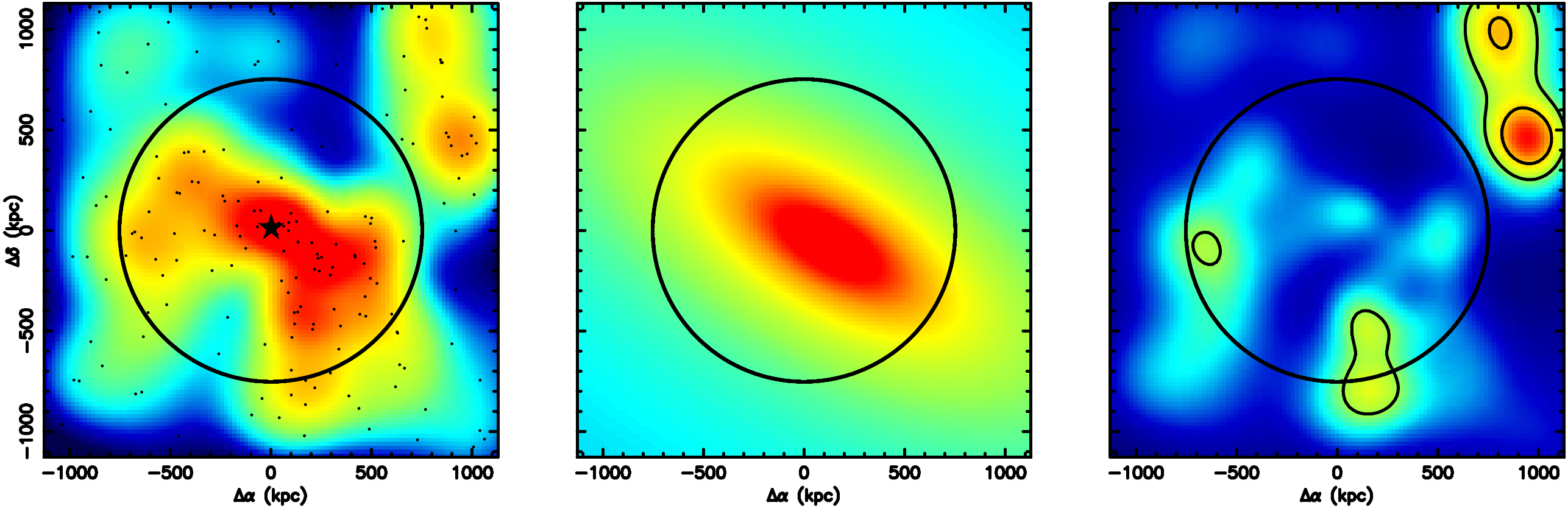}\\[3pt]
\includegraphics[width=5.cm, angle=-90]{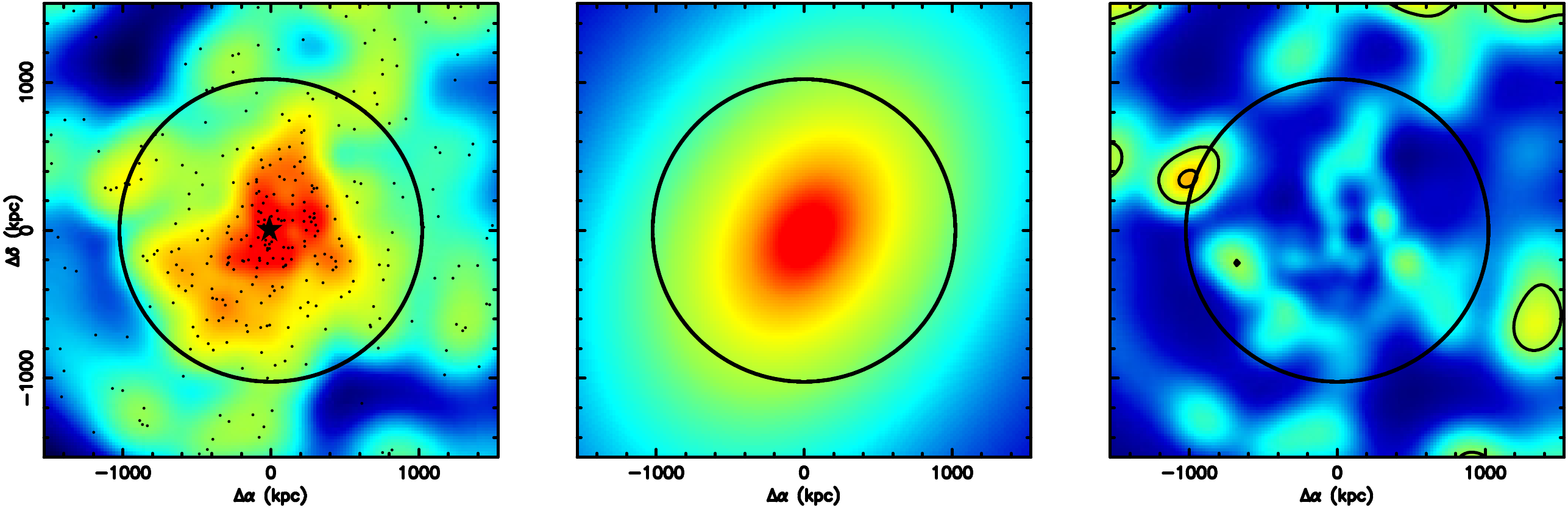}
\caption{Same as Figure \ref{fig:Nmaps}, for RXCJ2129.8-5048, RXCJ2149.1-3041, RXCJ2157.4-0747, and RXCJ2217.7-3543 from top to bottom, respectively.}
\end{figure}

\begin{figure}[h]
\center
\includegraphics[width=5.cm, angle=-90]{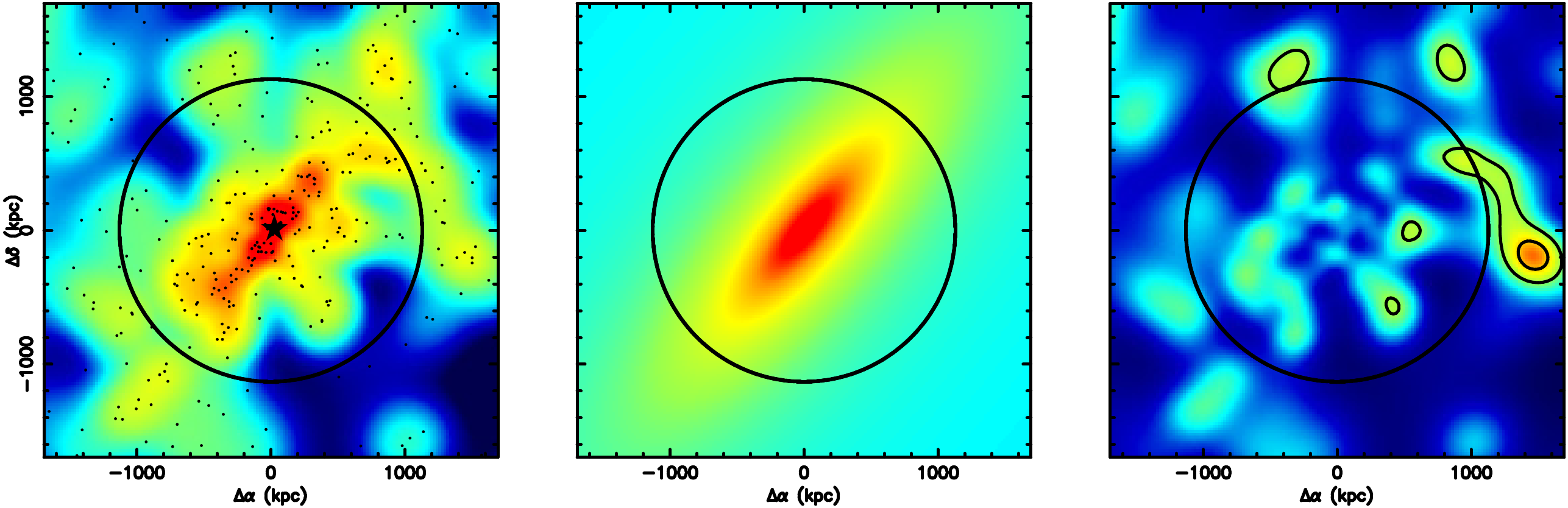}\\[3pt]
\includegraphics[width=5.cm, angle=-90]{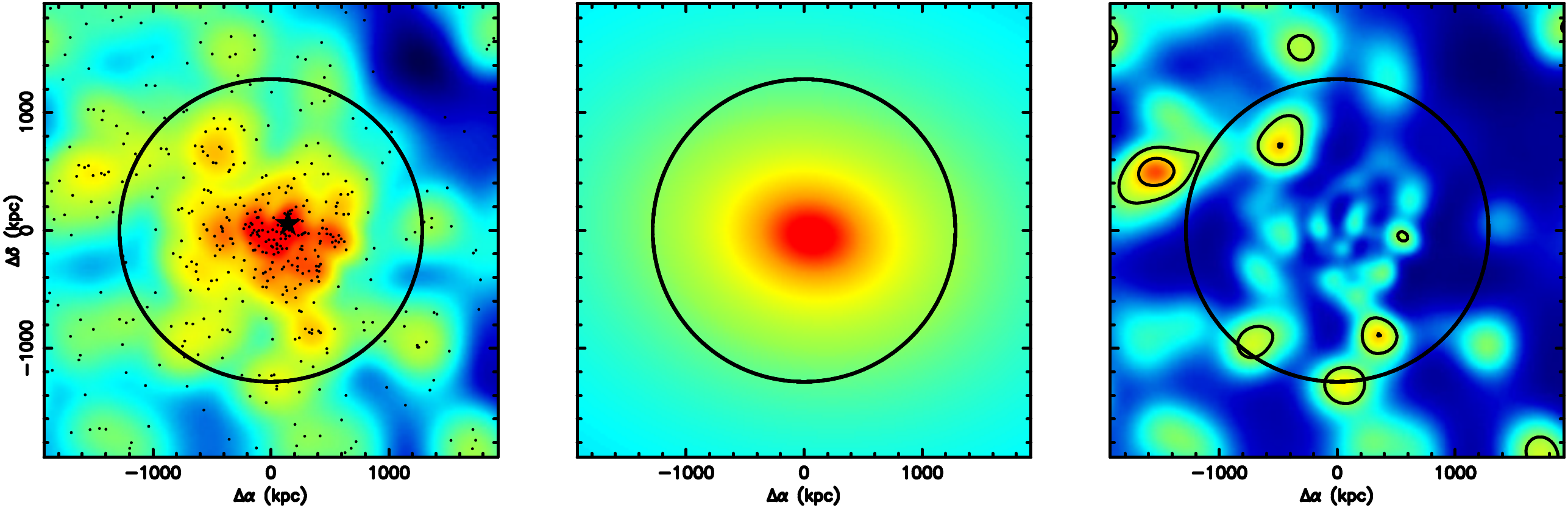}\\[3pt]
\includegraphics[width=5.cm, angle=-90]{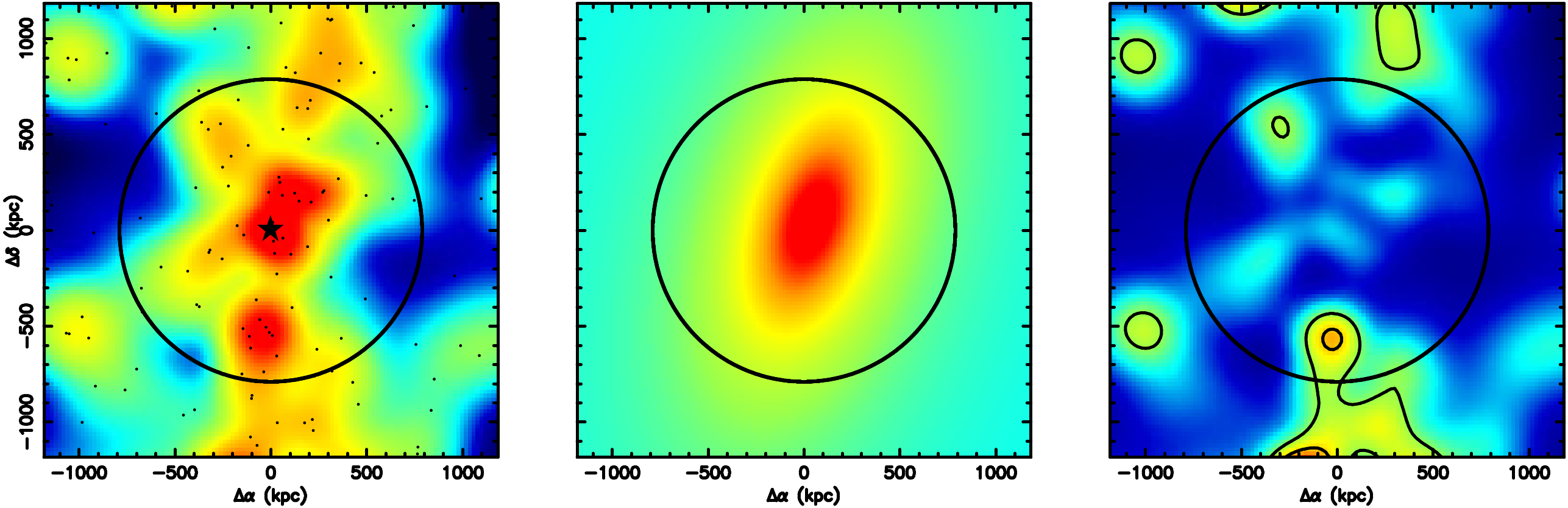}
\caption{Same as Figure \ref{fig:Nmaps}, for RXCJ2218.6-3853, RXCJ2234.5-3744, and RXCJ2319.6-7313 from top to bottom, respectively.}
\end{figure}

\end{document}